\documentclass[
letterpaper,prc,twocolumn,showpacs,floatfix,nofootinbib,preprintnumbers,superscriptaddress,amsmath,amssymb]{revtex4-2}
\usepackage[colorlinks=true, allcolors=blue]{hyperref}
\usepackage{graphicx}
\usepackage{dcolumn}
\usepackage{bm}
\usepackage{braket}
\usepackage{comment}
\usepackage{soul}  
\usepackage{enumitem}
\usepackage{color}
\usepackage{xcolor}
\usepackage{array}
\usepackage{float}
\usepackage{booktabs, xcolor}
\usepackage{colortbl}
\usepackage{multirow}
\usepackage{booktabs}

\newcommand{\gras}[1]{\boldsymbol{#1}}

\newcommand{\rCMF}{\rvec_{\rm cm}^{\rm \mathrm{f}}}
\newcommand{\rvec}{\mathbf{r}}
\newcommand{\jvec}{\mathbf{j}}
\newcommand{\nzmax}{\max\{n_z\}}

\newcommand{\Ecut}{E_{\rm cut}}

\newcommand{\TKE}{\mathrm{TKE}}
\newcommand{\TXE}{\mathrm{TXE}}
\newcommand{\ZF}{Z_{\rm f}}
\newcommand{\NF}{N_{\rm f}}
\newcommand{\AF}{A_{\rm f}}
\newcommand{\ZH}{Z_{\rm H}}
\newcommand{\NH}{N_{\rm H}}
\newcommand{\AH}{A_{\rm H}}
\newcommand{\ZL}{Z_{\rm L}}
\newcommand{\NL}{N_{\rm L}}
\newcommand{\AL}{A_{\rm L}}

\newcommand{\KL}{K_{\rm L}}
\newcommand{\KH}{K_{\rm H}}
\newcommand{\KF}{K_{\rm f}}
\newcommand{\vF}{\mathbf{v}_{\rm f}}
\newcommand{\zcmH}{z_{\rm cm}^{\mathrm{H}}}
\newcommand{\zcmL}{z_{\rm cm}^{\mathrm{L}}}

\newcommand{\EL}{E_{\rm L}}
\newcommand{\EH}{E_{\rm H}}
\newcommand{\ELs}{E_{\rm L}^{*}}
\newcommand{\EHs}{E_{\rm H}^{*}}

\newcommand{\EFs}{E_{\rm f}^{*}}

\newcommand{\EFtot}{E_{\rm f}^{(\rm tot.)}}

\newcommand{\ECtot}{E_{\rm Cou.}^{(\rm tot.)}}

\newcommand{\ECint}{E_{\rm Cou.}^{(\rm int.)}}

\newcommand{\dvol}{d^{3}\rvec}
\newcommand{\volF}{\mathcal{V}_{\rm f}}
\newcommand{\volH}{\mathcal{V}_{\rm H}}
\newcommand{\volL}{\mathcal{V}_{\rm L}}

\begin{document}
\preprint{LLNL-JRNL-2012260}

\title{Excitation energy of fission fragments within nuclear time-dependent density functional theory}

\author{Antonio Bjelčić}
\email{Corresponding author: bjelcic1@llnl.gov}
\affiliation{Nuclear data and Theory Group, Nuclear and Chemical Science Division, Lawrence Livermore National Laboratory, California, USA 94550}

\author{Nicolas Schunck}
\email{schunck1@llnl.gov}
\affiliation{Nuclear data and Theory Group, Nuclear and Chemical Science Division, Lawrence Livermore National Laboratory, California, USA 94550}

\author{Marc Verriere}
\email{marc.verriere@cea.fr}
\affiliation{Nuclear data and Theory Group, Nuclear and Chemical Science Division, Lawrence Livermore National Laboratory, California, USA 94550}
\affiliation{CEA, DAM, DIF, 91297 Arpajon, France}
\affiliation{ Université Paris-Saclay, CEA, Laboratoire Matière en Conditions Extrêmes, 91680 Bruyères-le-Châtel, France}

\begin{abstract}
The number and properties of the neutrons and photons emitted in nuclear fission are directly related to the excitation energy of the fission fragments when they are formed at scission. Though not observable experimentally because of the extremely short time scales, the excitation energy of fission fragments can be predicted by microscopic theory based on time-dependent density functional theory (TDDFT). Initial results on the value of the total kinetic energy of fission reactions were very promising, but could not probe all possible fragmentations. In this work, we perform large-scale TDDFT calculations in $^{240}$Pu enabled by the development of a new TDDFT solver. We obtain TDDFT trajectories covering nearly all possible fragmentations. We find that the total kinetic energy is close to experimental values only for the most likely fission while it is severely underestimated at both small and large asymmetries. This conclusion seems rather independent of the parameterization of the energy functional, both in its particle-hole and particle-particle channels.
\end{abstract}

\date{\today}

\maketitle


\section{Introduction}
\label{sec:intro}
A complete and accurate, end-to-end description of the nuclear fission process from the formation of the fissioning nucleus to the fission products is required for many applications ranging from nuclear astrophysics to nuclear engineering \cite{talou2023nuclear}. For a long time, the theoretical models used to simulate fission were largely phenomenological and heavily calibrated to nuclear data  \cite{schunck2022theory}. Since the early 2000s, the microscopic description of nuclear fission based on nuclear energy density functional theory (DFT) has been undergoing a renaissance \cite{schunck2016microscopic}. Besides providing a fundamental understanding of collective fission dynamics and timescale \cite{simenel2014formation,tanimura2015collective,bulgac2016induced,simenel2018heavyion,ren2022dynamical}, DFT methods can now accurately describe fission product yields \cite{goutte2005microscopic,regnier2016microscopic,scamps2018impact,zhao2021microscopic,zhao2022timedependent,schunck2023microscopic},  provide precise estimate of the number of particles in fission fragments \cite{verriere2019number,verriere2021microscopic}, or predict the characteristics of angular momentum distributions in fission fragments \cite{bulgac2021fission,marevic2021angular,scamps2022microscopic,scamps2023spatial}.

One of the most important ingredients needed to simulate the decay of fission fragment is the excitation energy of fission fragments and the total kinetic energy (TKE) of the fission reaction. The excitation energy directly determines how many prompt neutrons and photons can be emitted, hence has an outsized impact on the average number of neutrons $\bar{\nu}$ in the reaction, which in turn is a critical input to nuclear reactor technology. The standard way of estimating the energy of fission fragments relies on inferring the total excitation energy (TXE) from the measured values of TKE and invoking an energy sharing mechanism to distribute this TXE among the fragments (see \cite{schunck2022theory} for a review). 

At the same time, it is possible to directly compute the TKE and the excitation energy of fission fragments from time-dependent Hartree-Fock (TDHF) and Hartree-Fock-Bogoliubov (TDHFB) simulations. The initial results reported in \cite{bulgac2016induced,bulgac2019fission} and based on simulations of most likely fission were extremely promising, since TKE values were within a few MeV of the experimental data and predictions did not depend much on the parameterization of the energy functional. However, larger-scale calculations by the Beijing-Zagreb collaboration, which could probe more asymmetric fission configurations, pointed to significantly too small TKE values for asymmetric fission \cite{ren2022microscopic}, even though predictions for most likely fission remained in decent agreement with the data. This result was confirmed in a recent study based on configuration mixing of TDHFB trajectories \cite{li2025microscopic}. While the theoretical framework employed in \cite{li2025microscopic} represents the state of the art, it remained possible that the discrepancies with experimental data were caused by the relatively small number of TDHFB trajectories or the underlying features of the energy functional.

The goal of this paper is thus to clarify the predictive power of TDHFB for the excitation energy of fission fragments. We implemented several algorithmic improvements in a new TDHFB solver to enable fast and numerically well converged calculation of hundreds of trajectories per fissioning nucleus. For the first time, we present TDHFB solutions spanning the entire mass range of fission fragmentations. For the benchmark case of $^{240}$Pu, we confirm that TDHFB can reproduce experimental values of TKE for most likely fragmentation within a couple of MeV, while it underestimates the experimental values by up to 10-20 MeV for both smaller and larger asymmetries. We also prove that these conclusions are largely independent of the form of the energy functional and of the nature of pairing correlations, and are only marginally improved by modifying the time-odd channel of the energy functional.

Section \ref{sec:implementation} gives the details of the numerical implementation of our new TDHFB solver, especially the determination of the basis states and some of the algorithmic optimization techniques. A comprehensive study of the numerical convergence of the results, both static and time-dependent, is presented in Section \ref{sec:conv}. Section \ref{sec:tke} contains predictions of TKE for two different Skyrme energy functionals as well as a discussion of the possible origin of the discrepancy with experimental data. The excitation energy of fission fragments is discussed in Section \ref{sec:excitation}. Additional discussions realted to the convergence of calculations are provided in Appendix \ref{app:box}-\ref{app:continuity}.


\section{TDHFB Solver}
\label{sec:implementation}

This section outlines the numerical implementation and technical aspects of the newly developed \texttt{AxialHOHFB} solver for both static and time-dependent HFB solutions. We emphasize that axial symmetry is the only symmetry conserved throughout the calculations.


\subsection{Basis definition}
\label{subsec:BasisDef}

Both the static and time-dependent components of \texttt{AxialHOHFB} are implemented in the harmonic oscillator (HO) basis. However, the selection of basis states is different from traditional conventions adopted, e.g., in \texttt{HFBTHO} \cite{stoitsov2005axially} or \texttt{HFODD} \cite{dobaczewski1997solution}. In addition, the code implements numerous important algorithmic improvements that were essential to perform the high-precision simulations.


\subsubsection{Basis frequency and box size in one dimension}

Let $\phi_n^{(b)}(x)$ denote the $n$-th eigenfunction of the one-dimensional quantum harmonic oscillator with oscillator length $b$
\begin{equation}\label{Eq:phin1DbasisDef}
    \phi_n^{(b)}(x) = \frac{1}{\sqrt{b}} \frac{1}{\sqrt{\sqrt{\pi}2^n n!}} H_n(x/b) \exp{\left(-\frac{1}{2}(x/b)^2\right)}.
\end{equation}
For the harmonic potential $V(x)=\frac{1}{2}m\omega^2 x^2$ with eigen-energies $E_n=\hbar \omega (n+1/2)$, the classical turning point $\tilde{x}_n$ is given by the condition $V(\tilde{x}_n) = E_n$, i.e.
\begin{equation}
    \tilde{x}_n = \sqrt{\frac{\hbar}{m\omega}} \sqrt{2n+1} = b\sqrt{2n+1} \approx b\sqrt{2n}.
\end{equation}
This means that $\phi_n^{(b)}(x)$ is essentially nonzero only within a finite interval $\left[-b\sqrt{2n},+b\sqrt{2n}\,\right]$.

It is common practice to set the oscillator length $b$, either to a fixed value such as the shell model estimate $\hbar\omega = 41 A^{-1/3}\,\mathrm{MeV}$ or to a value dependent on the constraints on multipole moments, and use the \(N+1\) eigenfunctions \(\left\{ \phi_0^{(b)}(x), \phi_1^{(b)}(x), \ldots, \phi_N^{(b)}(x) \right\}\) with lowest energy as a truncated HO basis. The underlying assumption is that increasing \(N\) leads to better convergence of the basis expansion. However, this approach may converge slowly when approximating functions that exhibit features sharper than the characteristic length scale set by \(b\).
As \(N\) increases, for fixed $b$, the range over which the basis can represent functions accurately expands roughly as \(\big[ -b\sqrt{2N}, +b\sqrt{2N} \,\big]\). Nevertheless, in many applications, one can often estimate in advance a fixed spatial region where all relevant physics takes place, making it advantageous to adapt the basis accordingly.

\begin{figure*}[!htbp]
    \centering
    \includegraphics[width=1.0\textwidth]{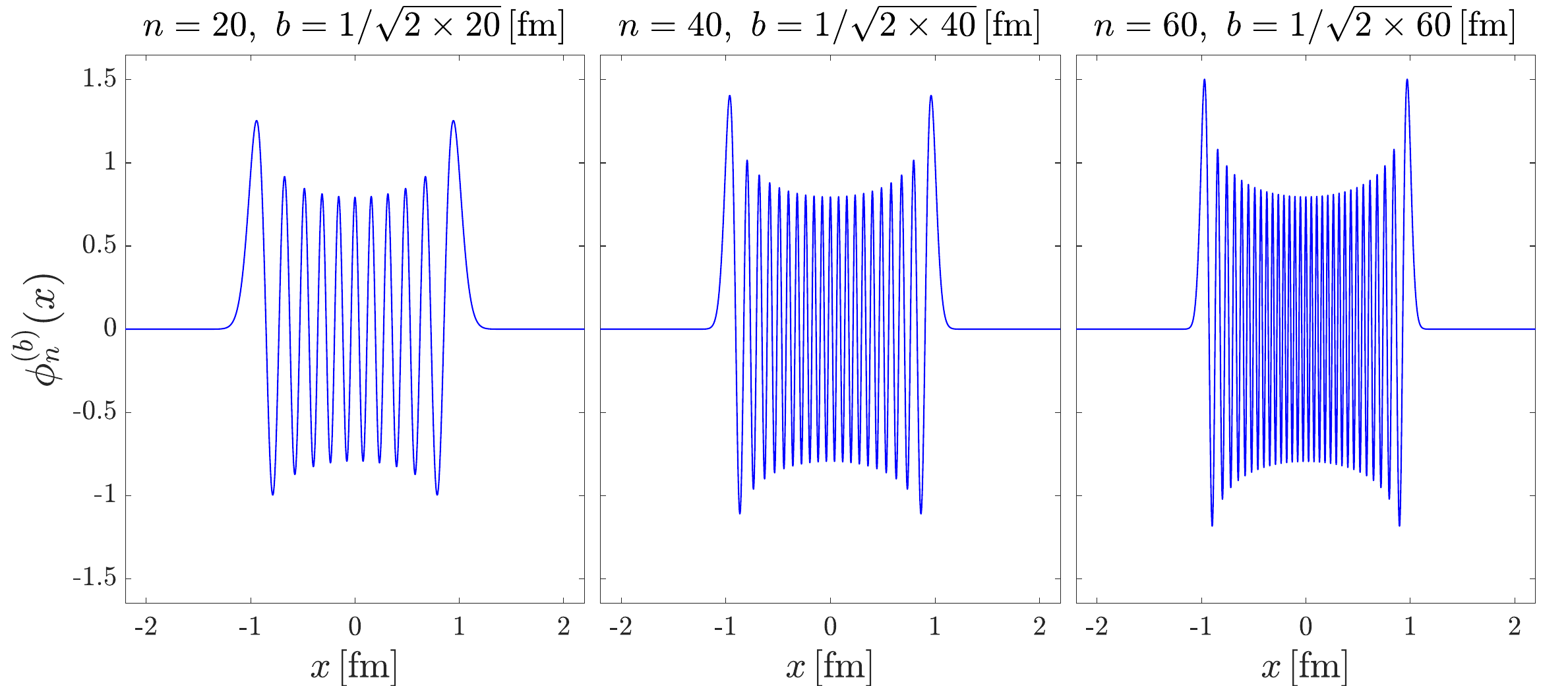}
    \caption{Harmonic oscillator basis functions $\phi_n^{(b)}(x)$ where the oscillator length $b$ is adjusted as the orders of harmonics $n$ are increased in such a way that all the basis functions are contained within an interval $[-1\,\mathrm{fm},+1\,\mathrm{fm}\,]$.}
    \label{fig:HObasis}
\end{figure*}

In our approach, we  thus first fix the box size $X_{\mathrm{box}}$, and for each $N$, we adjust the oscillator length $b_N$ such that $X_{\mathrm{box}} = b_N \sqrt{2N}$.
Therefore, increasing \(N\) is accompanied by a corresponding decrease in the oscillator length \(b_N\), thereby enhancing the resolution of the basis. This is analogous to Fourier analysis, where including higher-frequency modes improves the ability to approximate rapidly varying functions.
Figure \ref{fig:HObasis} illustrates an example where we fix $X_{\mathrm{box}}=1\,\mathrm{fm}$ and adjust accordingly oscillator length for $N+1=20, 40$ and $60$.

In the (TD)HFB theory, the physical quantities we want to contain within a box are expanded as a series of product of basis functions. For example, in the one-dimensional case the local density $\rho(x)$ is given by
\begin{equation}
    \rho(x) = \sum_{n_1 n_2=0}^N \rho_{n_1 n_2}^* \phi_{n_1}^{(b)*} \phi_{n_2}^{(b)}(x) \delta_{\sigma_1 \sigma_2}.
\end{equation}
It is well known that the product of two HO basis functions can be written as a linear combination of the same HO basis functions only with reduced oscillator length $\tilde{b} = b/\sqrt{2}$ \cite{talman1970properties}. 
More precisely, for any $0\leq n_1,n_2\leq N$,
\begin{equation}\label{eq:AxialTM}
    \phi_{n_1}^{(b)}(x) \phi_{n_2}^{(b)}(x) = \sum_{n=0}^{\widetilde{N}} c_n^{n_1n_2} \phi_{n}^{(\tilde{b})}(x),
\end{equation}
where $\widetilde{N} = 2N$, and some coefficients $c_n^{n_1n_2}$.
Therefore the density $\rho({x})$ will be essentially nontrivial in an interval $[-\widetilde{X}_{\mathrm{box}},+\widetilde{X}_{\mathrm{box}}]$, where
\begin{equation}
    \widetilde{X}_{\mathrm{box}} = \tilde{b} \sqrt{2\widetilde{N}} = {b} \sqrt{2{N}}.
\end{equation}
Consequently, if we want to contain the density within a box of size $X_{\mathrm{box}}$, the rule for changing the oscillator length as a function of the basis size $N$ is still given by the relation \(b_N=X_{\mathrm{box}}/\sqrt{2N}\) that applies to the basis functions. A similar reasoning holds for all other types of densities (e.g. spin density or kinetic density), currents and their derivatives, which appear in the Skyrme density functional.


\subsubsection{Axially symmetric case}

Motivated by the one-dimensional example, we now focus on the axially symmetric HO basis functions: $\phi_{n_z}^{(b_z)}(z) \phi_{n_r \Lambda}^{(b_r)}(r_\perp) \frac{1}{\sqrt{2\pi}}\exp{\left({i}\Lambda\varphi\right)}$ with oscillator lengths $b_z$ and $b_r$ in the axial and radial directions respectively. The longitudinal basis function $\phi_{n_z}^{(b_z)}(z)$ takes the same form \eqref{Eq:phin1DbasisDef} as in the one-dimensional case, while the radial basis function reads
\begin{multline}
    \phi_{n_r \Lambda}^{(b_r)}(r_\perp) = \frac{\sqrt{2}}{b_r} \sqrt{ \frac{n_r!}{(n_r+|\Lambda|)!} } \left(\frac{r}{b_r}\right)^{|\Lambda|} \\
    L_{n_r}^{|\Lambda|}((r/b_r)^2) \exp\left( -\frac{1}{2} \left(r/b_r\right)^2 \right).
\end{multline}
The one-body local density of an axially-symmetric system is $\rho(\rvec)=\rho(z,r_\perp)$ and involves products of two longitudinal basis functions and two radial basis functions with equal $\Lambda$ indices,
which can be expanded in a similar way as in Eq. \eqref{eq:AxialTM}
\begin{multline}
\label{eq:perpTM}
\delta_{\Lambda_1\Lambda_2} \phi^{(b_r)}_{n_{r_1}\Lambda_1}(r_\perp)
\phi^{(b_r)}_{n_{r_2}\Lambda_2}(r_\perp) \\
=
\sum_{N_r=0}^{\max\{2n_r+|\Lambda|\}} c_{N_r}^{n_{r_1}\Lambda_1 n_{r_2}\Lambda_2} \phi_{N_r 0}^{(b_r/\sqrt{2})}(r_\perp).
\end{multline}
This effectively means that the density $\rho(z,r_\perp)$ is expanded in terms of functions $\phi_{N_z}^{(b_z/\sqrt{2})}(z)\phi_{N_r 0}^{(b_r/\sqrt{2})}(r_\perp)$, for $0\leq N_z \leq 2\nzmax$ and $0\leq N_r \leq \max\{2n_r+|\Lambda|\}$.
Similarly to the previous 1D analysis, one can show that those functions
are non-trivial only for $(z,r)\in [-Z_{\mathrm{box}},+Z_{\mathrm{box}}] \times [0,R_{\mathrm{box}}]$,
where
\begin{align}
    Z_{\mathrm{box}} & =  b_z \sqrt{2\nzmax},\\
    R_{\mathrm{box}} & =  b_r \sqrt{2\max\{2n_r+|\Lambda|\}}.
\end{align}
The same conclusion holds for all
other densities, currents and their derivatives which appear in the Skyrme density functional.

In order to make the number of nodes of basis functions per unit length equal in the longitudinal and radial directions, we impose the equality
\begin{equation}
    \max\{2n_r+|\Lambda|\} = \left\lceil \nzmax \frac{R_{\mathrm{box}}}{Z_{\mathrm{box}}} \right\rceil.
\end{equation}
where as usual $\lceil \cdot \rceil$ denotes the ceiling function.
We can then set the oscillator lengths in each direction according to
\begin{align}
    b_z &= {Z_{\mathrm{box}}}/{\sqrt{2 \nzmax }}, \label{eq:bz} \\
    b_r &= {R_{\mathrm{box}}}/{\sqrt{2 \max\{2n_r+|\Lambda|\} }}, \label{eq:br}
\end{align}
and include in a truncated HO basis all the functions $\phi_{n_z}^{(b_z)}(z) \phi_{n_r \Lambda}^{(b_r)}(r_\perp) \frac{1}{\sqrt{2\pi}}\exp{\left({i}\Lambda\varphi\right)}$ which satisfy
\begin{equation}
\begin{gathered}
    0 \leq n_z \leq \nzmax \\
    0 \leq 2n_r+|\Lambda| \leq \max\{2n_r+|\Lambda|\}.
\end{gathered}
\end{equation}
In this framework, the products of two basis functions, hence all spatial densities and currents, are mostly confined within a cylinder of height \( 2Z_{\mathrm{box}} \) and diameter \( 2R_{\mathrm{box}} \). As the parameter \( \nzmax \) increases, higher oscillator harmonics are included in the expansion. The corresponding decrease in oscillator lengths enables improved resolution, allowing for more accurate approximation of sharp features in the density and current profiles.
The three parameters that uniquely define our truncated HO basis are therefore $Z_{\mathrm{box}}, R_{\mathrm{box}}$ and $\nzmax$. 


\subsection{Time evolution}
\label{subsec:TimeEvolution}

The real-time dynamics of a superfluid many-body system is described by the time-dependent Hartree-Fock-Bogoliubov (TDHFB) equation \cite{blaizot1985quantum},
\begin{equation}\label{eq:TDHFB}
    i\hbar \, \partial_t 
    \begin{bmatrix}
    U \\ V
    \end{bmatrix} = 
    \begin{bmatrix}
        h & \Delta \\ -\Delta^* & -h^*
    \end{bmatrix}
    \begin{bmatrix}
    U \\ V
    \end{bmatrix},
\end{equation}
where $U$ and $V$ are the quasiparticle wavefunctions, $h$ is the single-particle Hamiltonian, and $\Delta$ is the pairing field. This equation captures both the mean-field evolution and pairing correlations in a unified formalism. The TDHFB equation and many of its properties have been discussed extensively in the literature; see for example \cite{blaizot1985quantum,avez2008pairing,bulgac2008timedependent,ebata2010canonicalbasis,scamps2012pairing,nakatsukasa2012density,bulgac2013timedependent,jin2021lise}.

Here we discuss an important technical detail of implementing the TDHFB equation with zero-range pairing forces that is often overlooked. It is well known that contact pairing interactions lead to divergences both in nuclear matter simulations, and in calculations of finite nuclei with increasing basis sizes \cite{dobaczewski1984hartreefockbogolyubov,bruun1999bcs,bulgac2002renormalization,borycki2006pairing}. These divergences can be fixed by ``regularizing'' the pairing field \cite{bulgac2002renormalization} or by imposing an energy cutoff $\Ecut$ to filter the quasiparticles in the calculation of the densities. In practice, when computing the density matrix and pairing tensor
\begin{align}
\rho & = V^* V^T, \\
\kappa & = V^* U^T,
\end{align}
from the unitary Bogoliubov matrix $\mathcal{W}$
\begin{equation}
    \mathcal{W} = \begin{bmatrix}
        U & V^* \\ V & U^*
    \end{bmatrix},
\end{equation}
we retain only the columns of $U$ and $V$ corresponding to quasiparticles below $\Ecut$. These truncated matrices are denoted $\widetilde{U}$ and $\widetilde{V}$. Together they form a truncated Bogoliubov matrix $\widetilde{\mathcal{W}}$. Although $\widetilde{\mathcal{W}}$ still has orthonormal columns, i.e., $\widetilde{\mathcal{W}}^\dagger \widetilde{\mathcal{W}} = I$, it is no longer a projector onto the full space: $\widetilde{\mathcal{W}} \widetilde{\mathcal{W}}^\dagger \neq I$, since some part of the quasiparticle space is omitted due to the cutoff. As a consequence, the energy is not conserved during the TDHFB evolution.

To show this, we recall that the generalized, time-dependent density matrix is defined as
\begin{equation}
    \mathcal{R} = \mathcal{W}
    \begin{bmatrix}
        0 & 0 \\ 0 & I
    \end{bmatrix}
    \mathcal{W}^\dagger.
\end{equation}
The unitarity of the Bogoliubov matrix $\mathcal{W}$ leads to the identities $U^* V^T = - (V^* U^T)^* = -\kappa^*$ and $U^* U^T = I - (V^* V^T)^* = I - \rho^*$ and, therefore, to the well-known result 
\begin{equation} 
\label{eq:GeneralizedDensityMatrix}
    \mathcal{R} = \begin{bmatrix}
        V^* V^T & V^* U^T \\
        U^* V^T & U^* U^T
    \end{bmatrix}
    = \begin{bmatrix}
        \rho & \kappa \\
        -\kappa^* & I - \rho^*
    \end{bmatrix}.
\end{equation}
The variation of the energy can be written as \cite{blaizot1985quantum}
\begin{equation}
    \delta E = \frac{1}{2} \mathrm{Tr}\left[ \mathcal{H} \, \delta \mathcal{R} \right], \quad
    \text{where} \quad
    \mathcal{H} =
    \begin{bmatrix}
        h & \Delta \\
        -\Delta^* & -h^*
    \end{bmatrix}.
    \label{eq:EnergyBlaizotRipka}
\end{equation}
The TDHFB equation \eqref{eq:TDHFB} can be compactly written as $i\hbar\, \partial_t \mathcal{W} = \mathcal{H} \mathcal{W}$, yielding the time evolution for the generalized density matrix
\begin{equation}
i\hbar\, \partial_t \mathcal{R} = [\mathcal{H}, \mathcal{R}],
\end{equation}
which, in turn, implies
\begin{equation}
    i\hbar\, \partial_t E = \frac{1}{2} \mathrm{Tr}\left[ \mathcal{H} \, i\hbar\, \partial_t \mathcal{R} \right] = \frac{1}{2} \mathrm{Tr}\left[ \mathcal{H} [\mathcal{H}, \mathcal{R}] \right] = 0,
\end{equation}
i.e., the energy is conserved during the time evolution.

The time evolution of the truncated Bogoliubov matrix $\widetilde{\mathcal{W}}$ still reads $i\hbar\, \partial_t \widetilde{\mathcal{W}} = \mathcal{H} \widetilde{\mathcal{W}}$, and the generalized density matrix $\widetilde{\mathcal{R}}$,
\begin{equation}
    \widetilde{\mathcal{R}} = 
    \begin{bmatrix}
        \widetilde{V}^* \widetilde{V}^T & \widetilde{V}^* \widetilde{U}^T \\
        \widetilde{U}^* \widetilde{V}^T & \widetilde{U}^* \widetilde{U}^T
    \end{bmatrix},
\end{equation}
still satisfies
\begin{equation}
i\hbar\, \partial_t \widetilde{\mathcal{R}} = [\mathcal{H}, \widetilde{\mathcal{R}}].
\end{equation}
However: $\widetilde{U}^* \widetilde{V}^T \neq - (\widetilde{V}^* \widetilde{U}^T)^* = -\kappa^*$ and $\widetilde{U}^* \widetilde{U}^T \neq I - (\widetilde{V}^* \widetilde{V}^T)^*=I-\rho^*$. As a result, Eq.~\eqref{eq:GeneralizedDensityMatrix} is no longer valid, and  thus Eq.~\eqref{eq:EnergyBlaizotRipka} can only be approximately written as
\begin{equation}
\delta E 
\approx \frac{1}{2} \mathrm{Tr}\left[ \mathcal{H} \, \delta \widetilde{\mathcal{R}} \right]
.
\end{equation}
This implies that $\partial_t E \approx 0$ and, therefore, that the energy is only approximately conserved during the time evolution.

In our time-dependent calculations, we initialize the evolution with a truncated matrix $\widetilde{\mathcal{W}}$ containing only those quasiparticles below the static cutoff energy, and evolve only this fixed subset throughout the dynamic phase. As we will show in Sec.\ref{subsec:conv_TDHFB}, introducing a quasiparticle cutoff $\Ecut = 60\ \mathrm{MeV}$ results in typical energy conservation within $100\ \mathrm{keV}$, indicating that high-energy quasiparticles do not significantly impact the fission dynamics. Conversely, when we retain all quasiparticles (i.e., no cutoff), we observe energy conservation at the sub-eV level.


\subsection{Algorithmic improvements}
\label{subsec:Algo}

Our new \texttt{AxialHOHFB} code has been extensively optimized and parallelized to maximize computational efficiency.
The time-evolution pipeline has been carefully designed to preserve numerical stability and accuracy at every stage of the computation.
Instead of the commonly used Gauss–Hermite and Gauss–Laguerre quadrature rules, we employed a composite Gauss–Legendre quadrature with several hundreds of quadrature nodes in both the axial and radial directions.

Numerous algorithmic improvements have been implemented to enhance computational efficiency and numerical accuracy. While a detailed discussion lies beyond the scope of this paper, we provide a brief overview of the most significant advancements.


\subsubsection{Coulomb potential}

Special attention has been devoted to the accurate computation of the energy and mean field coming from the Coulomb direct potential,
\begin{equation}\label{Eq:CoulombVC}
    V_C(\rvec) = \hbar c \alpha \int\dvol'\, \frac{\rho^{(p)}(\rvec')}{|\rvec - \rvec'|},
\end{equation}
since the Coulomb force represents a crucial component for all fission simulations. Previous studies  highlighted that the standard substitution method to deal with the $1/|\rvec - \rvec'|$ term of the Coulomb potential \cite{girod1983triaxial} remains plagued by a numerical error caused by the singularity of the original potential \cite{stoitsov2013axially}. 

However, we notice that we actually never need to evaluate $V_C(\rvec)=V_C(z,r_\perp)$, rather we only need the coefficients
\begin{equation}\label{Eq:Vnznr_def}
    [V_C]_{N_z N_r} = \int \dvol \, V_C(z,r_\perp) \,\phi_{N_z}^{\left(\frac{b_z}{\sqrt{2}}\right)}(z) \phi_{N_r 0}^{\left(\frac{b_r}{\sqrt{2}}\right)}(r_\perp).
\end{equation}
To demonstrate this claim, recall that, in practical HFB calculations, the Coulomb potential appears only when computing the Coulomb direct energy
    \begin{equation}\label{Eq:EC}
        E_\mathrm{C} = \frac{1}{2} \int \dvol \,V_C(\rvec)\rho^{(p)}(\rvec),
    \end{equation}
or when evaluating the mean-field matrix elements
    \begin{equation}\label{Eq:VcintElements}
        V_{k_1 k_2}^{(\rm Cou.)} = \int \dvol\, \phi_{k_1}^*(\rvec) V_C(\rvec) \phi_{k_2}(\rvec).
    \end{equation}

As we have seen in Sec.~\ref{subsec:BasisDef}, products of two basis functions can be reduced by using HO functions of reduced oscillator length, for the $z$-dependent term with \eqref{eq:AxialTM} and for the perpendicular term with \eqref{eq:perpTM}. For example, the axially-symmetric proton density can be written as
\begin{equation}
\label{eq:rhoExpansion}
    \rho^{(p)}(z,r_\perp) = \sum_{N_z=0}^{\max N_z} \sum_{N_r=0}^{\max N_r }
    [\rho^{(p)}]_{N_z N_r} \phi_{N_z}^{\left(\frac{b_z}{\sqrt{2}}\right)}(z) \phi_{N_r 0}^{\left(\frac{b_r}{\sqrt{2}}\right)}(r_\perp) ,
\end{equation}
for some coefficients $[\rho^{(p)}]_{N_z N_r}$, where $\max N_z=2\nzmax$ and $\max N_r = \max\{ 2n_r + |\Lambda| \}$.
Plugging Eq.~\eqref{eq:rhoExpansion} into \eqref{Eq:EC} implies that the Coulomb energy can be written as
\begin{equation}
    E_C = \frac{1}{2}\sum_{N_z,N_r} [V_C]_{N_z N_r} [\rho^{(p)}]_{N_z N_r}.
\end{equation}
Similarly, by writing the products of two basis functions in Eq.~\eqref{Eq:VcintElements} as linear combination of HO basis functions with reduced oscillator lengths, one can easily show that integrals in Eq.~\eqref{Eq:VcintElements} can be evaluated as a linear combination of coefficients \eqref{Eq:Vnznr_def}.

We can therefore focus our attention on evaluating the coefficients $[V_C]_{N_z N_r}$. 
Inserting Eq.~\eqref{eq:rhoExpansion} into \eqref{Eq:CoulombVC} and then into the Eq.~\eqref{Eq:Vnznr_def}, we see that the key objects are integrals of the following form
\begin{equation}\label{eq:Cou4int}
    \int \dvol \int \dvol'\,
     \frac{\phi_{N_z}^{\left(\frac{b_z}{\sqrt{2}}\right)}(z)
    \phi_{N_r 0}^{\left(\frac{b_r}{\sqrt{2}}\right)}(r_\perp)\phi_{N_z'}^{\left(\frac{b_z}{\sqrt{2}}\right)}(z')
    \phi_{N_r' 0}^{\left(\frac{b_r}{\sqrt{2}}\right)}(r_\perp')}{|\rvec-\rvec'|},
\end{equation}
for all $0\leq N_z,N_z'\leq \max N_z$ and $0\leq N_r,N_r'\leq \max N_r$.

By invoking Talmi–Moshinsky brackets (see Appendix F in Ref.~\cite{bjelcic2020implementation}), one can express the products of HO functions in terms of coordinates $\rvec$ and $\rvec'$ in Eq.~\eqref{eq:Cou4int} as a linear combination of HO functions in terms of the relative coordinate $\boldsymbol{r}=\rvec-\rvec'$ and the center-of-mass coordinate $\boldsymbol{R}=(\rvec+\rvec')/2$.
Performing a change of variables $(\rvec,\rvec')\to(\boldsymbol{r},\boldsymbol{R})$, the integral over $\boldsymbol{R}$ can be trivially evaluated, while the integral over relative coordinate $\boldsymbol{r}$ leads to the following integrals
\begin{equation}\label{Eq:ICounznr}
    I_{n_z n_r} = \frac{1}{\sqrt{2}}\int \dvol\,\frac{\phi^{(b_z)}_{n_z}(\sqrt{2}z)\phi_{n_r 0}^{(b_r)}(\sqrt{2}r_\perp)}{|\rvec|},
\end{equation}
for $0\leq n_z \leq 4\max\{n_z\}$ and $0\leq n_r\leq2\max\{2n_r+|\Lambda|\}$.

Notice that $I_{n_z n_r}$ is zero for odd $n_z$, and only even $n_z$ are of interest.
Introducing the proper changes of variables, one can prove that for even $n_z$,
\begin{multline}
I_{n_z n_r} = 2\sqrt{2}\pi \int_{0}^{\pi/2} d\theta \frac{\cos\theta}{\frac{\sin^2\theta}{b_z^2} + \frac{\cos^2\theta}{b_r^2}} \\
\times \int_{0}^{+\infty} udu\, P^{\theta}_{\frac{1}{2}n_z+n_r}(u^2) e^{-u^2} ,
\end{multline}
where $P^{\theta}_{\frac{1}{2}n_z+n_r}(u^2)$ is a polynomial of degree $\frac{1}{2}n_z+n_r$ with respect to variable $u^2$ and a smooth $C^{\infty}([0,\pi/2])$ function of the angle $\theta$. This means that the integral over $u^2$ can be computed exactly with Gauss-Laguerre quadrature rules.
Since $\sin^2\theta/b_z^2 + \cos^2\theta/b_r^2 > 0$, the integrand of the integral over $\theta$, after the Gauss-Laguerre rule has been applied, is still a smooth $C^{\infty}([0,\pi/2])$ function. Therefore, the remaining one-dimensional integral over $\theta$ can in principle be computed with arbitrary accuracy with e.g. extended Gauss-Legendre quadrature rule or even Newton–Cotes rules.

After the integrals $I_{n_z n_r}$ are evaluated, by performing an SVD of the matrix $I_{n_z n_r}$, one can show that the final matrix of coefficients $[V_C]_{N_z,N_r}$ can be written as:
\begin{equation}
\label{eq:VCfromrho}
[V_C] = \sum_{i=1}^{2\min\{ \max N_z, \max N_r \}+1} [X^{(i)}] [\rho^{(p)}] [Y^{(i)}]^T,
\end{equation}
where $[\rho^{(p)}]$ is the matrix of coefficients $[\rho^{(p)}]_{N_z N_r}$ given in Eq.~\eqref{eq:rhoExpansion}, while $\max N_z \times \max N_z$ matrices $X^{(i)}$ and $\max N_r\times \max N_r$ matrices $Y^{(i)}$
are fixed matrices which depend only on oscillator lengths $b_z$ and $b_r$ respectively.
It is worth emphasizing that Eq. \eqref{eq:VCfromrho} is exact, i.e. it involves no approximations.
We developed an efficient algorithm which computes the matrices $X^{(i)}$ and $Y^{(i)}$ essentially to machine accuracy.
Once those matrices are computed and stored, one can easily compute the
coefficients $[V_C]_{N_z N_r}$ from coefficients $[\rho^{(p)}]_{N_z N_r}$ via Eq. \eqref{eq:VCfromrho}.


\subsubsection{Single-particle Hamiltonian matrix elements}

Another source of optimization is related to the computation of the matrix elements of the mean-field potential, 
\begin{equation}
\label{Eq:Ik1k2}
V_{ij} = \int \dvol \, \phi_{i}^*(\rvec) \, V(\rvec) \, \phi_{j}(\rvec).
\end{equation}
For the sake of simplicity, we will outline our algorithm in the case where $V(\mathbf r)$ is a spin-independent, non-gradient part of the Skyrme interaction.  Other Skyrme contributions involve gradients acting on the basis functions \(\phi_{i}^*(\mathbf r),\,\phi_{j}(\mathbf r)\), and additional spin dependence, but are handled in an analogous way. We will also focus on the two-dimensional case where, given a potential $V(x,y)$, we need to compute the integrals
\begin{equation}\label{Eq:I12}
I_{1,2}=
    \int\int dx dy \,
    \phi_{n_{x_1}}^{(b_x)}(x)
    \phi_{n_{y_1}}^{(b_y)}(y)
    \,
        V(x,y) 
    \,
    \phi_{n_{x_2}}^{(b_x)}(x)
    \phi_{n_{y_2}}^{(b_y)}(y).
\end{equation}
This type of integral is common in calculations with axially-symmetry Skyrme potentials, where most terms are (non-separable) functions of $(x,y) \equiv (z,r_\perp)$.
In $I_{1,2}$, $\phi_n^{(b)}(\cdot)$ is the one-dimensional HO basis function given in Eq.~\eqref{Eq:phin1DbasisDef}.
The integrals $I_{1,2}$ need to be computed for all indices $0\leq n_{x_1},n_{x_2},n_{y_1},n_{y_2}\leq N$, i.e. a total of $O(N^4)$ integrals.

In conventional implementations, one samples \(V(x,y)\) on an \(O(N)\times O(N)\) quadrature grid and approximates each integral $I_{1,2}$ with numerical quadrature rules.  This incurs a total computational cost of \(O(N^6)\) and may suffer accuracy degradation depending on the chosen quadrature rule.

In our approach, we first define the $(2N+1)\times (2N+1)$ matrix
\begin{equation}
    [V]_{N_x N_y} = \int \int dxdy\, V(x,y) \, \phi^{(b_x/\sqrt{2})}_{N_x}(x)
    \phi^{(b_y/\sqrt{2})}_{N_y}(y),
\end{equation}
for $0\leq N_x,N_y\leq 2N$, and compute its singular value decomposition (SVD)
\begin{equation}
[V]_{N_x N_y} = \sum_{i=0}^{2N} \gras{x}^{i}_{N_x}  \, \gras{y}^{i}_{N_y}.
\end{equation}
After expressing the products of basis functions as in Eq.~\eqref{eq:AxialTM} via the coefficients $c_N^{n_1 n_2}$, we can write the integral $I_{1,2}$ as
\begin{equation}
I_{1,2} =
\sum_{i=0}^{2N}
\left( \sum_{N_x=0}^{2N} c_{N_x}^{n_{x_1}n_{x_2}} \gras{x}_{N_x}^i \right)
\left( \sum_{N_y=0}^{2N} c_{N_y}^{n_{y_1}n_{y_2}} \gras{y}_{N_y}^i \right).
\end{equation}
Introducing the $(N+1)\times (N+1)\times (2N+1)$ tensors
\begin{align}
X^i_{n_{x_1}n_{x_2}} & = \sum_{N_x=0}^{2N} c_{N_x}^{n_{x_1}n_{x_2}} \gras{x}_{N_x}^i ,
\\
Y^i_{n_{y_1}n_{y_2}} & = \sum_{N_y=0}^{2N} c_{N_y}^{n_{y_1}n_{y_2}} \gras{y}_{N_y}^i ,
\end{align}
one can finally express $I_{1,2}$ as
\begin{equation}
\label{Eq:I12final}
I_{1,2} = \sum_{i=0}^{2N}\,  X^i_{n_{x_1}n_{x_2}}\,  Y^i_{n_{y_1}n_{y_2}}.
\end{equation}
Essentially, the initial expression for $I_{1,2}$ in Eq.~\eqref{Eq:I12} couples the variables $x$ and $y$ through the potential $V(x,y)$, even though the basis functions are factorized. In contrast, the final expression in Eq.~\eqref{Eq:I12final} fully separates the $x$ and $y$ directions, at the expense of introducing an additional summation over the singular value index $0\leq i \leq 2N$.

Evaluating the $O(N^2)$ matrix elements $[V]_{N_x N_y}$ on a quadrature grid of size $O(N) \times O(N)$ requires $O(N^4)$ operations, while computing the SVD of matrix $[V]_{N_x N_y}$ costs $O(N^3)$. With precomputed coefficients $c_N^{n_1 n_2}$, the evaluation of the tensors $X^i_{n_{x_1}n_{x_2}}$ and $Y^i_{n_{y_1}n_{y_2}}$ incurs a cost of $O(N^4)$. The evaluation of all $O(N^4)$ integrals $I_{1,2}$ using Eq.~\eqref{Eq:I12final} then requires $O(N^5)$ operations. Overall, the proposed algorithm computes all integrals $I_{1,2}$ with $O(N^5)$ complexity, in contrast to the conventional approach which requires $O(N^6)$. The added space complexity is negligible in practice.

Furthermore, the only step that requires numerical integration is the evaluation of the matrix $[V]_{N_x N_y}$. Since this matrix contains only $O(N^2)$ elements, one can afford to use a significantly finer quadrature grid than the conventional grid of $O(N) \times O(N)$ size without affecting the overall $O(N^5)$ computational complexity of the algorithm. This enables the evaluation of $[V]_{N_x,N_y}$ to essentially machine precision. All other steps of the algorithm are exact and can also be done with machine precision in practice.

Finally, it is worth noting that the proposed algorithm is feasible because the axially-symmetric 3D case can be effectively reduced to the 2D formulation presented here. In contrast, a fully 3D treatment breaking axial symmetry would require a generalization of the SVD from two to three dimensions, which is notoriously nontrivial. 


\subsubsection{HFB matrix diagonalization}

Finally, the specific structure of the set of HFB eigenvalues provides an additional source of optimization when solving the HFB eigenvalue problem that has very rarely been exploited. 
In configuration space, we recall that the HFB eigenvalue equation reads
\begin{equation}
\label{Eq:HFBeigen}
  \begin{bmatrix}
    U & V^* \\
    V & U^*
  \end{bmatrix}^\dagger
  \begin{bmatrix}
    h & \Delta \\
    -\Delta^* & -h^*
  \end{bmatrix}
  \begin{bmatrix}
    U & V^* \\
    V & U^*
  \end{bmatrix}
  =
  \begin{bmatrix}
    E & 0 \\
    0 & -E
  \end{bmatrix},
\end{equation}
where \(E\) is the diagonal matrix of positive quasiparticle energies. As is well known, if axial symmetry is enforced and the HO basis vectors are ordered into two blocks, first the states with positive total angular momentum $\Omega=\Lambda+\tfrac\sigma2>0$
\begin{equation}
\phi_{n_z}^{(b_z)}(z)\,\phi_{n_r \Lambda}^{(b_r)}(r_\perp)\,\frac{e^{i\Lambda\varphi}}{\sqrt{2\pi}}\,|\sigma\rangle,  
\end{equation}
then their time-reversed partners
\begin{equation}
\sigma\;\phi_{n_z}^{(b_z)}(z)\,\phi_{n_r -\Lambda}^{(b_r)}(r_\perp)\,\frac{e^{-\,i\Lambda\varphi}}{\sqrt{2\pi}}\,|-\sigma\rangle,
\end{equation}
then, the single-particle Hamiltonian and pairing fields acquire a simple block structure
\begin{equation}
  h = 
  \begin{bmatrix}
    h_1 & 0 \\
    0   & h_1
  \end{bmatrix},
  \qquad
  \Delta = 
  \begin{bmatrix}
    0        & \Delta_1 \\
    -\Delta_1 & 0
  \end{bmatrix},
\end{equation}
where \(h_1\) and \(\Delta_1\) are real symmetric matrices in the case of a Skyrme density functional and contact pairing force. It follows that the Bogoliubov matrix and quasiparticle energies also acquire the following block structure
\begin{equation}
  U = 
  \begin{bmatrix}
    u_1 & 0 \\
    0   & u_1
  \end{bmatrix},
  \quad
  V = 
  \begin{bmatrix}
    0        &- v_1 \\
    v_1 & 0
  \end{bmatrix},
  \quad
  E = 
  \begin{bmatrix}
    e_1 & 0 \\
    0   & e_1
  \end{bmatrix},
\end{equation}
and the original HFB eigenvalue problem \eqref{Eq:HFBeigen} is reduced to a real symmetric eigenvalue problem
\begin{equation}\label{Eq:HFBreduced}
    \mathcal{W}_1^T
    \mathcal{H}_1
    \mathcal{W}_1
    =
    \left[
    \begin{matrix}
        e_1 & 0 \\ 0 & -e_1 
    \end{matrix}
    \right],
\end{equation}
with
\begin{equation}
\mathcal{H}_1 = \left[
    \begin{matrix}
        h_1 & \Delta_1 \\ \Delta_1 & -h_1 
    \end{matrix}
    \right],
\quad
\mathcal{W}_1 = \left[
    \begin{matrix}
        u_1 & -v_1 \\ v_1 & u_1 
    \end{matrix}
    \right].
\end{equation}

In a typical implementation, one would explicitly construct the full real symmetric matrix $\mathcal{H}_1$ and diagonalize it, say using one of LAPACK's eigensolver subroutines \cite{stoitsov2005axially}. This ignores the optimization opportunity that arises from the fact that the eigenvalues $e_1$ and $-e_1$ are known a priori to occur in pairs with opposite signs. Indeed, we observe that the reduced eigenvalue problem \eqref{Eq:HFBreduced} is equivalent to computing the Autonne–Takagi factorization \cite{horn2013matrix} of the complex symmetric matrix $h_1 - i \Delta_1$:
\begin{equation}\label{Eq:AutonneTakagi}
    (u_1 + i v_1)^T (h_1 - i \Delta_1)(u_1 + i v_1) = e_1,
\end{equation}
where the orthogonality of the real matrix $\mathcal{W}_1$ is equivalent to the unitarity of the complex matrix $u_1 + i v_1$.

We have developed an Autonne–Takagi solver, and numerical tests show that the computation time to solve the Eq.~\eqref{Eq:AutonneTakagi} is typically three times shorter compared to the straightforward diagonalization of \eqref{Eq:HFBreduced} using state-of-the-art Intel MKL eigensolver.  
This results in a significant performance boost when calculating the static HFB solution.


\section{Convergence study}
\label{sec:conv}

In this section, we conduct a detailed study of the convergence properties of the \texttt{AxialHOHFB} solver with respect to the characteristics of the basis. The code is first benchmarked against the open-source \texttt{HFBTHO} code of \cite{marevic2022axiallydeformed} in Sec.~\ref{subsec:hfbtho}. We then perform a very comprehensive analysis of convergence properties, both for the static part of \texttt{AxialHOHFB}, in Sec.~\ref{subsec:conv_HFB}, and for its time-dependent part in Sec.~\ref{subsec:conv_TDHFB}-\ref{subsec:conv_Ex}.


\subsection{Comparison to HFBTHO code}
\label{subsec:hfbtho}

The static part \texttt{AxialHOHFB} can define the same basis and quadrature rules as the ones implemented in \texttt{HFBTHO}. Furthermore, the new algorithm for the Coulomb's field explained in Sec.~\ref{subsec:Algo} can also be replaced by the \texttt{HFBTHO} code implementation for computing $V_C(\rvec)$; see \cite{stoitsov2013axially} for a discussion of the limitations of this implementation. When those adjustments are made, the outputs of the two codes are in excellent agreement, with total energies differing by less than a fraction of eV.

To demonstrate this consistency, we have compared the contributions to the Skyrme energy and expectation values $q_{\lambda 0} = \braket{\hat{Q}_{\lambda 0}}$ of axial multipole moments obtained from both codes for ${}^{240}\mathrm{Pu}$. We use the SLy5 parameterization \cite{chabanat1998skyrme} of the Skyrme functional and a mixed surface-volume, density-dependent pairing interaction of the form $V(\rvec) = V_0^{(q)} \big(1 - \eta\frac{\rho(\rvec)}{\rho_c}\big)$ with $\rho_c = 0.16\,\mathrm{fm}^{-3}$, $\eta=0.5$, pairing strengths $V_0^{(n,p)}=-300\,\mathrm{MeV}\mathrm{fm}^3$, and a sharp cutoff of $E_{\rm cut} = 60$ MeV on quasiparticles.

We consider two different deformed configurations, one with $(q_{20},q_{30}) = (30\,\mathrm{b},0\,\mathrm{b}^{3/2})$ close to the ground state of this nucleus, the other one $(q_{20},q_{30}) = (200\,\mathrm{b},20\,\mathrm{b}^{3/2})$ located past the saddle point. Tables \ref{tab:comparison_30_0} and \ref{tab:comparison_200_20} give the results of the benchmark for both configurations.

\begin{table}[!htb]
    \caption{Comparison of axial multipole moments and components of the Skyrme energy for $(q_{20},q_{30})=(30\,\mathrm{b},0\,\mathrm{b}^{3/2})$ \label{Table:30_0}. All energies are given in MeV; multipole moment  $ q_{\lambda 0} $ of order $\lambda$ are given in b$^{\lambda/2}$.}
    \label{tab:comparison_30_0}
    \begin{ruledtabular}
    \begin{tabular}{l r r}
        Observable & \texttt{AxialHOHFB} & \texttt{HFBTHO} \\
        \hline
        $ q_{10} $ &  0.0000000000 & 0.0000000000 \\
        $ q_{20} $ & 30.0000000000 & 30.0000000000 \\
        $ q_{30} $ &  0.0000000000 & 0.0000000000 \\
        $ q_{40} $ &  3.070583200\textbf{\color{red}0} & 3.070583200\textbf{\color{red}9} \\
        $ q_{50} $ &  0.0000000000 & 0.0000000000 \\
        $ q_{60} $ &  0.525153321\textbf{\color{red}6} & 0.525153321\textbf{\color{red}7} \\
        $ q_{70} $ &  0.0000000000 & 0.0000000000 \\
        $ q_{80} $ &  0.036074771\textbf{\color{red}3} & 0.036074771\textbf{\color{red}5} \\
        \hline
        $E_{\rho\rho}$      & -8996.572027\textbf{\color{red}430} & -8996.572027\textbf{\color{red}056} \\
        $E_{\rho\Delta\rho}$&   348.0290619\textbf{\color{red}40} & 348.0290619\textbf{\color{red}18} \\
        $E_{\rho\tau}$      &  1496.4038533\textbf{\color{red}15}  & 1496.4038533\textbf{\color{red}39}  \\
        $E_{J\cdot J}$      & 2.75592182\textbf{\color{red}4}  & 2.75592182\textbf{\color{red}6}  \\
        $E_{\rho\nabla J}$  & -97.318534\textbf{\color{red}082} & -97.318534\textbf{\color{red}104}  \\
        $E^{\mathrm{Coulomb}}_{\mathrm{direct}}$ & 1026.4132260\textbf{\color{red}90} & 1026.4132260\textbf{\color{red}73}  \\
        $E^{\mathrm{Coulomb}}_{\mathrm{exchange}}$ & -35.74791500\textbf{\color{red}0}  & -35.74791499\textbf{\color{red}9}  \\
        $E_{\mathrm{pairing}}$ & -8.6217417\textbf{\color{red}95} & -8.6217417\textbf{\color{red}80}  \\
        $E_{\mathrm{kinetic}}$ & 4465.170729\textbf{\color{red}010} & 4465.170729\textbf{\color{red}270} \\
        $E_{\mathrm{total}}$ & -1799.48742\textbf{\color{red}6126}  & -1799.48742\textbf{\color{red}5512}  \\
    \end{tabular}
    \end{ruledtabular}
\end{table}

\begin{table}[!htb]
    \caption{Same as Table \ref{Table:30_0}, but for $(q_{20},q_{30})=(200\,\mathrm{b},20\,\mathrm{b}^{3/2})$.}
    \label{tab:comparison_200_20}
    \begin{ruledtabular}
    \begin{tabular}{l r r}
        Observable & \texttt{AxialHOHFB} & \texttt{HFBTHO} \\
        \hline
        $ q_{10} $ &   0.000000001\textbf{\color{red}1} &   0.000000001\textbf{\color{red}4} \\
        $ q_{20} $ & 199.9999999\textbf{\color{red}347} & 200.0000000\textbf{\color{red}050} \\
        $ q_{30} $ &  19.99999999\textbf{\color{red}65} &  20.00000000\textbf{\color{red}06} \\
        $ q_{40} $ &  52.9655562\textbf{\color{red}638} &  52.9655562\textbf{\color{red}110} \\
        $ q_{50} $ &  33.585391\textbf{\color{red}2863} &  33.585391\textbf{\color{red}1817} \\
        $ q_{60} $ &  49.945634\textbf{\color{red}4298} &  49.945634\textbf{\color{red}1918} \\
        $ q_{70} $ &  42.321648\textbf{\color{red}7055} &  42.321648\textbf{\color{red}3373} \\
        $ q_{80} $ &  51.05531\textbf{\color{red}84669} &  51.05531\textbf{\color{red}78446} \\
        \hline
        $E_{\rho\rho}$ & -8756.72282\textbf{\color{red}6352}  & -8756.72282\textbf{\color{red}7077} \\
        $E_{\rho\Delta\rho}$ & 410.197349\textbf{\color{red}672}  & 410.197349\textbf{\color{red}765}  \\
        $E_{\rho\tau}$ & 1415.945110\textbf{\color{red}234}  & 1415.945110\textbf{\color{red}531} \\
        $E_{J\cdot J}$ & 3.218325\textbf{\color{red}591} &   3.218325\textbf{\color{red}601} \\
        $E_{\rho\nabla J}$ & -113.733402\textbf{\color{red}777} & -113.733402\textbf{\color{red}875} \\
        $E^{\mathrm{Coulomb}}_{\mathrm{direct}}$ & 915.258402\textbf{\color{red}343}  & 915.258402\textbf{\color{red}400} \\
        $E^{\mathrm{Coulomb}}_{\mathrm{exchange}}$ & -35.4167023\textbf{\color{red}79} & -35.4167023\textbf{\color{red}82} \\
        $E_{\mathrm{pairing}}$ &   -14.971973\textbf{\color{red}847} &   -14.971973\textbf{\color{red}630} \\
        $E_{\mathrm{kinetic}}$ &  4385.132641\textbf{\color{red}160} &  4385.132641\textbf{\color{red}750} \\
        $E_{\mathrm{total}}$   & -1791.09307\textbf{\color{red}6354} & -1791.09307\textbf{\color{red}5917} \\
    \end{tabular}
    \end{ruledtabular}
\end{table}


\subsection{Static HFB Solution}
\label{subsec:conv_HFB}

We investigate the convergence behavior of various deformed configurations using the static component of the \texttt{AxialHOHFB} solver. Specifically, we study the nucleus $^{240}$Pu with the Skyrme SkM${}^*$ energy density functional \cite{bartel1982better}. We keep the form of the pairing functional to a mixed surface-volume pairing with $\eta = 0.5$. However, the neutron and proton pairing strengths are now taken as \(V_0^{(n)} = -265.25\, \mathrm{MeV}\mathrm{fm}^3\) and \(V_0^{(p)} = -340.0625\, \mathrm{MeV}\mathrm{fm}^3\), respectively. This parameterization is the one adopted in \cite{schunck2014description} and several subsequent studies of nuclear fission. 

\begin{figure}[!htb]
    \centering
    \includegraphics[width=1.0\columnwidth]{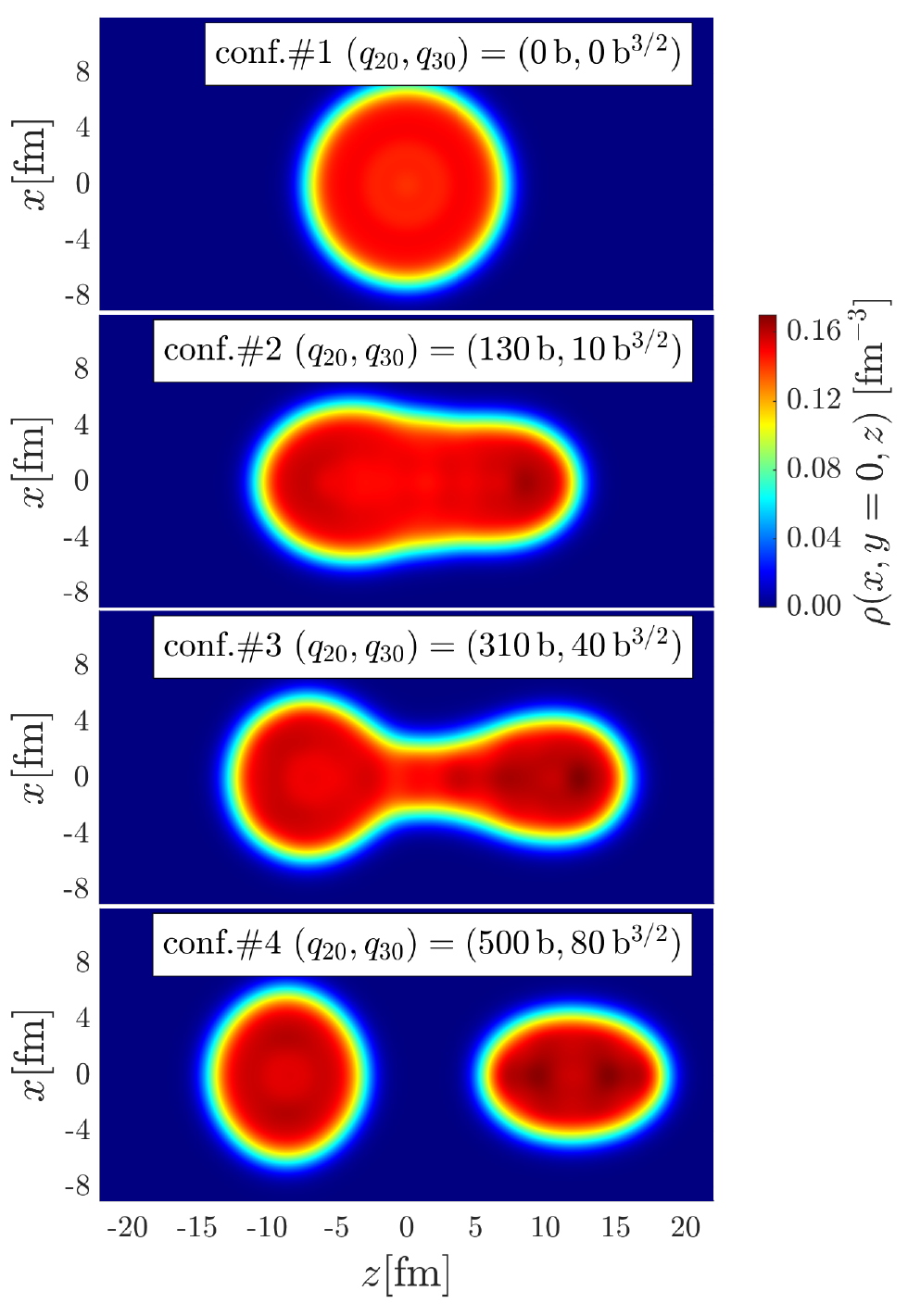}
    \caption{Axially symmetric density profiles $\rho(x, y = 0, z)$ of the four deformed configurations of ${}^{240}$Pu listed in Table \ref{tab:HFB_configs} used to test the convergence of the \texttt{AxialHOHFB} static solver.}
    \label{fig:4configurations}
\end{figure}

The four representative configurations in the $(q_{20}, q_{30})$ collective space that we consider range from spherical to highly deformed shapes to shapes corresponding to two, fully separated fragments. These are summarized in Table~\ref{tab:HFB_configs}. To visualize these configurations, Figure~\ref{fig:4configurations} shows their axially symmetric density profiles.

\begin{table}[!htb]
    \caption{Selected static HFB configurations for $^{240}$Pu and their axial quadrupole ($q_{20}$) and octupole ($q_{30}$) moments.}
    \label{tab:HFB_configs}
    \begin{ruledtabular}
    \begin{tabular}{clrr}
    Configuration & Description  & $q_{20}\,[\mathrm{b}]$ & $q_{30}\,[\mathrm{b}^{3/2}]$ \\
    \hline
    \#1 & Spherical      & 0   & 0  \\
    \#2 & Saddle point   & 130 & 10 \\
    \#3 & Near-scission  & 310 & 40 \\
    \#4 & Fragmented     & 500 & 80 \\
    \end{tabular}
    \end{ruledtabular}
\end{table}

Recall that the basis is defined by three input parameters: $Z_{\mathrm{box}}$, $R_{\mathrm{box}}$, and $\nzmax$. We set the box dimensions to $Z_{\mathrm{box}} = 25\,\mathrm{fm}$ and $R_{\mathrm{box}} = 12.5\,\mathrm{fm}$, which effectively constrains the basis functions within a cylinder of height $50\,\mathrm{fm}$ and diameter $25\,\mathrm{fm}$. The parameter $\nzmax$ is varied from 28 to 64 in steps of 4.

\begin{figure*}[!htbp]
    \centering
    \includegraphics[width=1.0\textwidth]{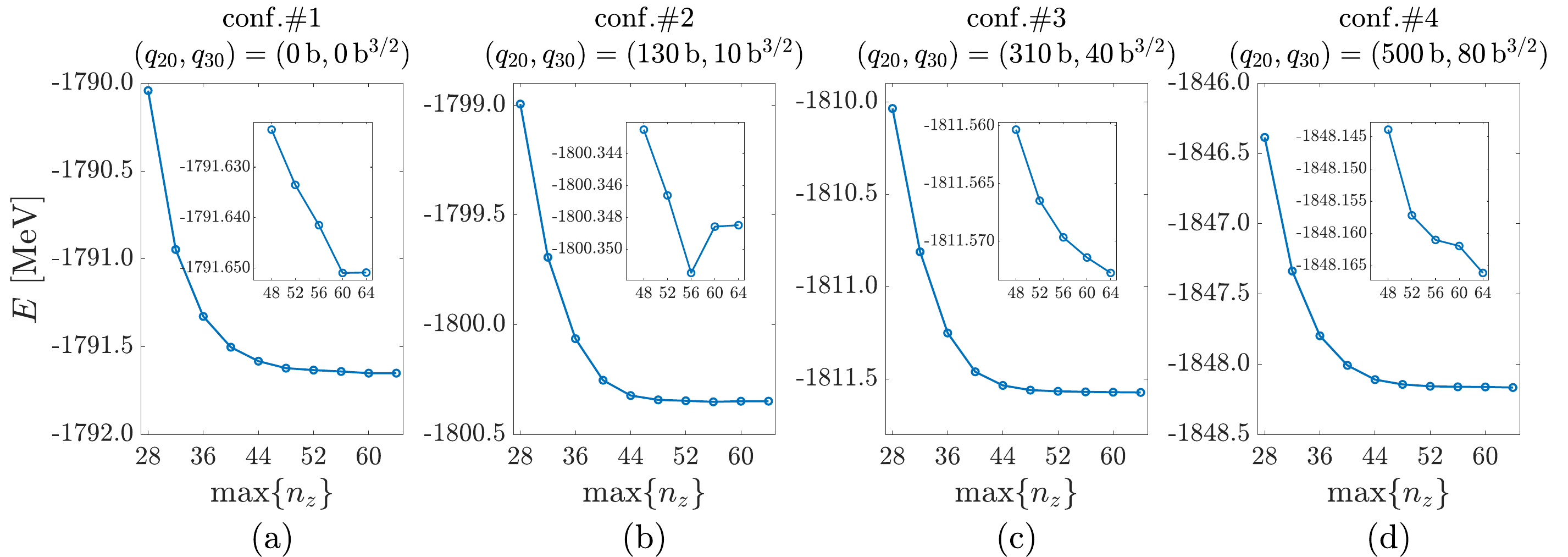}
    \caption{Convergence of the total HFB energy as $\nzmax$ increases for four (a)-(d) static configurations of $^{240}$Pu shown in Fig.~\ref{fig:4configurations}.
    Inset plots focus on higher values of $\nzmax$.}
    \label{fig:E_vs_maxnz}
\end{figure*}

To quantify the basis size, Table~\ref{tab:Ntotal_vs_nz} lists the total number of basis states $N_{\mathrm{total}}$, i.e., the number of valid combinations of $(n_z \geq 0,\, n_r \geq 0,\, \Lambda)$ for each $\nzmax$. For reference, Table~\ref{tab:Ntotal_vs_nz} also lists the equivalent number of complete spherical harmonic oscillator shells, \( N_{\mathrm{sh}} \), which is extracted from the well-known formula $ N_{\mathrm{total}} = \frac{1}{6}(N_{\mathrm{sh}} + 1)(N_{\mathrm{sh}} + 2)(N_{\mathrm{sh}} + 3)$, which gives the total number of basis states in a basis that includes all combinations satisfying the condition: \( n_z + 2n_r + |\Lambda| \leq N_{\mathrm{sh}} \).

\begin{table}[!htb]
    \caption{Total number of basis states $N_{\mathrm{total}}$ as a function of $\nzmax$ for a basis with $Z_{\mathrm{box}} = 25\,\mathrm{fm}$ and $R_{\mathrm{box}} = 12.5\,\mathrm{fm}$. Also shown is the corresponding number of complete shells $N_{\mathrm{sh}}$ which yield the equivalent basis size.}
    \label{tab:Ntotal_vs_nz}
    \begin{ruledtabular}
    \begin{tabular}{c r r}
        $\nzmax$ & $N_{\mathrm{total}}$ & $N_{\mathrm{sh}}$ \\
        \hline
        28 & 3\,480  & 25.55 \\
        32 & 5\,049  & 29.18 \\
        36 & 7\,030  & 32.82 \\
        40 & 9\,471  & 36.45 \\
        44 & 12\,420 & 40.09 \\
        48 & 15\,925 & 43.72 \\
        52 & 20\,034 & 47.36 \\
        56 & 24\,795 & 50.99 \\
        60 & 30\,256 & 54.63 \\
        64 & 36\,465 & 58.26 \\
    \end{tabular}
    \end{ruledtabular}
\end{table}

Figure \ref{fig:E_vs_maxnz} shows the convergence behavior of the total HFB energy for each of the four deformed configurations of $^{240}$Pu of Table \ref{tab:HFB_configs} as the parameter $\nzmax$ is increased. As expected, increasing $\nzmax$ to large values can yield energies converged to within approximately $10\,\mathrm{keV}$ accuracy, albeit at the cost of a substantial increase in the basis size. Most importantly, this convergence pattern is very similar for all four configurations. In particular, a basis with $\nzmax=32$ achieves sub-$800\,\mathrm{keV}$ accuracy for all four configurations, including the conf \#4 that corresponds to two well-separated fragments. This choice offers a reasonable compromise between numerical accuracy and computational cost. For applications requiring higher precision, choosing $\nzmax=40$ yields convergence within approximately $150\,\mathrm{keV}$. For the sake of completeness, we also illustrate in Appendix~\ref{app:box} how enlarging the box size parameters $Z_\mathrm{box}$ and $R_\mathrm{box}$ slows down the convergence. Appendix~\ref{app:shift} shows a very stringent convergence test especially relevant for fission studies, as it demonstrates how the convergence of the total energy is impacted after shifting the position of the origin used to define the constraints on multipole moments. In both cases, however, we show that the total energy converges to the same value.

Once a value of $\nzmax$ is selected -- based on the desired accuracy -- we emphasize that the {\em same} basis is used uniformly across all deformed configurations explored in this work. This involves static configurations used to build a potential energy surface as well as time-dependent configurations along TDHFB trajectories. While it stands in contrast to the common practice of adjusting basis parameters at each constrained deformation, our convergence analysis shows that our numerical results are in fact most likely better converged than most previous studies published in the literature. 


\subsection{Dynamic TDHFB solution}
\label{subsec:conv_TDHFB}

In this section, we now examine the convergence behavior of TDHFB solutions, starting from three different deformed static HFB configurations. Each of them has an HFB energy that is 1 MeV below the saddle-point energy, but they differ in the values of their quadrupole and octupole deformations. Trajectory \#1 starts at $(q_{20},q_{30}) = (150\,\mathrm{b},\, 12\,\mathrm{b}^{3/2})$, trajectory \#2 at $(q_{20},q_{30}) = (220\,\mathrm{b},\,  6\,\mathrm{b}^{3/2})$ and trajectory \#3 at $(q_{20},q_{30}) = (245\,\mathrm{b},\, 40\,\mathrm{b}^{3/2})$.

\begin{figure*}
    \centering
    \setlength{\tabcolsep}{0pt} 
    \begin{tabular}{@{}ccc@{}}
        \includegraphics[width=0.33\textwidth]{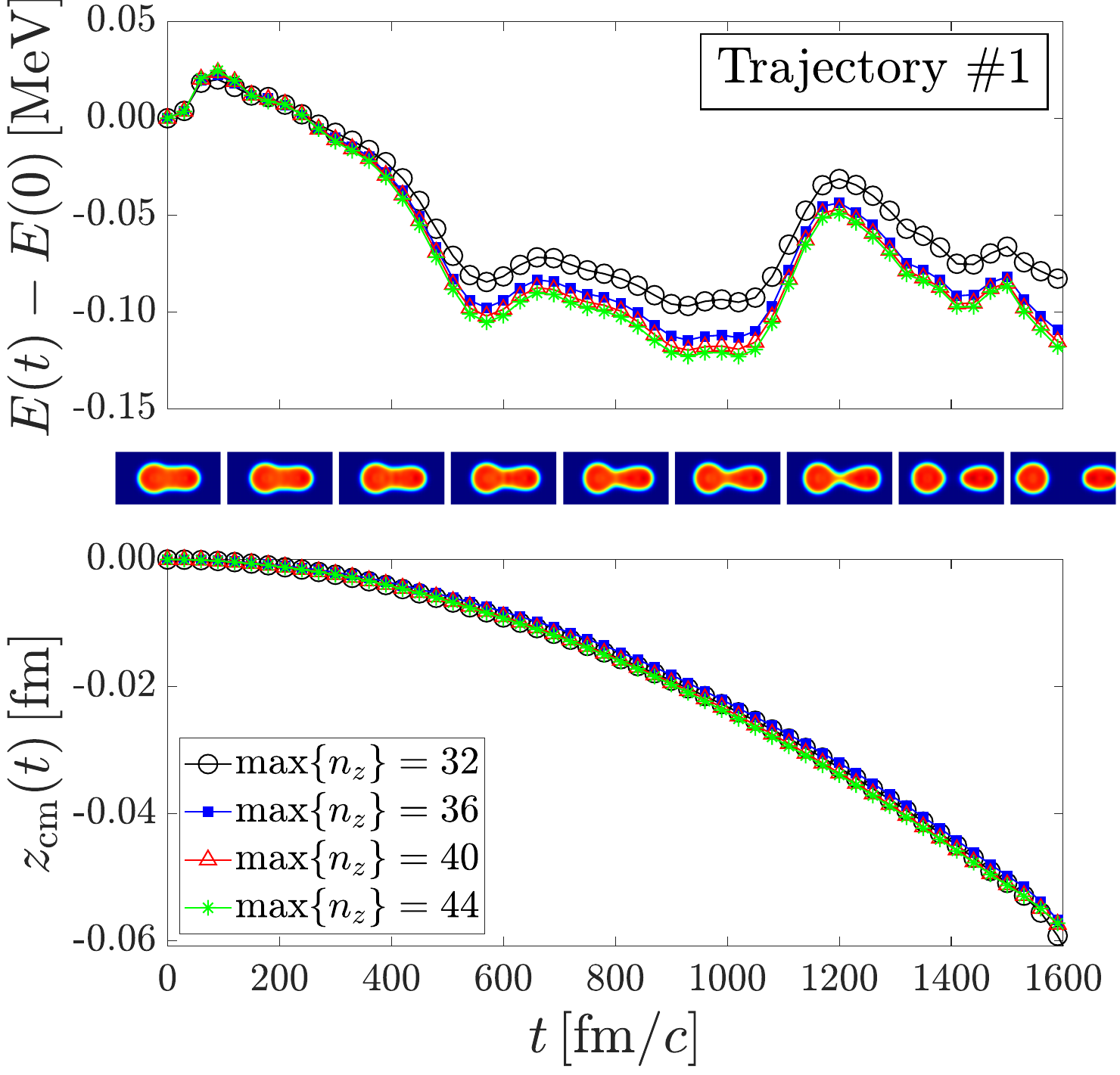} &
        \includegraphics[width=0.33\textwidth]{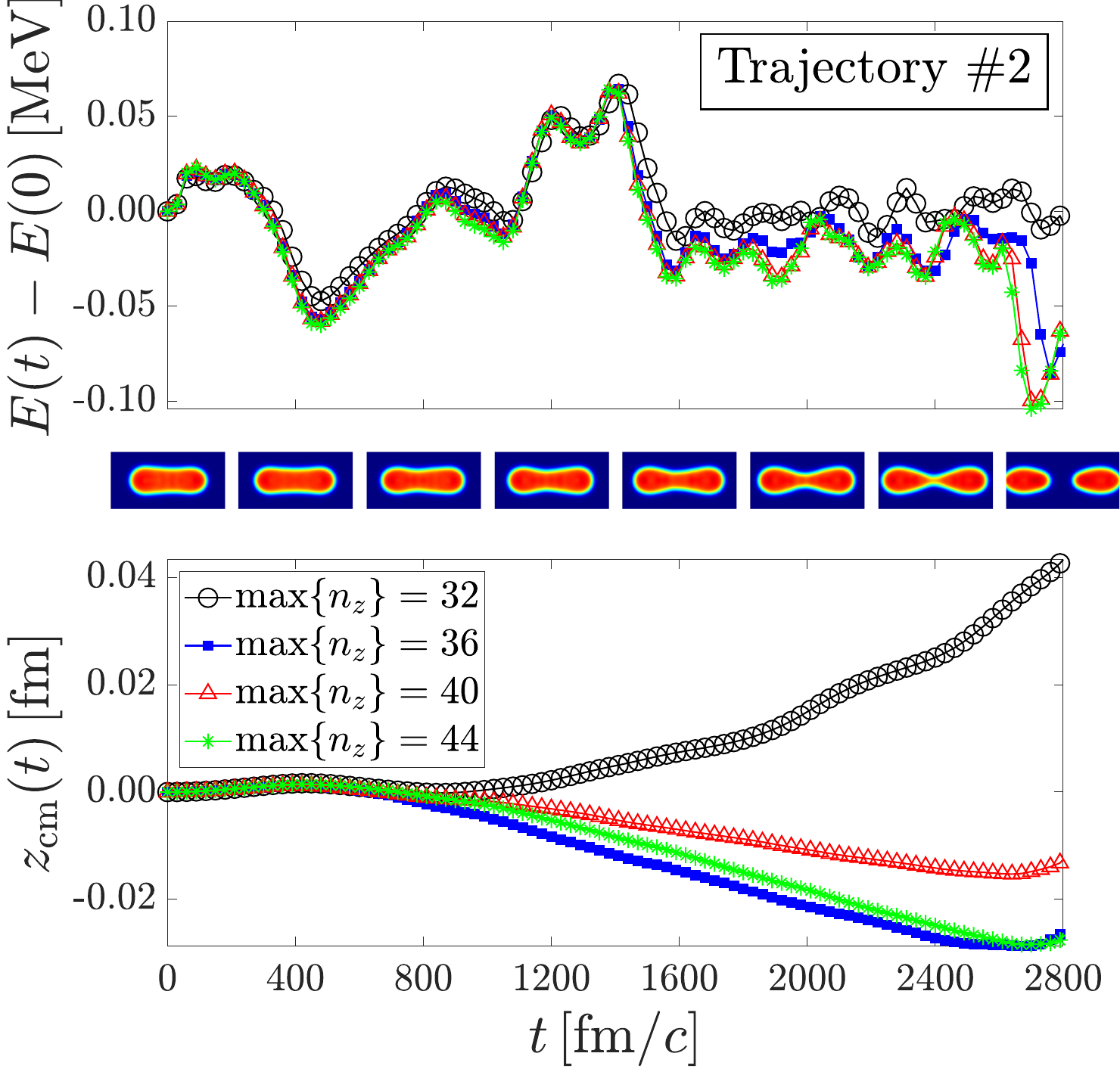} &
        \includegraphics[width=0.33\textwidth]{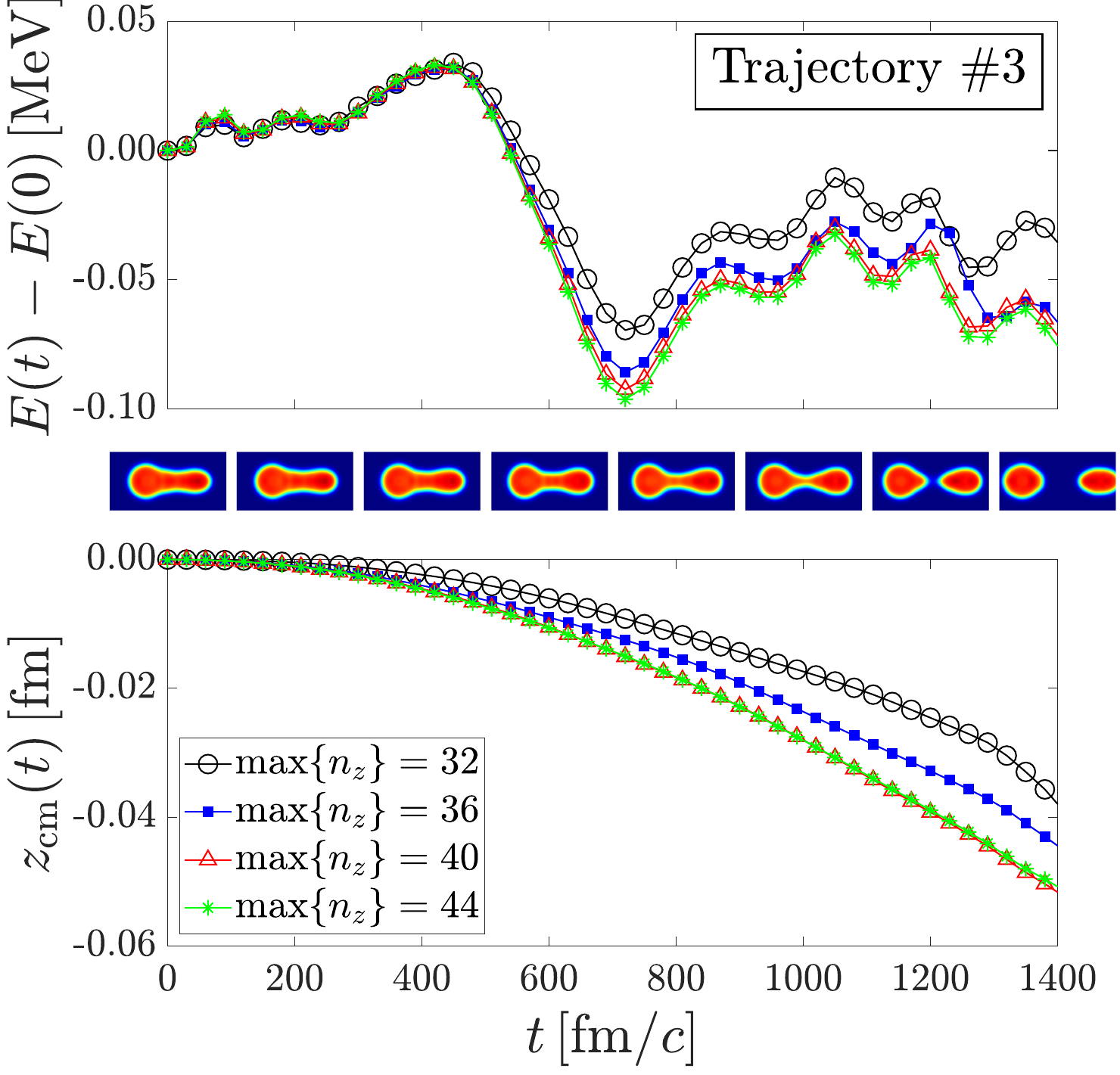} \\
        (a) & (b) & (c)
    \end{tabular}
    \caption{Energy conservation and center-of-mass position for the three TDHFB trajectories shown in Fig.~\ref{fig:3trajectories}, with increasing basis size parameter $\nzmax$. Panels (a), (b), and (c) correspond to Trajectories \#1, \#2, and \#3, respectively. Color maps for the density profiles match those in Fig.~\ref{fig:4configurations}.
    }
    \label{fig:EnergyZcm3trajectories}
\end{figure*}

As mentioned before, the TDHFB equation~\eqref{eq:TDHFB} is solved using the same basis setup described in Sec.~\ref{subsec:conv_HFB}, namely with $Z_{\rm box} = 25$ fm and $R_{\rm box} = 12.5$ fm. Each trajectory is evolved in time with increasing values of the basis truncation parameter $\nzmax$, ranging from 32 to 44. Further numerical details, including the integration scheme and time step selection, are discussed in Appendix~\ref{app:time}. Figure~\ref{fig:3trajectories} shows the potential energy surface and each of the three trajectories for different values of $\nzmax$. We observe that even with $\nzmax=32$, the trajectories exhibit good convergence in the $(q_{20}, q_{30})$ deformation space.

\begin{figure}[!htb]
    \centering
    \includegraphics[width=1.0\columnwidth]{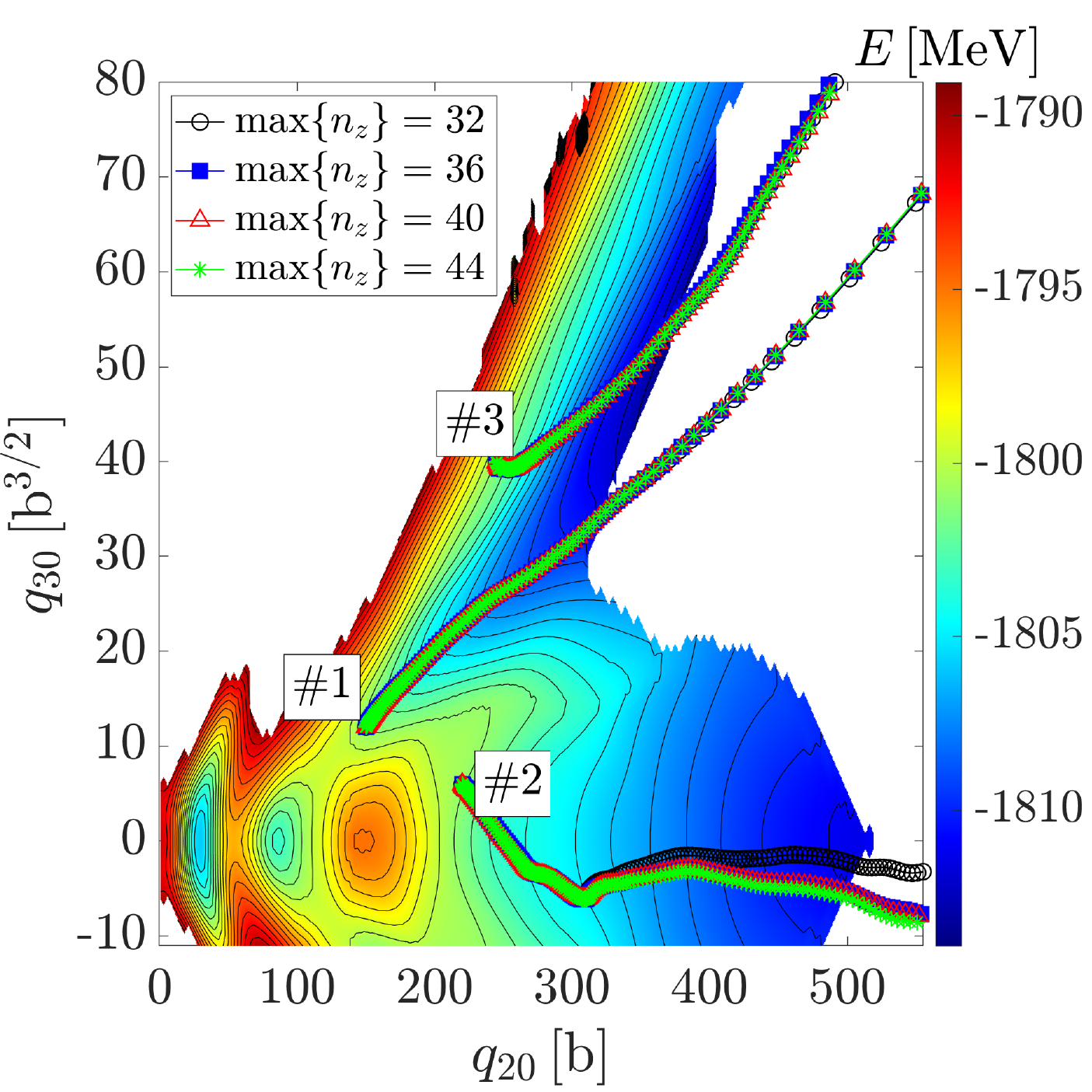}
    \caption{Potential energy surface of $^{240}$Pu calculated with the Skyrme SkM$^*$ EDF, overlaid with three TDHFB trajectories initialized at the same isoenergy contour 1~MeV below the saddle point energy. Points along the trajectories are separated in time by $15\,\mathrm{fm}/c$. Each trajectory is computed with increasing values of $\nzmax$.
    Isoenergy lines are spaced by 1 MeV.
    }
    \label{fig:3trajectories}
\end{figure}

Figure \ref{fig:EnergyZcm3trajectories} shows the energy conservation $E(t)-E(0)$ and the center-of-mass position $z_\mathrm{cm}(t) = \frac{1}{A} \int \dvol \, z \rho(\rvec,t)$ for each trajectory. The total energy is conserved within approximately $120\,\mathrm{keV}$ for all three trajectories. Note that improving the conservation of energy would require increasing the quasiparticle cutoff energy $E_{\rm cut}$ beyond the $60\,\mathrm{MeV}$ value used here, as discussed in Sec.~\ref{subsec:TimeEvolution}. Appendix~\ref{app:time} shows that, in the absence of such cutoff, sub-eV energy conservation is numerically achievable. The observed 120~keV deviation thus confirms that high-energy quasiparticle states excluded by the cutoff have negligible influence on the fission dynamics.

The center-of-mass position remains stable within a maximum drift of $0.06\,\mathrm{fm}$ over the time evolution. Increasing $\nzmax$ does not reduce this drift, since the finite harmonic oscillator basis restricted in a box, breaks translational invariance. Consequently, the total linear momentum $\braket{\hat{P}_z(t)}$ is not conserved. Even though initially $\braket{\hat{P}_z(0)} = 0$, the non conservation of $\braket{\hat{P}_z(t)}$ leads to a small, but non-zero, spurious center-of-mass motion. Despite this artifact, the magnitude of the drift is practically negligible. A rough estimate of the associated center-of-mass kinetic energy for a nucleus with $A=240$ nucleons and $z_\mathrm{cm}(t)$ displacement of $0.06\,\mathrm{fm}$ over a time interval of $t \approx 1500\,\mathrm{fm}/c$ yields only $\sim 0.2\,\mathrm{keV}$.


\subsection{Total kinetic energy of fission fragments}
\label{subsec:conv_TKE}

Once the two fragments have split and are ``sufficiently apart'' from one another, we divide the full spatial domain into two half-spaces, $\mathbb{R}^3 = \volH\cup \volL$ joined at $z = z_0$. The point $z_0$ can be chosen such that the total density of the entire nucleus has a local minimum. We assume that the heavy fragment is in $\volH$ while the light fragment is in $\volL$. These are not rigorous mathematical definitions: in practice, it simply means that the properties of the heavy fragments will involve integrals over $\volH$ while the properties of the light fragments will involve integrals over $\volL$.

After this separation, the interaction between the fragments is caused by the long-range Coulomb force and the nuclear contribution vanishes. Note that this is an approximation caused by the fact that the Skyrme interaction has zero range (= the Skyrme functional depends on the local density and its gradients only) and that we define the energy of the light (heavy) fragment as the integral of the full functional over the subspace $\volL$ ($\volH$). The Gogny force, which is equivalent to a functional of the non-local density, behaves like the Coulomb potential. Upon integrating over subdomains, it would automatically yield an interaction term between the two fragments. Also note that the current approach is fundamentally different from what was proposed in \cite{younes2011nuclear} and implemented in \cite{schunck2014description,schunck2015description}. In these studies, the subspaces were used to define separate sets of quasiparticles that defined (pseudo)densities associated with each fragment. This enabled defining a nuclear interaction energy even for zero-range forces such as Skyrme.

Since the nuclear interaction vanishes, the total kinetic energy (TKE) of the fission process reads
\begin{equation}
\TKE(t) = \KH(t) + \KL(t) + \ECint(t).
\label{eq:tke}
\end{equation}
where $\KH(t)$ ($\KL(t)$) is the kinetic energy of the heavy (light) fragment and $\ECint(t)$ the Coulomb interaction energy between them,
\begin{equation}
\label{eq:ECint}
\ECint(t) = \hbar c \alpha \int_{\volH}\dvol \int_{\volL} \dvol' \, \frac{\rho^{(p)}(\rvec,t) \rho^{(p)}(\rvec',t)}{|\rvec-\rvec'|},
\end{equation}
In principle, all terms in \eqref{eq:tke} are functions of time. In a semi-classical picture, the acceleration of the fission fragments, that is, the change in $\KH(t)$ and $\KL(t)$, is caused exclusively by the Coulomb repulsion, hence $\TKE$ should be constant in time  (``fully accelerated fragments''). However, this assertion needs to be numerically verified in the quantum-mechanical framework of density functional theory.

In fact, the first question to ask is: How to evaluate $\KF(t)$ for the fragment 'f'? Classically, the velocity $\vF$ of a fragment of mass number $\AF$ is given by $\vF = \partial_t{\textbf{r}}_\mathrm{cm}^\mathrm{f}$ where $\rCMF$ is the position of the center of mass, $\rCMF = (0,0,z_{\rm cm}^{\rm \mathrm{f}})$, and 
\begin{align}
\AF(t) & = \int_{\volF} \dvol\, \rho(\rvec,t) , \\
z_\mathrm{cm}^\mathrm{f}(t) & = \frac{1}{\AF(t)} \int_{\volF} \dvol \, z\rho(\rvec,t) ,
\end{align}
with $\rho(\rvec,t)$ the total isoscalar density of the whole nucleus and integrations are performed over the spatial domain $\volF$ enclosing the fragment of interest. The kinetic energy of the fragment is then given by
\begin{equation}
\label{eq:T1}
\KF^{(1)}(t) = \frac{1}{2} m \AF(t) |\vF(t)|^2,
\end{equation}
with $m$ being the nucleon mass.

A more standard approach commonly used in TDDFT studies \cite{nakatsukasa2016timedependent,jin2021lise,bulgac2019fission,ren2022microscopic} relies on the total current density $\jvec = \jvec_n + \jvec_p$ \cite{engel1975timedependent,bender2003selfconsistent,schunck2019energy}, 
\begin{equation}
\jvec_{q}(\rvec,t) = \frac{1}{2i} \big[ \left( \gras{\nabla} - \gras{\nabla}'\right) \rho_{q}(\rvec,\rvec') \big]_{\rvec=\rvec'} ,
\end{equation}
with the non-local density $ \rho_{q}(\rvec,\rvec') = \sum_{\sigma} \rho_q(\rvec,\sigma,\rvec',\sigma')$.
This second form of the kinetic energy reads
\begin{equation}
\label{eq:T2}
    \KF^{(2)}(t) = \frac{\hbar^2}{2m}\, \frac{1}{\AF(t)} \left| \int_{\volF} \dvol\, \jvec(\rvec,t) \right|^2 .
\end{equation}
We note that if the kinetic energy density term $\frac{\hbar^2}{2m}\tau$ in the EDF is corrected by a factor $(1 - 1/A)$ to account for the one-body center-of-mass correction, then the prefactor $\frac{\hbar^2}{2m}$ in Eq.~\eqref{eq:T2} must be multiplied by an additional factor $(1 - 1/A)^2$ to ensure consistency.

\begin{figure}[!htb]
    \centering
    \includegraphics[width=1.0\columnwidth]{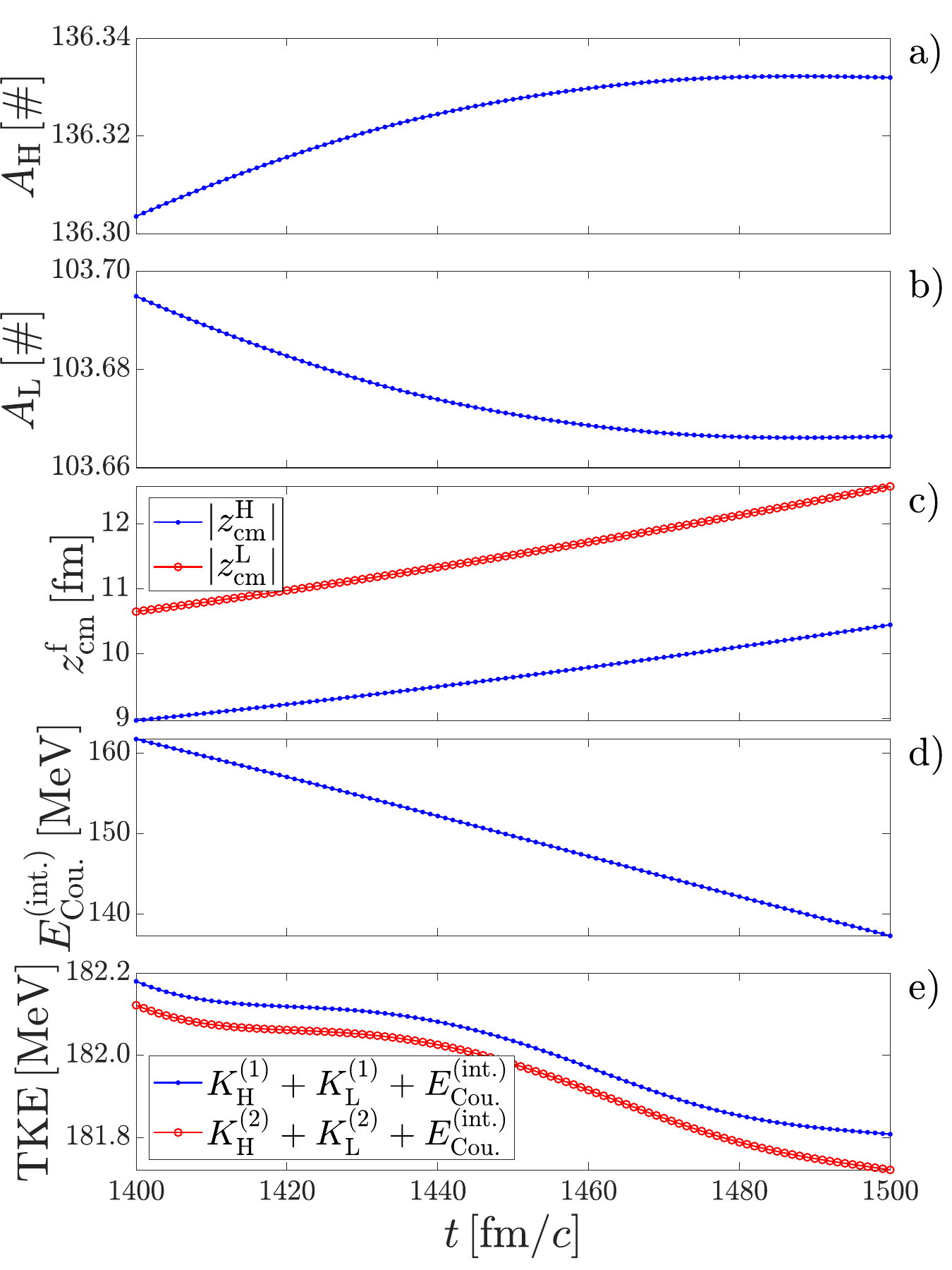}
    \caption{Observables of trajectory \#1 from Sec.~\ref{subsec:conv_TDHFB} during the first $100\,\mathrm{fm}/c$ after scission (at $t=1400\,\mathrm{fm}/c$). Panels a) and b) display the particle numbers in each fission fragment; panel c) shows the center-of-mass positions of the fragments; panel d) shows the Coulomb interaction energy between the fragments; and panel e) shows the total kinetic energy (TKE) calculated using two different methods.}
    \label{fig:TKEtest}
\end{figure}

If the average total number of particles $\AF(t)$ is constant in time, the two formulae \eqref{eq:T1} and \eqref{eq:T2} are equivalent, i.e. $\KF^{(1)} = \KF^{(2)}$. This follows directly from the continuity equation \cite{engel1975timedependent}
\begin{equation}\label{eq:mainContinutiy}
    \partial_t\rho(\rvec,t) + \frac{\hbar}{m}\nabla \cdot \jvec(\rvec,t) = 0.
\end{equation}
In Appendix~\ref{app:continuity} we provide additional technical details regarding the numerical validity of the continuity equation. Panels (a) and (b) of Figure \ref{fig:TKEtest} show that the conservation of particle number over time is indeed satisfied to high degree of accuracy in actual calculations. Since both formulas for the kinetic energy are equivalent, we will use Eq.~\eqref{eq:T1} to compute $\KH(t)$ and $\KL(t)$ in the rest of this paper.

It is worth mentioning that some practitioners~\cite{goddard2015fission,tanimura2015collective,simenel2018heavyion} adopt the following definition,
\begin{equation}
    \KF^{(3)}(t) = \frac{\hbar^2}{2m} \int_{\volF} \dvol \, \frac{|\jvec(\rvec,t)|^2}{\rho(\rvec,t)} .
\end{equation}
This expression relies on the fact that the fragment kinetic energy is expressed in terms of the local drift velocity $\boldsymbol{v}(\rvec,t)$, which is defined through
\begin{equation}
    \frac{\hbar}{m}\,\jvec(\rvec,t) = \rho(\rvec,t)\,\boldsymbol{v}(\rvec,t) .
\end{equation}
Using the continuity equation~\eqref{eq:mainContinutiy}, one can then show that
\begin{equation}
    \KF^{(3)}(t) =
    \KF^{(1)}(t)
    + \frac{1}{2} m \int_{\volF} \dvol \, \rho(\rvec,t)\,|\boldsymbol{v}(\rvec,t)-\vF(t)|^2 .
\end{equation}
Here the quantity $\boldsymbol{v}(\rvec,t)-\mathbf{v}_{\mathrm{f}}(t)$ represents the local nucleon drift velocity relative to the fragment's center-of-mass velocity.  
This contribution should naturally be interpreted as part of the fragment’s internal excitation energy, since it accounts only for relative motion (e.g., collective vibrations).
Because the goal is to separate the center-of-mass motion of the departing fragments from their internal kinetic energy, we do not employ $\KF^{(3)}$, but instead use $\KF^{(1)}$ (which is equivalent to $\KF^{(2)}$).

We now focus our attention on trajectory~\#1 from Sec.~\ref{subsec:conv_TDHFB} and analyze various conservation laws. For simplicity, we use $\nzmax=32$. For this trajectory, the two fragments are well defined and separated at time $t_i=1400\,\mathrm{fm}/c$, which we take as the initial (scission) time. We track the fragments until $t_f=1500\,\mathrm{fm}/c$, and at each time step within the interval $\Delta t = t_f - t_i$, we calculate the number of particles in the fragments, $\AH(t)$ and $\AL(t)$, the position of the two centers of mass, $\zcmH(t)$ and $\zcmL(t)$, the Coulomb interaction energy \eqref{eq:ECint}, and the $\TKE$ calculated via \eqref{eq:T1} and \eqref{eq:T2}. We use symmetric finite differences to compute $\partial_t\,z_\mathrm{cm}^\mathrm{f}(t)$ in Eq.~\eqref{eq:T1}. Figure~\ref{fig:TKEtest} shows all these observables as a function of time. 

The first two panels a) and b) display the number of particles in each fragment. As mentioned earlier, these values remain constant to within approximately $\pm0.02$ particles, confirming that $\AF(t)$ is a well-defined constant. For this particular trajectory, the values are: $\AH = 136.32 \pm 0.02$ and $\AL = 103.67 \pm 0.02$. 

Panel c) shows the position of the centers of mass of both fragments. During the time interval $\Delta t = 100\,\mathrm{fm}/c$ immediately after scission, the distance between the fragments increases by $\Delta z = 3.4\,\mathrm{fm}$. This gives an estimate of their relative post-scission velocity of around $0.034\,c$. Panel d) shows a decrease in the Coulomb interaction energy of about $25\,\mathrm{MeV}$ as the separation between fragments increases by $3.4\,\mathrm{fm}$. If the two fragments were point charges $\ZL$ and $\ZH$, the increase of the distance by $\Delta z$ would result in a decrease of the Coulomb energy by $\Delta\ECint \approx 24$ MeV, which is very close to our calculated value.

The last panel e) shows the TKE calculated using both Eq.~\eqref{eq:T1} and Eq.~\eqref{eq:T2}. We observe that the two approaches yield TKE values that agree to within a difference of $0.07\,\mathrm{MeV}$. This small discrepancy is attributed to the use of a finite basis, for which the continuity equation~\eqref{eq:mainContinutiy} -- the necessary condition for the two formulas to give the same results -- is satisfied only approximately. Moreover, although the Coulomb interaction energy $\ECint(t)$ decreases by $25\,\mathrm{MeV}$, the total quantity $\TKE(t) = \KH(t) + \KL(t) + \ECint(t)$ remains constant up to a discrepancy of just $\pm 0.2\,\mathrm{MeV}$.
This confirms that the classical picture, in which TKE remains constant after scission, is justified, and that the TKE value is well defined.
Averaging $\TKE(t)$ over the post-scission time window yields: $\TKE = 182.0 \pm 0.2\,\mathrm{MeV}$.

It is worth mentioning that the slight decreasing trend in $\TKE(t)$ seen in panel e) of Figure~\ref{fig:TKEtest} can be explained by scission neutrons, which are ejected along both axial directions following neck rupture with high velocities ($\approx 0.2\,c$), an order of magnitude larger than those of the fragments. Since the basis functions are confined within a box, these neutrons artificially reflect from the box boundaries $\sim 30\,\mathrm{fm}/c$ after the neck rupture. This reflection leads to an effective reduction in $|\partial_t z_\mathrm{cm}^\mathrm{f}(t)|$, as the fast rebounding neutrons move in the opposite direction of the corresponding fragment, which in turn slightly decreases the computed $\TKE(t)$. We confirm this behavior by examining the trajectory's density profile on a logarithmic scale. More technical details on this phenomenon are discussed in Ref.~\cite{abdurrahman2024neck}.

\begin{table}[!htb]
{
  \caption{Total kinetic energy and particle numbers of fission fragments for the three trajectories of Sec.~\ref{subsec:conv_TDHFB}, as a function of the basis size parameter $\nzmax$.}
    \label{Tab:TKEconvergence}
    \begin{ruledtabular}
    \begin{tabular}{c c c c c}
       {Traject.} & $\nzmax$ & $\AH\,[\#]$ & $\AL\,[\#]$ & $\TKE\,[\mathrm{MeV}]$\\
        \hline
        \multirow{4}{*}{\#1} 
        & 32 & $136.32\pm0.02$ & $103.67\pm0.02$ & $182.0\pm0.2$ \\
        & 36 & $136.35\pm0.02$  & $103.65\pm0.02$ & $181.8\pm0.2$ \\
        & 40 & $136.41\pm0.02$  & $103.58\pm0.02$ & $181.6\pm0.2$ \\
        & 44 & $136.42\pm0.02$  & $103.58\pm0.02$ & $181.6\pm0.2$ \\
        \hline
        \multirow{4}{*}{\#2} 
        & 32 & $121.85\pm0.01$  & $118.14\pm0.01$ & $149.1\pm0.1$ \\
        & 36 & $121.96\pm0.01$  & $118.03\pm0.01$ & $148.3\pm0.1$ \\
        & 40 & $122.75\pm0.01$  & $117.25\pm0.01$ & $148.2\pm0.1$ \\
        & 44 & $122.75\pm0.01$  & $117.26\pm0.01$ & $148.2\pm0.1$ \\
        \hline
        \multirow{4}{*}{\#3} 
        & 32 & $144.54\pm0.01$  & $95.46\pm0.01$ & $159.3\pm0.3$ \\
        & 36 & $144.19\pm0.01$  & $94.81\pm0.01$ & $158.7\pm0.2$ \\
        & 40 & $144.75\pm0.01$  & $95.25\pm0.01$ & $159.0\pm0.2$ \\
        & 44 & $144.80\pm0.01$  & $95.19\pm0.01$ & $158.9\pm0.2$ \\
    \end{tabular}
    \end{ruledtabular}
}
\end{table}

Repeating the same procedure for the other two trajectories from Sec.~\ref{subsec:conv_TDHFB} and for different values of the basis size parameter $\nzmax$, the resulting values of $\AH$, $\AL$, and the $\TKE$ are given in Table~\ref{Tab:TKEconvergence}.  Overall, we conclude that a basis with $\nzmax=32$ yields $\TKE$ values accurate to within $\pm1\,$MeV, while the fragment particle numbers are accurate within $\pm1$ nucleon—including uncertainties due to the time dependence.  This level of precision is sufficient for practical studies of the TKE dependence on fragment masses and represents a favorable compromise between computational efficiency (basis size) and accuracy.


\subsection{Excitation energy of fission fragments}
\label{subsec:conv_Ex}

At the HFB approximation, the total energy of the fissioning nucleus reads
\begin{equation}
E_{\rm tot} = 
\int \dvol\, \mathcal{E}_{\rm Skyrme}(\rvec) + 
\int \dvol\, \mathcal{E}_{\rm pair}(\rvec) + 
\ECtot,
\label{eq:Etot}
\end{equation}
where $\mathcal{E}_\mathrm{Skyrme}(\rvec)$ includes the kinetic energy as well as the contribution from the time-even and time-odd terms of the Skyrme functional, $ \mathcal{E}_{\rm pair}(\rvec)$ is the pairing functional built out of the zero-range pairing force we use, and $\ECtot$ is the total Coulomb energy (direct and exchange). After scission, when the fragments occupy distinct half-spaces, the total energy naturally separates as
\begin{equation}\label{eq:Etot_split}
E_{\rm tot} = \EH^{(\rm tot.)} + \EL^{(\rm tot.)} + \ECint,
\end{equation}
where $\ECint$ is the Coulomb interaction energy between the two fragments defined in Eq.~\eqref{eq:ECint}. The total energy of each fragment $\EFtot$ is formally analogous to Eq.\eqref{eq:Etot}, only the integration is not performed over the entire space but only over the half-space enclosing the fragment
\begin{equation}
\EFtot = 
\int_{\volF} \dvol\, \mathcal{E}_{\rm Skyrme}(\rvec) + 
\int_{\volF} \dvol\, \mathcal{E}_{\rm pair}(\rvec) + 
E_{\rm Cou}^{(\rm f)} .
\end{equation}

Since each fragment has a well defined average number of neutrons $\NF\equiv\langle\NF\rangle$ and proton $\ZF\equiv\langle\ZF\rangle$, we can perform a HFB calculation with constraints on these particle numbers and extract the ground-state energy $B(\ZF,\NF)$. The fragment excitation energy $\EFs$ is then defined as
\begin{equation}
\label{eq:ExF}
\EFs = \EFtot - \KF - B(\ZF,\NF),
\end{equation}
where $\KF$ is the fragment center-of-mass kinetic energy defined in Eq.~\eqref{eq:T1}. Notice that substituting Eq.~\eqref{eq:ExF} into Eq.~\eqref{eq:Etot_split} yields the standard energy balance formula
\begin{equation}
E_{\rm tot} = \TXE + \TKE + B(\ZH,\NH) + B(\ZL,\NL),
\label{eq:Econs}
\end{equation}
where the sum $\TXE = \EHs + \ELs$ is the total excitation energy.

Because both $\KF(t)$ and $\EFtot(t)$ evolve as the fragments accelerate, the excitation energy $\EFs(t)$ is inherently time dependent. Although physically, $\EFs(t)$ should become constant once the fragments are fully separated and internal excitations decouple from collective motion (and disregarding deexcitation through particle emission), this assumption must be numerically verified.

We again focus our attention on trajectory~\#1 from Sec.~\ref{subsec:conv_TDHFB} with basis parameter $\nzmax=32$, and track the fragments for a time interval of duration $100\,\mathrm{fm}/c$ following the scission. Analogous to Sec.~\ref{subsec:conv_TKE}, we compute the average number of neutrons and protons in each fragment
\begin{equation}
\begin{aligned}
\NH &= 83.47 \pm 0.02, &\quad \ZH &= 52.8509 \pm 0.0006, \\
\NL &= 62.53 \pm 0.02, &\quad \ZL &= 41.1488 \pm 0.0006.
\end{aligned}
\label{eq:n_parts}
\end{equation}
For these fragments, we find that the ground-state energies are
\begin{equation}
\begin{aligned}
    B(\ZH,\NH) & = - \,1143.195 \  \mathrm{MeV}, \\ 
    B(\ZL,\NL) & = - \, 882.386 \  \mathrm{MeV}.
\end{aligned}
\end{equation}
In practice, these values were obtained with constraints on the average, non-integer number of protons and neutrons \eqref{eq:n_parts} but without any constraint on deformations.

\begin{figure}[!htb]
    \centering
    \includegraphics[width=1.0\columnwidth]{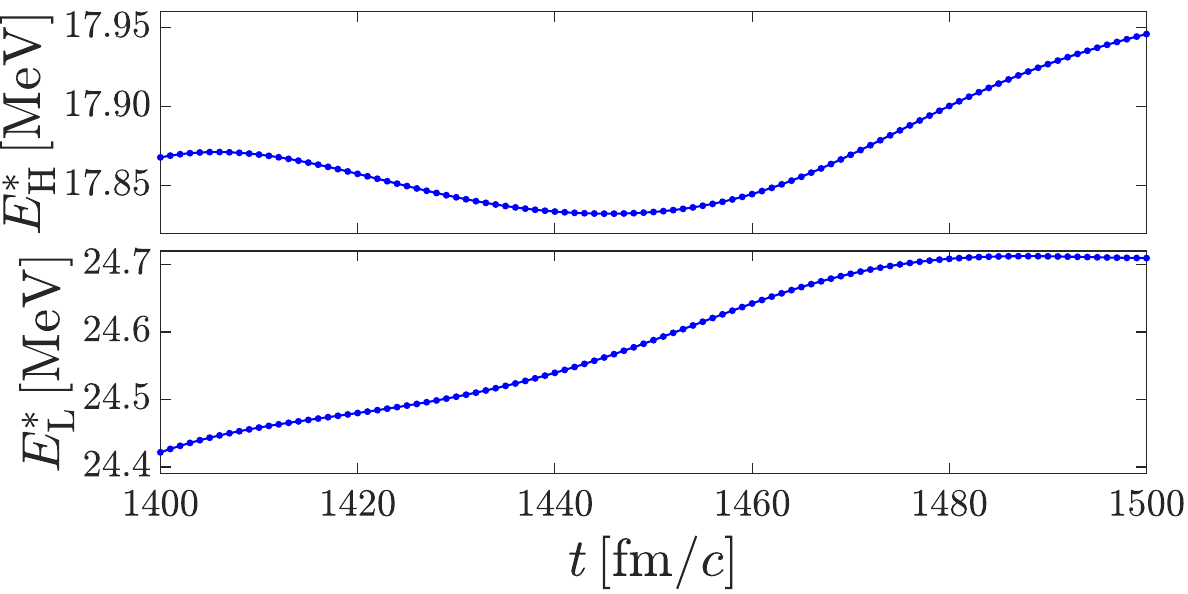}
    \caption{Same as Fig.~\ref{fig:TKEtest}, but for the excitation energies of the heavy and light fragments, $E_\mathrm{H}^*$ and $E_\mathrm{L}^*$, respectively.}
    \label{fig:Extest}
\end{figure}

Figure \ref{fig:Extest} shows the time dependence of fission fragments excitation energies calculated via Eq.~\eqref{eq:ExF}. We note that the excitation energies are indeed constant within a discrepancy interval of about $\pm 0.1\,\mathrm{MeV}$ for the heavy fragment and about $\pm 0.2\,\mathrm{MeV}$ for the light one. This demonstrates that $\EHs$ and $\ELs$ are well defined, and for this particular trajectory are: $\EHs = 17.9 \pm 0.1 \,\mathrm{MeV}$, and $\ELs = 24.6 \pm 0.2 \,\mathrm{MeV}$.

Repeating the same procedure for the other two trajectories from Sec.~\ref{subsec:conv_TDHFB} and for different values of the basis size parameter $\nzmax$, the resulting values of $\EHs$ and $\ELs$ are given in Table~\ref{Tab:Ex_convergence}. We notice that a basis with $\nzmax=32$ yields the values of excitation energies accurate to within roughly $\pm1\,$MeV. This level of accuracy offers a good balance between computational cost (basis size) and result precision.

\begin{table}[!htb]
{
  \caption{
  Excitation energies of fission fragments for the three trajectories of Sec.~\ref{subsec:conv_TDHFB}, as a function of the basis size parameter $\nzmax$.}
    \label{Tab:Ex_convergence}
    \begin{ruledtabular}
    \begin{tabular}{cccc}
       Traject. & $\nzmax$ & $\EHs\,[\mathrm{MeV}]$ & $\ELs\,[\mathrm{MeV}]$ \\
        \hline
        \multirow{4}{*}{\#1} 
        & 32 & $ 17.9 \pm 0.1 $ & $ 24.6 \pm 0.2 $ \\
        & 36 & $ 18.0 \pm 0.1 $ & $ 24.7 \pm 0.2 $ \\
        & 40 & $ 18.0 \pm 0.1 $ & $ 24.8 \pm 0.2 $ \\
        & 44 & $ 18.0 \pm 0.1 $ & $ 24.8 \pm 0.2 $ \\
        \hline
        \multirow{4}{*}{\#2} 
        & 32 & $ 39.5 \pm 0.1 $ & $ 34.9 \pm 0.1 $ \\
        & 36 & $ 39.9 \pm 0.1 $ & $ 34.2 \pm 0.1 $ \\
        & 40 & $ 41.0 \pm 0.1 $ & $ 33.7 \pm 0.1 $ \\
        & 44 & $ 41.0 \pm 0.1 $ & $ 33.8 \pm 0.1 $ \\
        \hline
        \multirow{4}{*}{\#3} 
        & 32 & $ 19.5 \pm 0.2 $ & $ 32.6 \pm 0.1  $ \\
        & 36 & $ 19.1 \pm 0.1 $ & $ 32.7 \pm 0.1 $ \\
        & 40 & $ 19.2 \pm 0.1 $ & $ 32.6 \pm 0.1 $ \\
        & 44 & $ 19.2 \pm 0.1 $ & $ 32.6 \pm 0.1 $ \\
    \end{tabular}
    \end{ruledtabular}
}
\end{table}


\section{Evaluation of TKE of ${}^{240}\mathrm{Pu}$ fission fragments}
\label{sec:tke}

In this section we present the results of our calculations of the total kinetic energy of fission fragments from the nucleus ${}^{240}\mathrm{Pu}$. Results are shown for both the SkM$^*$ and SLy5 Skyrme energy density functionals. We also examine the sensitivity of the predicted TKE to variations in EDF parameters, such as the pairing parameters and selected time-odd terms. 
All calculations are carried out using a basis parameters: 
$Z_\mathrm{box}=25\,\mathrm{fm}$, $R_\mathrm{box}=12.5\,\mathrm{fm}$, and $\nzmax=32$. 
These parameters were chosen in light of the convergence study presented in the previous section, 
which demonstrated that this basis size is sufficiently large to ensure numerical stability of the results.


\subsection{TKE for the SkM$^{*}$ functional}
\label{subsec:tke_skms}

\begin{figure}[!htb]
    \centering
    \includegraphics[width=1.0\columnwidth]{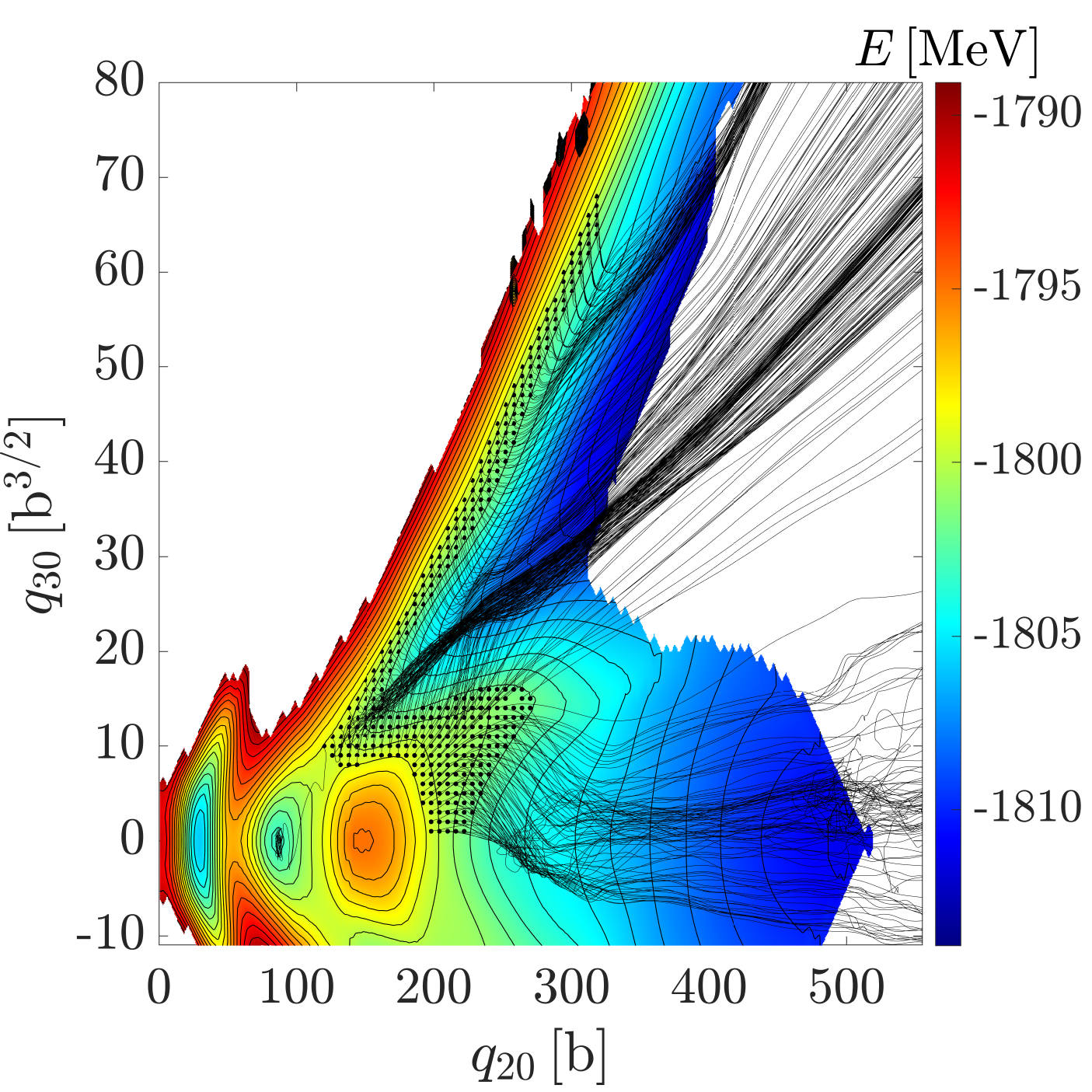}
\caption{Potential energy surface from Fig.~\ref{fig:3trajectories} with the location of the 358 initial HFB solutions (black dots) and their TDHFB trajectories (black curves); see text for details.}
    \label{fig:allTrajectoriesSkM}
\end{figure}

We begin with the SkM$^{*}$ EDF and a mixed surface–volume pairing interaction with pairing strengths
$V_0^{(n)}=-265.25\,\mathrm{MeV}\mathrm{fm}^3$, 
$V_0^{(p)}=-340.0625\,\mathrm{MeV}\mathrm{fm}^3$. 
The resulting potential energy surface is shown in Figure~\ref{fig:allTrajectoriesSkM}. 
From this PES we determine the saddle-point deformation $(q_{20}^s,q_{30}^s)=(125\,\mathrm{b},\,8\,\mathrm{b}^{3/2})$ and the corresponding 
saddle-point energy $E^s=-1799\,\mathrm{MeV}$. 

Next, we generate a mesh of starting points $(q_{20},q_{30})$ beyond the saddle-point deformation, 
restricted to the energy window $E^s - 3\,\mathrm{MeV} \leq E \leq E^s$ 
and to quadrupole deformations $q_{20} \leq 320\,\mathrm{b}$. 
These starting points are indicated as black dots in Fig.~\ref{fig:allTrajectoriesSkM}. 
For each $(q_{20},q_{30})$ point, the constrained HFB solution is obtained and then propagated in time, producing a trajectory in deformation space. 
The full set of trajectories is shown as black curves in Fig.~\ref{fig:allTrajectoriesSkM}. 
For this example, the mesh contains 358 starting $(q_{20},q_{30})$ points, leading to 358 trajectories in total. 
For each trajectory that undergoes fission, the TKE is computed as described in Sec.~\ref{subsec:conv_TKE}.

\begin{figure}[!htb]
    \centering
    \includegraphics[width=1.0\columnwidth]{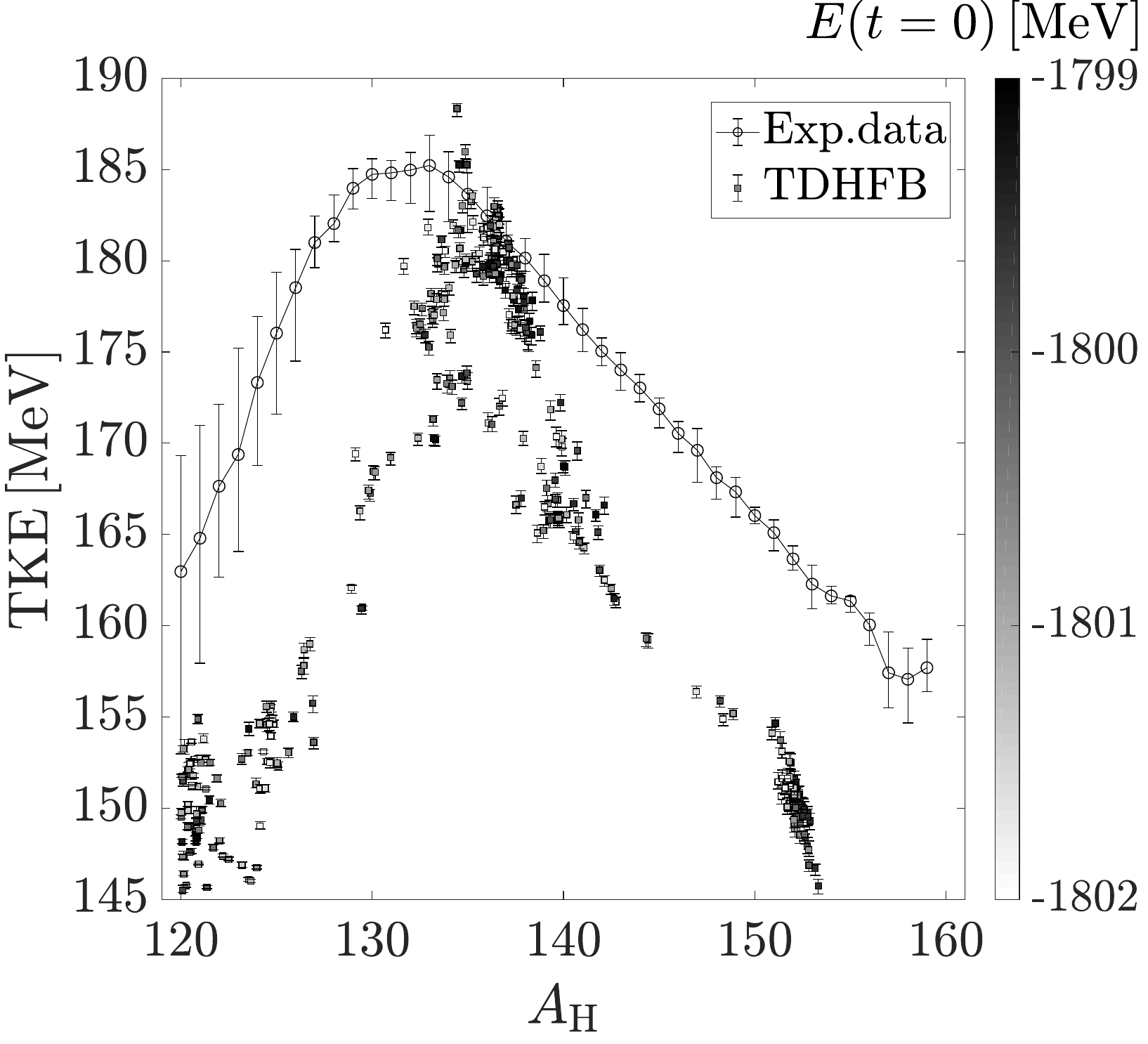}
\caption{Calculated TKE values for all 358 TDHFB trajectories from Fig.~\ref{fig:allTrajectoriesSkM}. 
Error bars represent the range $[\min \TKE(t),\max \TKE(t)]$ for each trajectory, while the TKE is calculated as average $\TKE=\langle \TKE(t)\rangle$. 
Each point is color coded by the initial constrained HFB energy $E(t=0)$. 
Experimental data are obtained by averaging results from \cite{wagemans1984comparison,geltenbort1986precision,nishio1995measurement,tsuchiya2000simultaneous}.}
    \label{fig:TKE_SkM}
\end{figure}

The calculated TKE values for all fissioning trajectories are shown in Fig.~\ref{fig:TKE_SkM} 
as a function of heavy fragment mass $\AH$, together with the average experimental data taken from EXFOR. 
All fragmentations with $\AH$ between 120 and 153 appear in the TDHFB calculations. 
Each TDHFB data point $(\AH,\TKE)$ is shown with an error bar indicating 
$\min \TKE(t)$ and $\max \TKE(t)$, confirming that $\TKE(t)$ remains essentially constant 
throughout the time evolution for all trajectories as discussed in Sec.~\ref{subsec:conv_TKE}. 
In addition, the points are color-coded according to their initial HFB energy $E(t=0)$. 
No systematic dependence of $\TKE$ on $E(t=0)$ is observed, demonstrating that the precise 
value of the starting HFB energy below the saddle-point energy does not play a significant role 
in the calculation of TKE.

For the most probable fragmentation, $\AH \approx 136$, the calculated $\TKE$ is in good agreement with the experimental data. 
In contrast, for more symmetric fragmentations with $\AH \leq 130$ and more asymmetric ones with $\AH \geq 140$, 
the calculated $\TKE$ is consistently underestimated by around $10\,\mathrm{MeV}-20 \,\mathrm{MeV}$ relative to experiment. 
To address this discrepancy, we next explore the sensitivity of the calculated $\TKE$ to variations in specific EDF parameters.


\subsection{Role of pairing correlations}

Since pairing is known to play a crucial role in fission dynamics 
(e.g., stronger pairing reduces the saddle-to-scission time, while weaker pairing increases it, 
and fission may even fail in the absence of pairing), 
we first examine the sensitivity of the calculated $\TKE$ to the pairing strength parameters 
$V_0^{(n)}$ and $V_0^{(p)}$ of our pairing interaction
\begin{equation}
V_{\rm pair}(\rvec,\rvec') = V_{0}^{(q)} \left[ 1 - \eta\frac{\rho(\rvec)}{\rho_c} \right] \delta(\rvec-\rvec').
\end{equation}
while keeping the surface–volume mixing parameter fixed at $\eta = 0.5$.

\begin{figure}[!htb]
    \centering
\includegraphics[width=1.0\columnwidth]{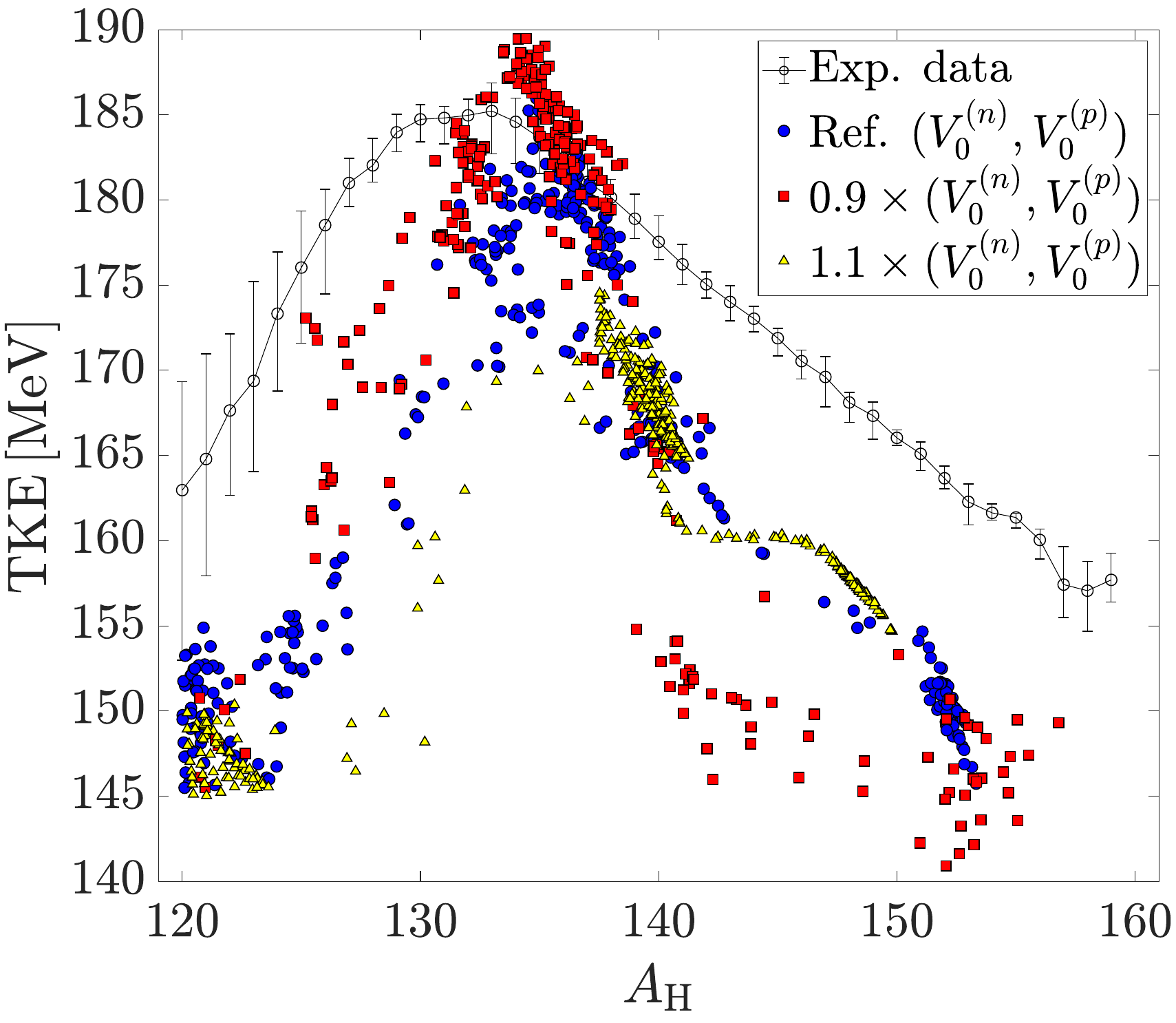}
    \caption{TKE obtained with the SkM$^{*}$ functional and three different pairing strength parameters, $V_0^{(n)}$ and $V_0^{(p)}$. Results correspond to the reference pairing strengths same as in Fig.~\ref{fig:TKE_SkM}, as well as to 10\% reduced and 10\% enhanced values.}
    \label{fig:TKE_SkM_pairingStrengths}
\end{figure}

We repeat the same workflow presented in the previous section: computing PES, selecting constrained HFB starting points within $3\,\mathrm{MeV}$ below the the saddle-point energy, and evolving trajectories in time, but for three different pairing strength parameters 
$V_0^{(n)}$ and $V_0^{(p)}$. The reference calculation corresponds to the initial pairing strengths used in Fig.~\ref{fig:TKE_SkM}, i.e.
$V_0^{(n)} = -265.25\,\mathrm{MeV}\mathrm{fm}^3$, $V_0^{(p)} = -340.0625\,\mathrm{MeV}\mathrm{fm}^3$. The two other calculations correspond to 10\% reduced strengths: $0.90\times V_0^{(n)}$, $0.90\times V_0^{(p)}$, and 10\% enhanced strengths: $1.10\times V_0^{(n)}$, $1.10\times V_0^{(p)}$, respectively.
The three sets of $\TKE$ are shown in Figure~\ref{fig:TKE_SkM_pairingStrengths}.
Near the most probable fragmentation $\AH\approx 136$, we notice a consistent trend that reduced pairing increases the $\TKE$ while enhanced pairing reduces it. However, near-symmetric fragmentations ($120\leq \AH\leq 125$) and very asymmetric fragmentations $(150\leq \AH\leq 155)$ are not significantly affected by variations of pairing strengths.

\begin{figure}[!htb]
    \centering
\includegraphics[width=1.0\columnwidth]{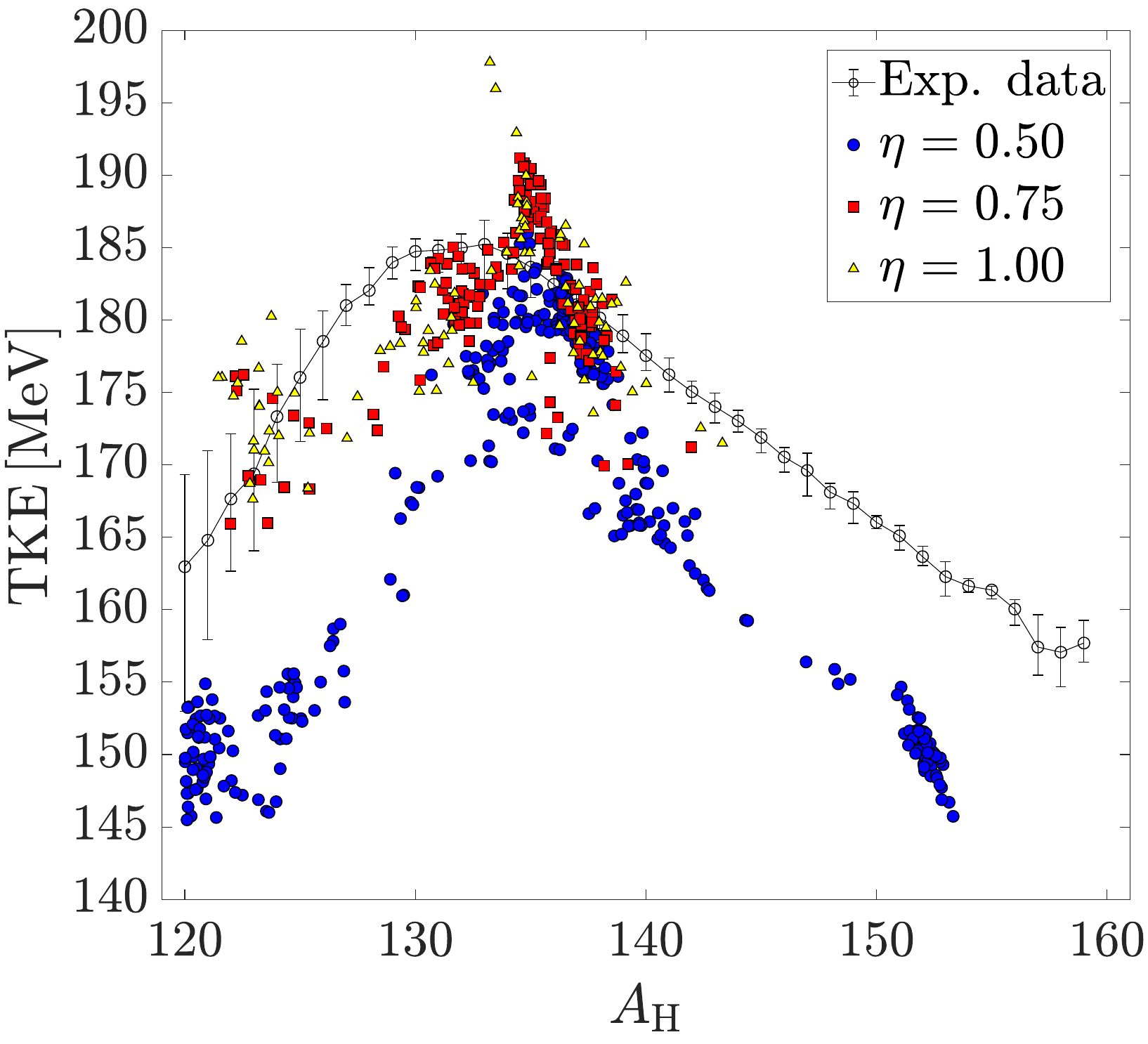}
\caption{$\TKE$ obtained with the SkM$^{*}$ functional for three surface-volume mixing parameters, $\eta=0.50$, $\eta=0.75$, and $\eta=1.00$ and fixed pairing strengths $V_0^{(n)}=-265.25\,\mathrm{MeV}\mathrm{fm}^3$ and $V_0^{(p)}=-340.0625\,\mathrm{MeV}\mathrm{fm}^3$.}

    \label{fig:TKE_SkM_SurfacePairing}
\end{figure}

In addition to the pairing strengths $V_0^{(q)}$, the pairing force is also characterized by the surface–volume mixing parameter $\eta$. This parameter is most often set to $\eta=0.5$ to reflect insights from many-body theory \cite{baldo1999surface,lesinski2009nonempirical}.
We fix the pairing strengths at $V_0^{(n)} = -265.25,\mathrm{MeV}\mathrm{fm}^3$ and $V_0^{(p)} = -340.0625,\mathrm{MeV}\mathrm{fm}^3$, but now consider three values of the surface–volume mixing parameter: $\eta=0.50$ (reference), $\eta=0.75$, and $\eta=1.00$ (pure surface pairing). The results are shown in Figure~\ref{fig:TKE_SkM_SurfacePairing}. For more surface-dominated pairing ($\eta=0.75$ and $\eta=1.00$), trajectories initialized with very symmetric or very asymmetric constrained HFB deformations fail to fission even after $15{,}000\,\mathrm{fm}/c$ of time evolution, becoming trapped in local minima instead. In contrast, trajectories leading to near-symmetric fragmentations in the range $124 \leq \AH \leq 132$ exhibit a significant increase of about $10\,\mathrm{MeV}$ in $\TKE$ when $\eta=0.75$ or $\eta=1.00$ is used, relative to the reference case with $\eta=0.50$. We therefore conclude that, for near-symmetric fragmentations, the surface–volume character of the pairing interaction has a significant impact on the predicted $\TKE$, highlighting the importance of properly constraining the parameter $\eta$ in future EDF optimization efforts.


\subsection{TKE for the SLy5 functional}
\label{subsec:tke_sly5}

To assess the impact of the parameterization of the Skyrme functional itself on the predicted $\TKE$, we 
repeat the calculations from the previous section for the SLy5 parameterization of the EDF \cite{chabanat1998skyrme}. Recall that the SLy5 functional contains a tensor term and has substantially different nuclear matter and deformation properties from SkM* \cite{jodon2016constraining}. Following global calculations of \cite{li2024multipole}, the reference pairing strength parameters are now set to
$V_0^{(n)}=-297.149\,\mathrm{MeV}\,\mathrm{fm}^3$
and
$V_0^{(p)}=-334.353\,\mathrm{MeV}\,\mathrm{fm}^3$.

\begin{figure}[!htb]
    \centering
    \includegraphics[width=1.0\columnwidth]{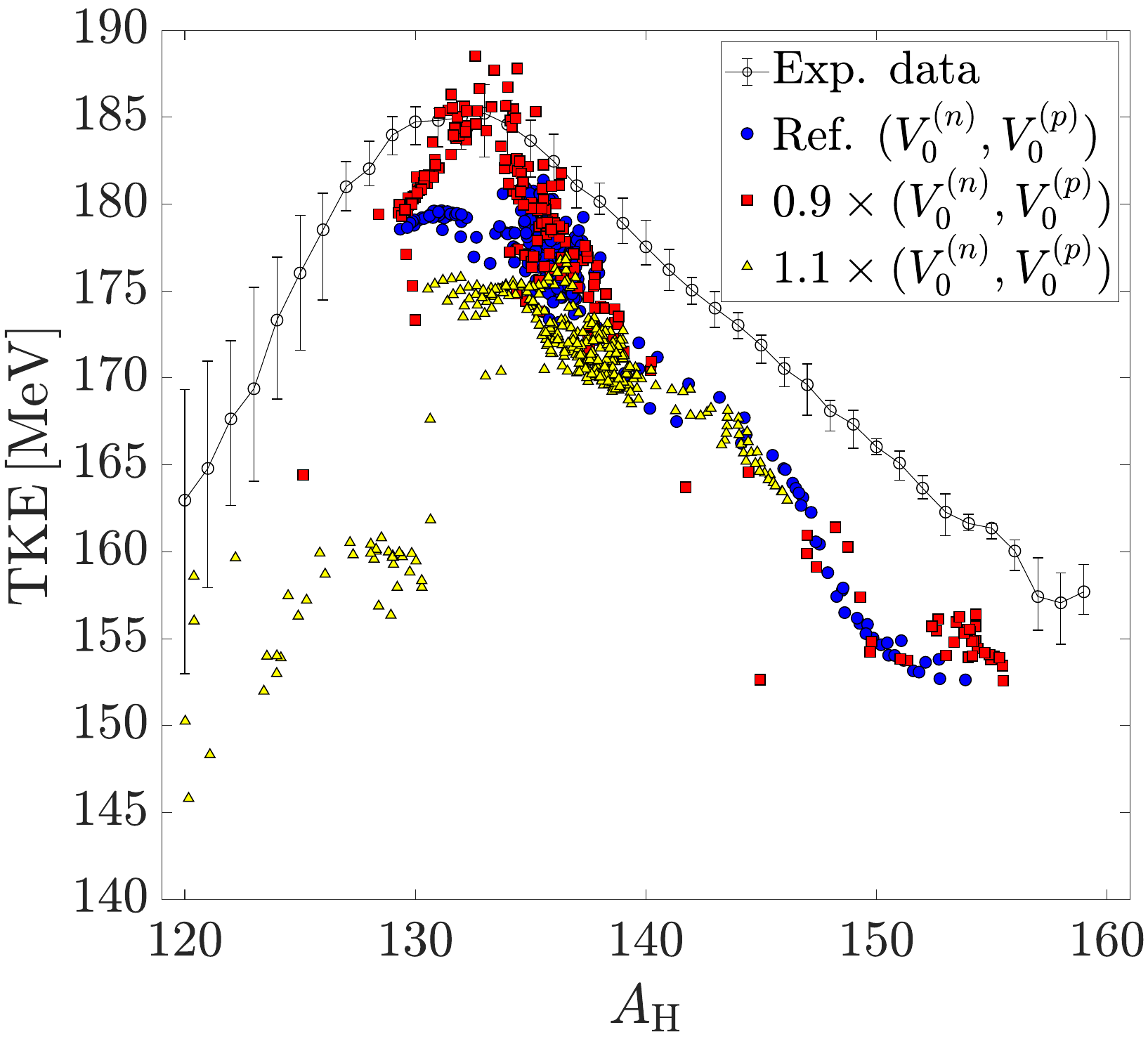}
    \caption{Same as Fig.~\ref{fig:TKE_SkM_pairingStrengths}, but for the SLy5 functional. }
    \label{fig:TKE_SLy5_pairing}
\end{figure}

Figure \ref{fig:TKE_SLy5_pairing} is the analogue for SLy5 of Fig.~\ref{fig:TKE_SkM_pairingStrengths} for SkM*, namely it shows the predicted $\TKE$ for the reference parameterization of the pairing interaction as well as for the two cases where both pairing strengths are multiplied by a factor 0.9 and a factor 1.1. With this choice of pairing correlations, the PES of SLy5 exhibits many local minima in the symmetric low-$q_{30}$ region. As a result, nearly all trajectories starting with low $q_{30}$ become trapped in local minima and never undergo fission, which results in missing fragmentations for $\AH\leq 130$ for both the reference case (blue circles in Fig.~\ref{fig:TKE_SLy5_pairing}) and the reduced-pairing case (red squares in Fig.~\ref{fig:TKE_SLy5_pairing}). Only in the case of increased pairing do we obtain a significant number of fully-scissioned trajectories.

We first note that the TKE predictions with SLy5 are qualitatively and quantitatively very similar to the ones with SkM*.
For the most probable fragmentation around $\AH\approx 136$, we again observe that reduced (enhanced) pairing strengths yield larger (smaller) TKE that are closer to experimental measurements. 
However, as in the case of the SkM* functional, very asymmetric fragmentations with $\AH\geq 145$ are not significantly affected by varying the pairing strength parameters. 

\begin{figure}[!htb]
    \centering
    \includegraphics[width=1.0\columnwidth]{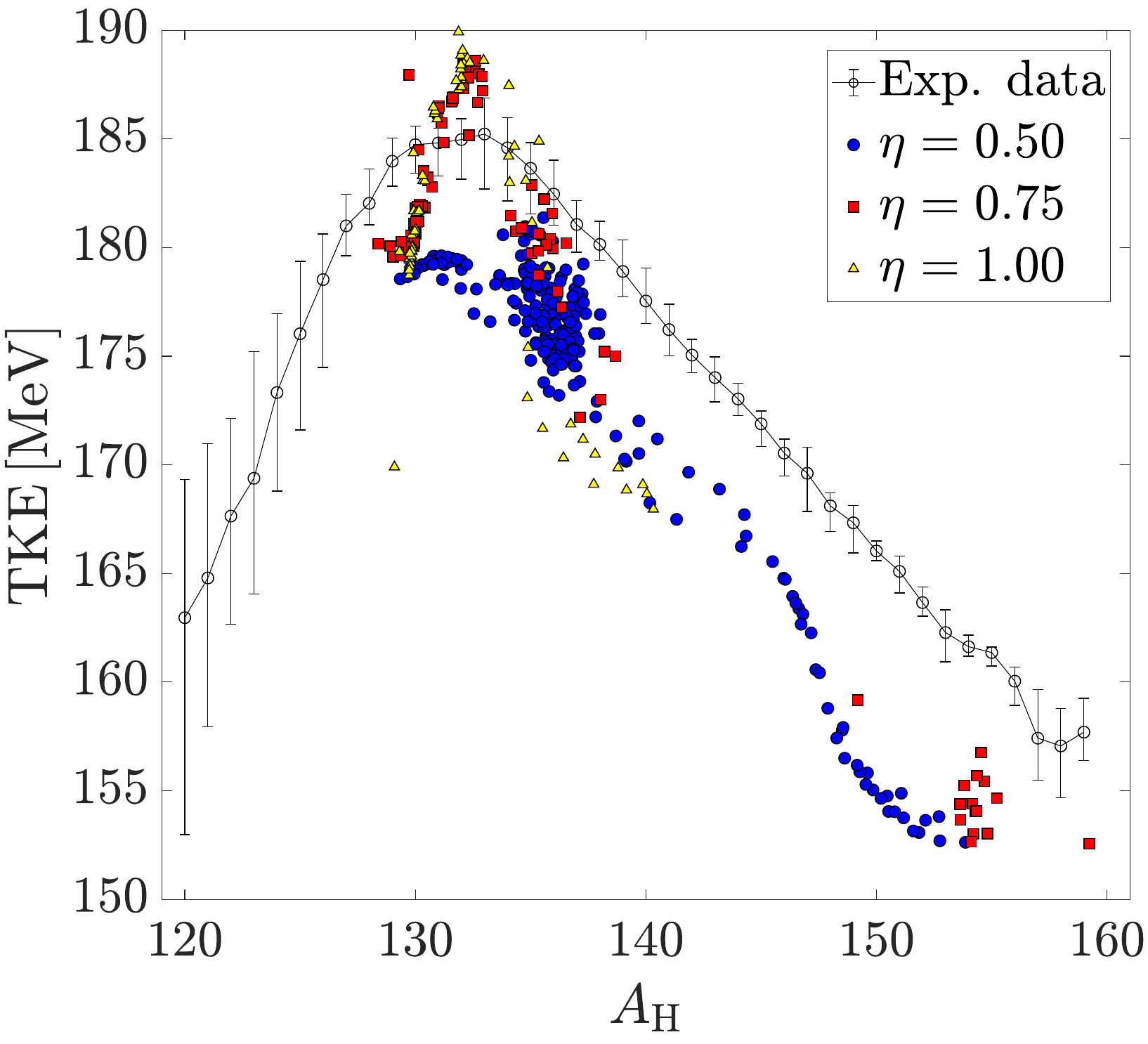}
    \caption{Same as Fig.~\ref{fig:TKE_SkM_SurfacePairing}, but for the SLy5 functional.}
    \label{fig:TKE_SLy5_SurfacePairing}
\end{figure}

The situation is analogous for the role of the surface-volume pairing parameter $\eta$. Figure \ref{fig:TKE_SLy5_SurfacePairing} shows the TKE for three different values of $\eta$: $\eta=0.50, 0.75$ and 1.0. We found that symmetric low-$q_{30}$ trajectories are still trapped in local minima irrespective of the value of $\eta$ and do not undergo fission. In fact, for pure surface-pairing interactions ($\eta=1.00$), most trajectories fail to lead to scission. Like in the case of SkM*, surface-dominated pairing leads to better agreement with TKE near most likely fission.

Overall, the results of this and the previous section suggest that the TKE are not overly sensitive to the parametrization of the energy functional, both its particle-hole and particle-particle (pairing) channel. In all cases, the TKE near most likely fission are relatively well reproduced. This confirms earlier results \cite{bulgac2016induced,bulgac2019fission,ren2022microscopic,li2025microscopic} obtained with both the SLy4 and SeaLL1 Skyrme functionals and covariant functionals. However, our TKE predictions for near-symmetric or very asymmetric fission fall systematically short by up to 20 MeV as already suggested in \cite{ren2022microscopic}.

\subsection{Role of the current density}

Since the TKE depends explicitly on the current density $\jvec$, see Eq.~\eqref{eq:T2}, one may also be able to improve the agreement with experimental data by artificially increasing or reducing the term of the energy functional that is proportional to it.
Recall that the time-odd component of the Skyrme EDF reads
\begin{equation}
E^{(\rm{odd})}_\mathrm{Skyrme} = \sum_{t=0,1} \int \dvol \, \mathcal{E}_{t}^{(\rm odd)}(\rvec),
\label{eq:Eodd_def}
\end{equation}
where
\begin{align}
\mathcal{E}_{t}^{(\rm odd)}(\rvec) = 
&+ C^{ss}_{t} \big[\rho(\rvec)\big] \, \mathbf{s}_{t}(\rvec)\cdot\mathbf{s}_{t}(\rvec)  \nonumber \\ 
&+ C^{s\Delta s}_{t} \, \mathbf{s}_{t}(\rvec) \cdot \Delta\mathbf{s}_{t}(\rvec) \nonumber \\
&+ C^{sT}_{t} \, \mathbf{s}_{t}(\rvec) \cdot \mathbf{T}_{t}(\rvec) \nonumber \\
&+ C^{jj}_{t} \, \mathbf{j}_{t}(\rvec) \cdot \mathbf{j}_{t}(\rvec) \nonumber \\
&+ C^{s\nabla\times j}_{t} \, \mathbf{s}_{t}(\rvec) \cdot \nabla\times\mathbf{j}_{t}(\rvec) .
\label{eq:Eq:timeOdd}
\end{align}
In our simulations, we found that all time-odd terms contribute to the total energy by less than $0.5$~MeV 
with the exception of the $C^{jj}_{t} |\mathbf{j}_{t}(\rvec)|^2$ term, which typically contributes about 15 MeV to the total TDHFB energy. Note that since this is a time-odd term, it does not affect the potential energy surface
or the calculation of the starting HFB configuration, but only plays a role in the TDHFB time evolution.

After scission this term becomes significantly large. We therefore expect that by reducing the absolute value of the coupling constant $C_{t}^{jj}$, the current density $|\mathbf{j}_t(\rvec)|^2$ will increase in order to compensate for the decrease of $|C_{t}^{jj}|$, which could in turn lead to an increased TKE and reduce the observed discrepancy between the calculated TKE and  the experimental data. To verify this hypothesis, we perform a set of two calculations, both for the SkM$^{*}$ and SLy5 functionals, where we multiply the $C_{t}^{jj}$ by either a factor 1/2 or a factor 2 compared to its initial value.

\begin{figure}[!htb]
    \centering
\includegraphics[width=1.0\columnwidth]{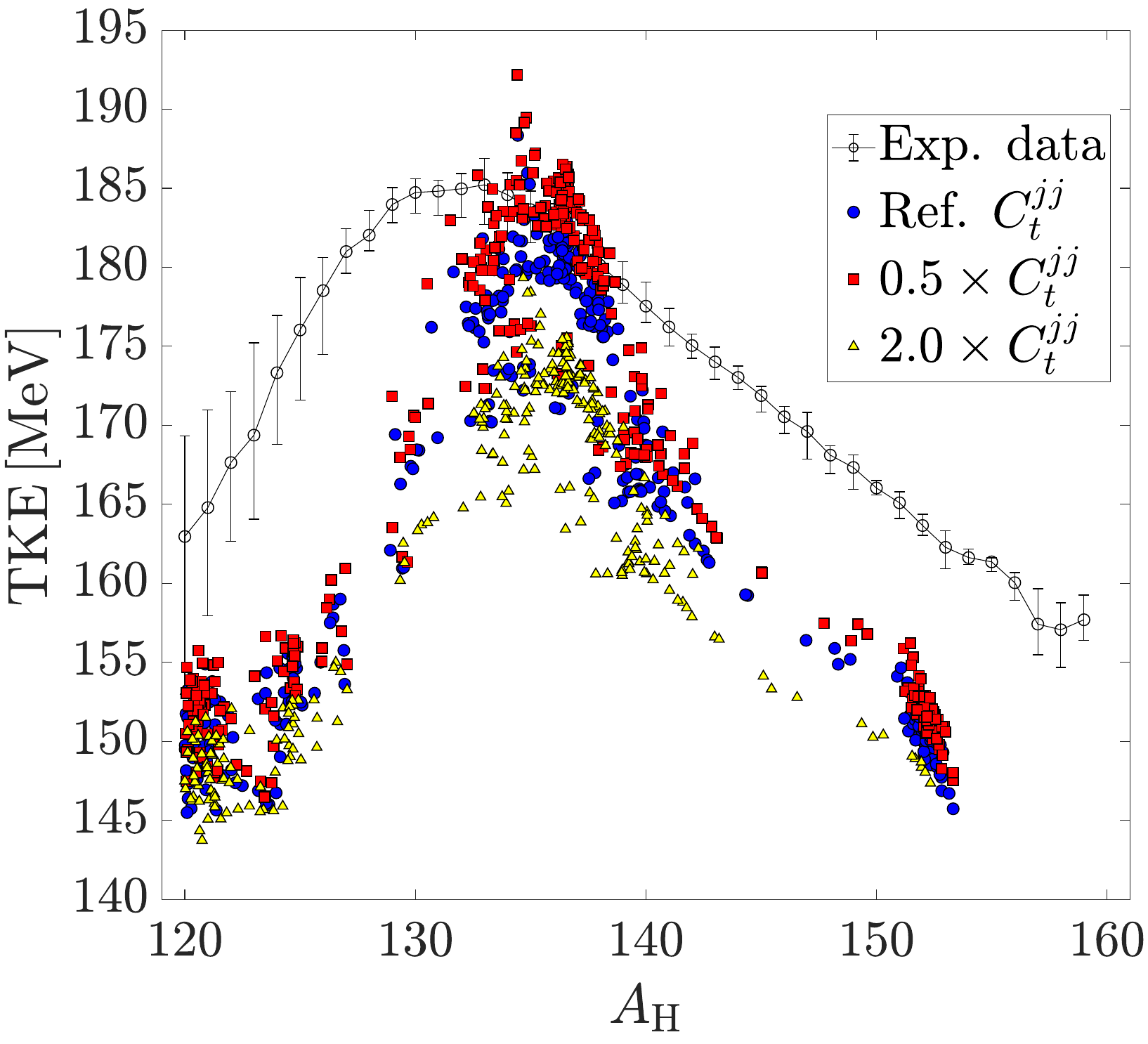}
    \caption{TKE obtained with the SkM$^{*}$ functional using its reference value of the $C_{t}^{jj}$ coupling constant (blue circles), with the absolute value of $C_{t}^{jj}$ reduced by a factor of two (red squares), and with the absolute value increased by a factor of two (yellow triangles).}
    \label{fig:TKE_SkM_Cjj}
\end{figure}

\begin{figure}[!htb]
    \centering
\includegraphics[width=1.0\columnwidth]{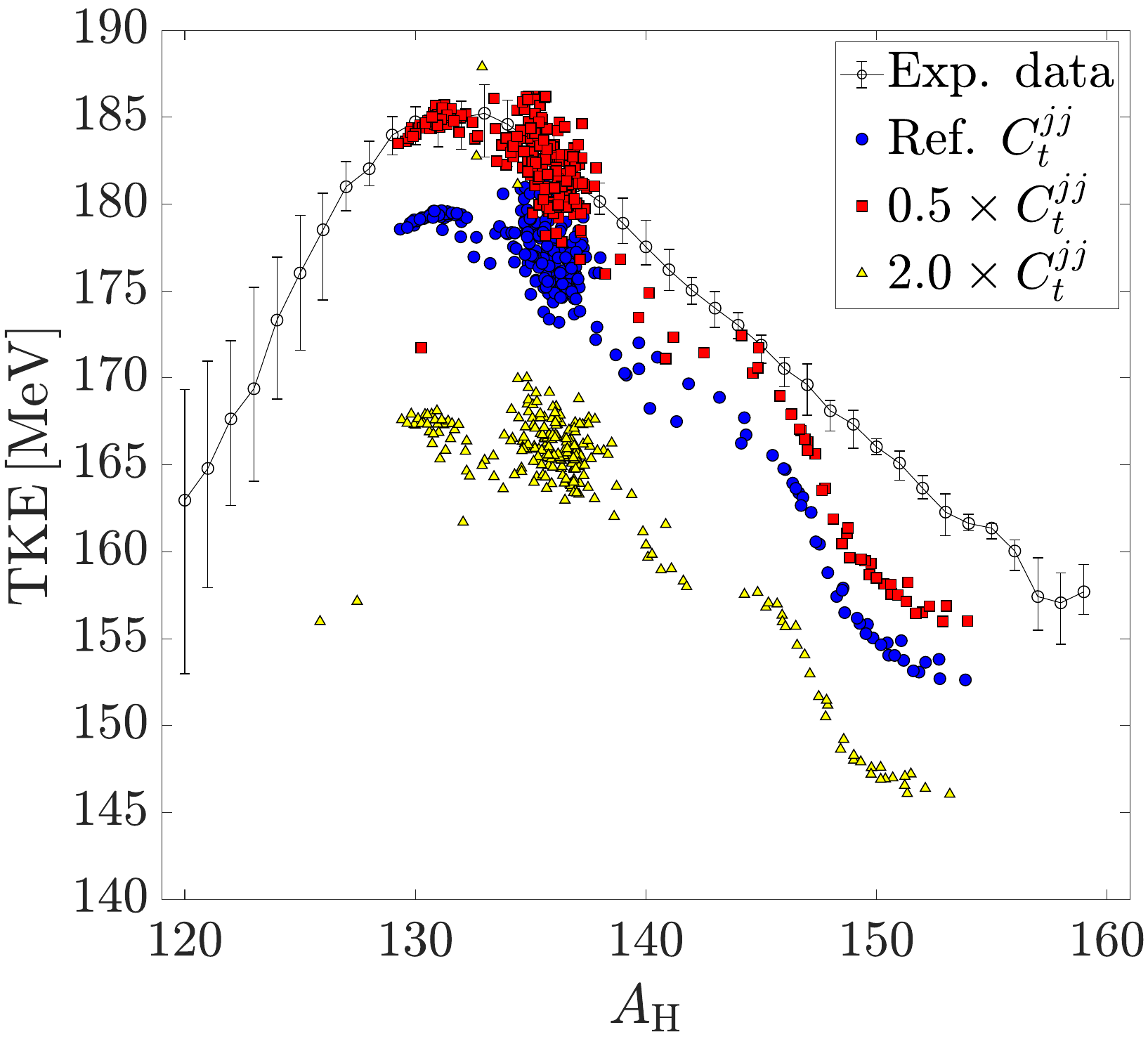}
    \caption{Same as Fig.~\ref{fig:TKE_SkM_Cjj} but for the SLy5 functional.}
    \label{fig:TKE_SLy5_Cjj}
\end{figure}

Figures ~\ref{fig:TKE_SkM_Cjj} and \ref{fig:TKE_SLy5_Cjj} show the results of these calculations for the SkM$^{*}$ and SLy5 functionals, respectively. As anticipated, we find that reducing the absolute value of the $C_{t}^{jj}$ coupling constant leads to a substantial increase in the TKE across all fragmentations, while increasing it has the opposite effect. For example, in the case of SLy5 with a reduced $C_{t}^{jj}$, the calculated TKE is in very good agreement with experiment for all fragmentations with $130 \leq \AH \leq 146$, and shows only a $\sim$5~MeV discrepancy for fragmentations with $\AH \geq 147$. 
In fact, the overall shape of the $\TKE(\AH)$ curves is largely independent of the $C_{t}^{jj}$ coupling constants. This demonstrates that tuning a single time-odd Skyrme EDF parameter, which affects only the dynamical part of the calculation, can significantly improve agreement with experimental data.

The drawback of this simple adjustment of a single time-odd coupling constant $C_{t}^{jj}$ in the Skyrme EDF is that the terms $\rho(\rvec)\tau(\rvec)$ and $|\jvec(\rvec)|^2$ no longer appear exclusively in the Galilean-invariant combination $\rho(\rvec)\tau(\rvec)-|\jvec(\rvec)|^2$. This property is needed to guarantee both Galilean invariance of the total energy and, most importantly, the validity of the continuity equation \cite{engel1975timedependent}. As discussed in Sec.~\ref{subsec:conv_TKE}, if the continuity equation is not satisfied, the two definitions of the fragment kinetic energy, $\KF^{(1)}$ and $\KF^{(2)}$, are not equivalent. Nevertheless, the present analysis highlights which EDF parameters most strongly influence the TKE and shows that even a single time-odd parameter can be tuned to reduce the discrepancy between theoretical predictions and experimental data.


\section{Excitation energies of ${}^{240}\mathrm{Pu}$ fission fragments}
\label{sec:excitation}

\begin{figure}[!htb]
    \centering
\includegraphics[width=1.0\columnwidth]{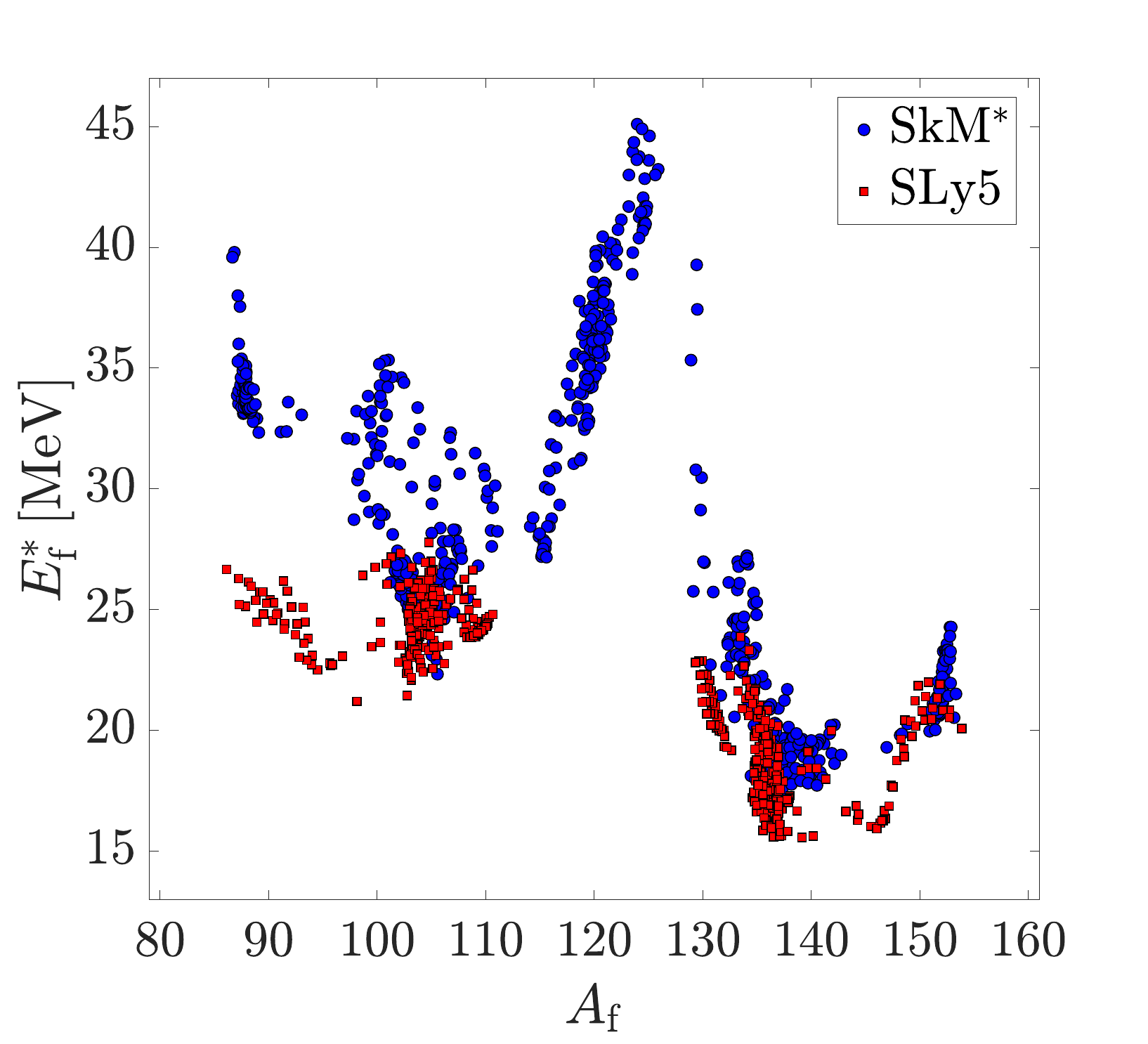}
    \caption{Excitation energy of fission fragments for $^{240}$Pu extracted fom the TDHFB evolutions for the SkM*  (blue circles) and SLy5 (red squares) functional. The results are only presented for the reference cases with mixed surface-volume pairing with the initial parameterization of the pairing channel and initial value of $C_{t}^{jj}$ coupling constant.}
    \label{fig:Ex}
\end{figure}

In this last section, we present the excitation energies of fission fragments based on the formula \eqref{eq:ExF}. Because of the conservation of total energy and the fact that our predicted TKE are -- except for most likely fission -- overestimated by 10-20 MeV, we expect the excitation energy to be overestimated accordingly. Figure \ref{fig:Ex} shows the excitation energy of all fission fragments for both the SkM* and SLy5 functional. In both cases, we used the reference parameterizations where the value of the $C_{t}^{jj}$ coupling constant has not been modified, and where the pairing channel is represented by a mixed surface-volume ($\eta=0.5$) interaction with the pairing strength parameters given at the beginning of Sec.~\ref{subsec:tke_skms} and \ref{subsec:tke_sly5}, respectively. Only the results near $\AH \approx 130-140$ (and, therefore, $\AL\approx 100-110$) should be somewhat reliable because both functionals give similar results that are also quite close to experimental values.

The predictive power of the TDHFB theory to estimate the excitation energy of fission fragments can also be tested by computing the ratio of the excitation energies between light and heavy fragments. As mentioned in the introduction, traditional deexcitation models of fission fragments such as CGMF \cite{talou2021fission}, FREYA \cite{verbeke2018fission} or FIFRELIN \cite{thulliez2019neutron} always rely on sharing the total excitation energy. This is often parameterized by the ratio $R_T = T_\mathrm{L} / T_\mathrm{H}$ where $T_\mathrm{L}$ ($T_\mathrm{H}$) is the internal temperature of the light (heavy) fission fragment. In the Fermi gas model, internal temperature and excitation energy are related through $E^{*} = a(E^*)T^2$ with $a(E^*)$ the energy-dependent level-density parameter.

\begin{figure}[!htb]
    \centering
\includegraphics[width=1.0\columnwidth]{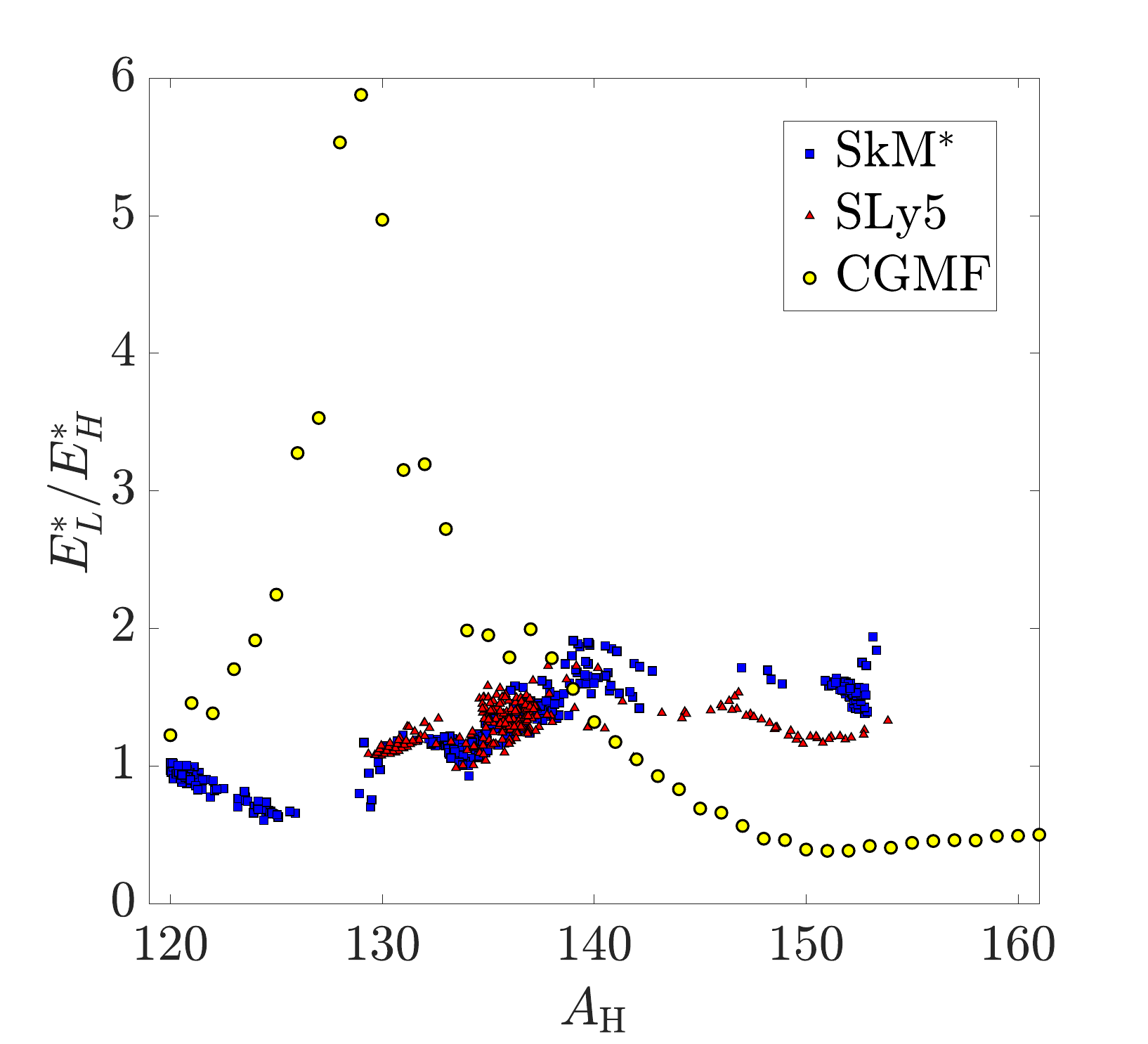}
    \caption{Ratio of excitation energy between the light and heavy fragment. Results from TDHFB simulations with the reference SkM* (blue squares) and SLy5 (red triangles) functionals are compared with the phenomenological fit from the CGMF code (yellow circles).}
    \label{fig:Ex_ratios}
\end{figure}

In CGMF the $R_T(\AH)$ parameter was fitted independently for each heavy fragment mass $\AH$ to reproduce the prompt neutron spectrum and should thus provide a rather reliable estimate of the ratios of temperatures in the two fragments. Starting from the conservation of energy given in Eq.~\eqref{eq:Econs}, we can extract the TXE of the reaction. Upon introducing the $R_T(\AH)$ fit available from the CGMF distribution list \cite{talou2021fission} and using the model for level density implemented in that code to obtain the form $a(E^*)$, we can compute the excitation energy for the heavy fragment by simply solving
\begin{equation}
\EHs \left ( 1 + R_T^2(\AH) \frac{a_L(\TXE-\EHs)}{a_H(\EHs)} \right) = \TXE.
\end{equation}
Figure \ref{fig:Ex_ratios} compares the excitation energy ratio $\ELs / \EHs$ computed from the TDHFB solutions with the phenomenological ratio extracted from the CGMF code. 

The large discrepancy between the TDHFB results and the CGMF fit is worrying, because it suggests that the problem with TDHFB is not restricted to underestimating the TKE: one could still have a theory that overestimates TXE but gives the correct {\em ratios} of excitation energies. Instead, if we take the CGMF fit at face value, it seems that TDHFB does not share the excitation energy correctly. 


\section{Conclusion}
\label{sec:conclusion}

In this article, we have provided a comprehensive analysis of the predictive power of the time-dependent Hartree-Fock-Bogoliubov (TDHFB) theory to estimate the total kinetic energy (TKE) of fission fragments and the excitation energy of fission fragments. Our large-scale TDHFB simulations covered a very broad range of fission fragments from near-symmetric to very asymmetric fission while being numerically extremely well converged. We confirmed earlier results that the TKE for most likely fission is rather well predicted in TDHFB. However, we proved that this result is an exception to the general rule: for most other fragmentations, either near-symmetric fragments or for very asymmetric fragments, the TKE is underestimated by 10--20 MeV. This conclusion is rather independent of the parameterization of the energy functional, with SkM* and SLy5 giving similarly bad results or of the form of the pairing interaction or the intensity of pairing correlation. Only direct modification to the current-density coupling constant ($C_{t}^{jj}$) could significantly improve agreement with experiment -- at the price of breaking Galilean invariance, the validity of the continuity equation and the equivalence between the standard prescriptions to compute the TKE. Not only are the TKE poorly reproduced, the TDHFB theory does not seem to predict the correct sharing of excitation energy between the fragments.

All calculations in this work were performed assuming axial symmetry. It would be rather surprising if our conclusions changed dramatically upon including the effect of triaxiality, but this cannot be disregarded entirely. While the results with two rather different Skyrme functionals are qualitatively very similar, one cannot exclude either that there exists some functional that would give a better agreement with experiment. In particular, it would be interesting to see the effect of a fully finite-range interaction such as the Gogny force. Since it is implemented in the HO basis, our new solver would naturally lend itself to such an extension. Overall, the most likely explanation for the poor 
performance of TDHFB is that the theory is intrinsically limited since it does not allow for configuration mixing and, in its current implementation, does not allow for symmetry restoration. Current non-variational configuration mixing of precomputed TDHFB trajectories \cite{marevic2023quantum,li2025microscopic} is unlikely to bring major improvements. More ambitious extensions such as the time-dependent RPA \cite{balian1992correlations,simenel2011particlenumber}, coupled configuration mixing and time evolution, or a formalism mixing large-amplitude collective motion and thermal excitations \cite{dietrich2010microscopic} may be more promising avenues.


\begin{acknowledgments}
Discussions with Z. Drmač regarding the
implementation of the Autonne-Takagi solver are gratefully acknowledged.
This work was partly performed under the auspices of 
the US Department of Energy by the Lawrence Livermore National Laboratory under 
Contract DE-AC52-07NA27344. 
This material is partially based upon work supported by the U.S. Department of Energy, Office of Science, Office of Advanced Scientific Computing Research and Office of Nuclear Physics, Scientific Discovery through Advanced Computing (SciDAC) program. Computing support for this work came from the Lawrence 
Livermore National Laboratory Institutional Computing Grand Challenge program.
\end{acknowledgments} 

\appendix


\section{Convergence with respect to box size}
\label{app:box}

In Sec.~\ref{subsec:conv_HFB}, we studied the convergence of the HFB energy for four deformed configurations of $^{240}\mathrm{Pu}$ using a fixed box size of $Z_\mathrm{box}=25,\mathrm{fm}$ and $R_\mathrm{box}=12.5,\mathrm{fm}$. Here, we focus specifically on configuration \#3 with deformation $(q_{20}, q_{30}) = (310\,\mathrm{b}, 40\,\mathrm{b}^{3/2})$, and analyze the convergence of the HFB energy as $\nzmax$ is increased from 28 to 64 in steps of 4, for three different box sizes:
\begin{itemize}
\item $Z_\mathrm{box} = 25\,\mathrm{fm}, R_\mathrm{box} = 12.5\,\mathrm{fm}$,
\item $Z_\mathrm{box} = 27\,\mathrm{fm}, R_\mathrm{box} = 13.5\,\mathrm{fm}$,
\item $Z_\mathrm{box} = 29\,\mathrm{fm}, R_\mathrm{box} = 14.5\,\mathrm{fm}$.
\end{itemize}

\begin{figure}[!htb]
    \centering
    \includegraphics[width=1.0\columnwidth]{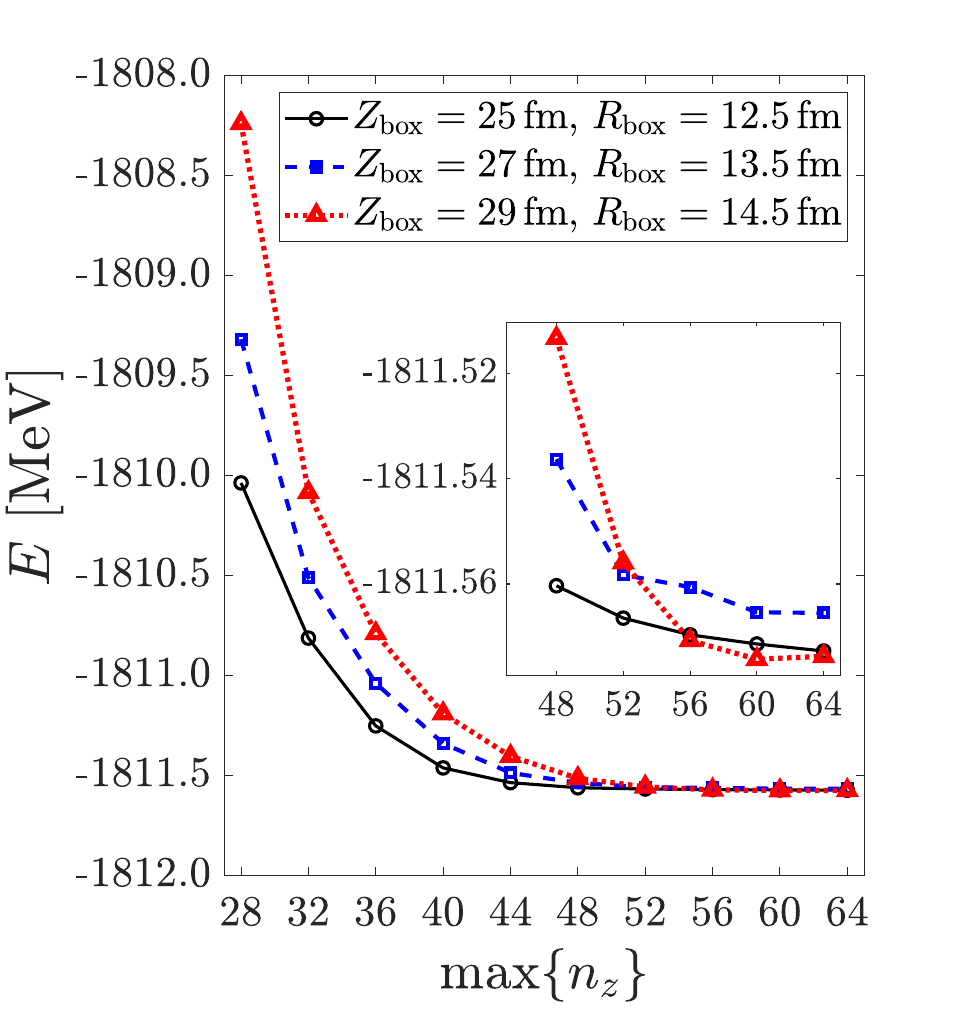}
    \caption{Convergence of the total HFB energy as $\nzmax$ is increased for static HFB calculations at fixed deformation $(q_{20}, q_{30}) = (310\,\mathrm{b},\,40\,\mathrm{b}^{3/2})$. Three different box sizes are used.}
    \label{fig:Evsmaxnz3boxes}
\end{figure}

Figure~\ref{fig:Evsmaxnz3boxes} shows the convergence of the total HFB energy. As expected, for all three box sizes, the energy converges to the same value within a discrepancy of approximately $\sim 7\,\mathrm{keV}$ when using a large basis with $\nzmax=64$. We also observe that the convergence is slower for larger box size, which is again expected, since the corresponding oscillator lengths defined in Eqs.~\eqref{eq:bz} and \eqref{eq:br} are larger for fixed value of $\nzmax$ compared to the smaller box size.

\begin{figure}[!htb]
    \centering
    \includegraphics[width=1.0\columnwidth]{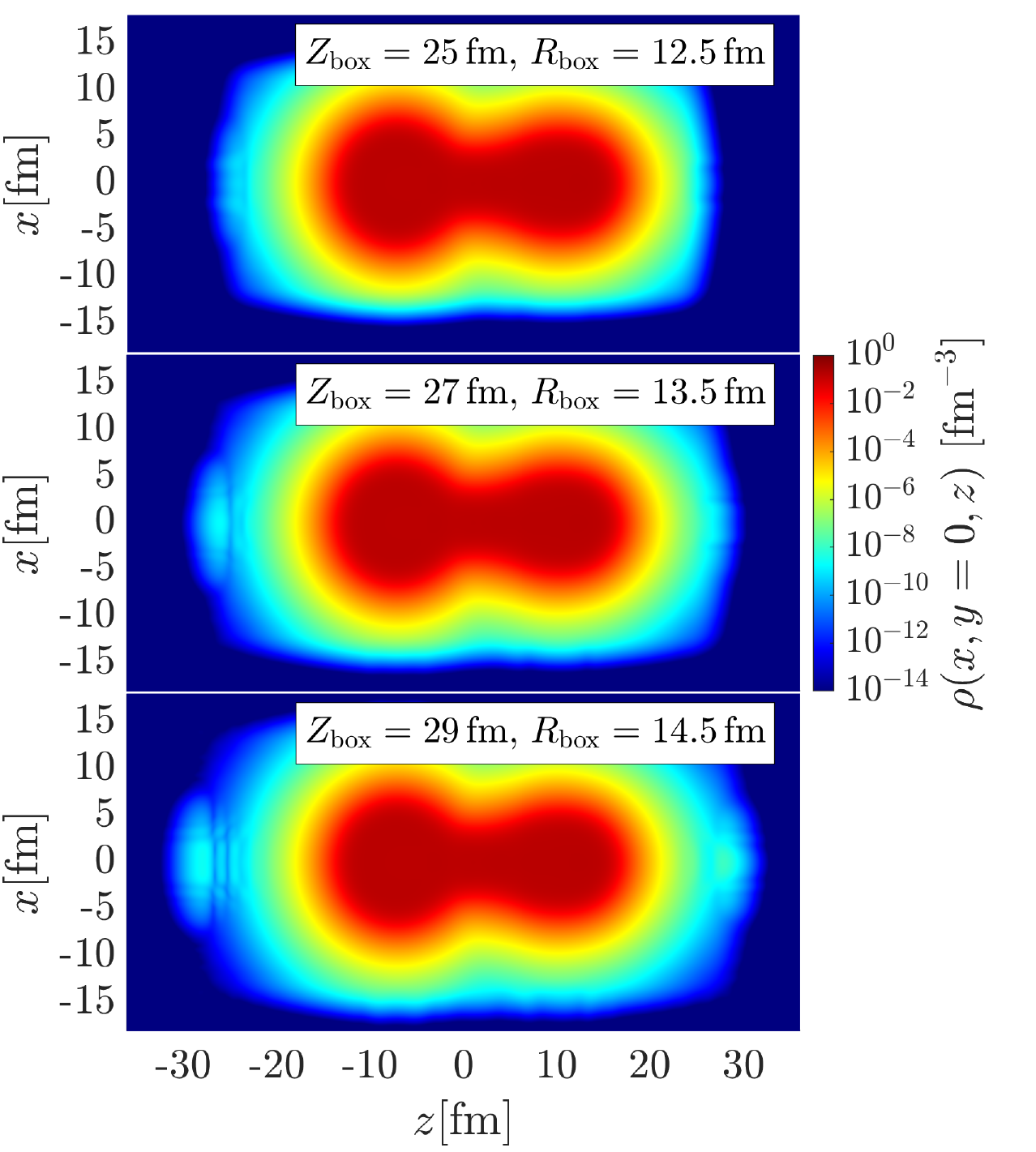}
    \caption{Density profiles (logarithmic scale) obtained from three static HFB calculations at fixed deformation $(q_{20}, q_{30}) = (310\,\mathrm{b},\,40\,\mathrm{b}^{3/2})$, where the only difference lies in the choice of the box parameters $Z_\mathrm{box}$ and $R_\mathrm{box}$.}
    \label{fig:DensitiesBoxes}
\end{figure}

To aid visualization, Figure~\ref{fig:DensitiesBoxes} shows the density profiles on a logarithmic scale for three different box sizes, all computed with $\nzmax = 64$. As we increase the size of the box, the density hardly changes. Only at the tips of the nucleus, around $z < -25 $ fm and $z > 25$ fm, is there a visible impact of the box size, but these changes correspond to values of the density lower than $10^{-9}$ fm$^{-3}$ and will have a negligible impact on the physics we seek to describe.

\section{Convergence with respect to origin shift}
\label{app:shift}

Recall that the multipole moment $q_{\ell 0}$ is defined as the expecation value $q_{\ell 0} = \braket{\hat{Q}_{\ell 0}(x,y,z)}$ of the operator
\begin{equation}
\hat{Q}_{\ell 0}(x,y,z) = |\rvec|^\ell \,Y_{\ell,0}(\theta,\varphi),    
\end{equation}
with $Y_{\ell,0}(\theta,\varphi)$ the spherical harmonics \cite{varshalovich1988quantum} and the spatial vector $\rvec$ reads, in spherical coordinates,
\begin{equation}
\rvec = (x,y,z) = (\,|\rvec| \sin\theta\cos\varphi \,,\, |\rvec| \sin\theta \sin\varphi \,,\, |\rvec|\cos\theta\,).
\end{equation}
The operators $\hat{Q}_{\ell 0}(x,y,z)$ are defined with an arbitrary origin, which coincides with the origin of the basis functions. In principle, if we constrain the values of the multipole moments defined with respect to some other origin, we should obtain the same solution for scalar HFB observables provided the basis is large enough. In other words, in the limit of infinite bases, scalar observables should be translationally invariant. In particular, if we constrain the expectation value of the operators $\braket{\hat{Q}_{\ell 0}(x,y,z-z_0)}$ shifted along the $z$ axis by the distance $z_0$, then the calculated HFB energy should converge to the same value, regardless of the value of $z_0$. 

We will again focus our attention on configuration \#3 from Sec.~\ref{subsec:conv_HFB} with deformation $(q_{20},q_{30}) = (310\,\mathrm{b},40\,\mathrm{b}^{3/2})$. We use the same box parameters: $Z_\mathrm{box}=25\,\mathrm{fm}$, $R_\mathrm{box}=12.5\,\mathrm{fm}$, and increase the $\nzmax$ parameter from 28 to 64 in steps of 4. Three values of the shift distance $z_0$ are considered: $z_0=-5.0\,\mathrm{fm}$, $z_0=0\,\mathrm{fm}$ and $z_0=+2.5\,\mathrm{fm}$. The $z_0=0\,\mathrm{fm}$ case is the one discussed in Sec.~\ref{subsec:conv_HFB}.

\begin{figure}[!htb]
    \centering
    \includegraphics[width=1.0\columnwidth]{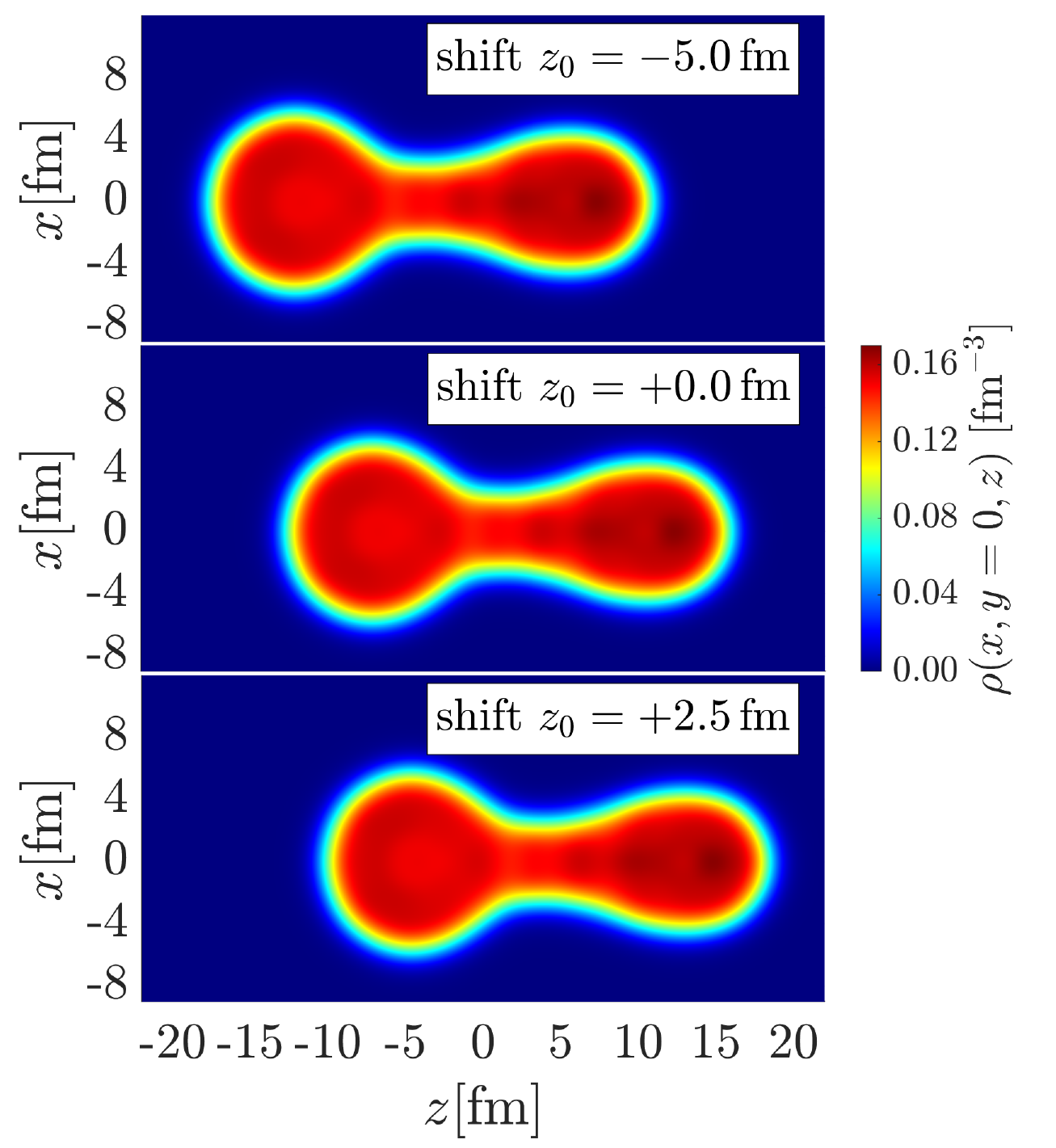}
    \caption{Density profiles obtained from the three static HFB calculations at fixed deformation $(q_{20}, q_{30}) = (310\,\mathrm{b},\,40\,\mathrm{b}^{3/2})$, where the only difference lies in the choice of origin $z_0$ in the definition of the multipole operators, $\hat{Q}_{\ell 0}(x,y,z - z_0)$.}
    \label{fig:DensitiesShifted}
\end{figure}

\begin{figure}[!htb]
    \centering
    \includegraphics[width=1.0\columnwidth]{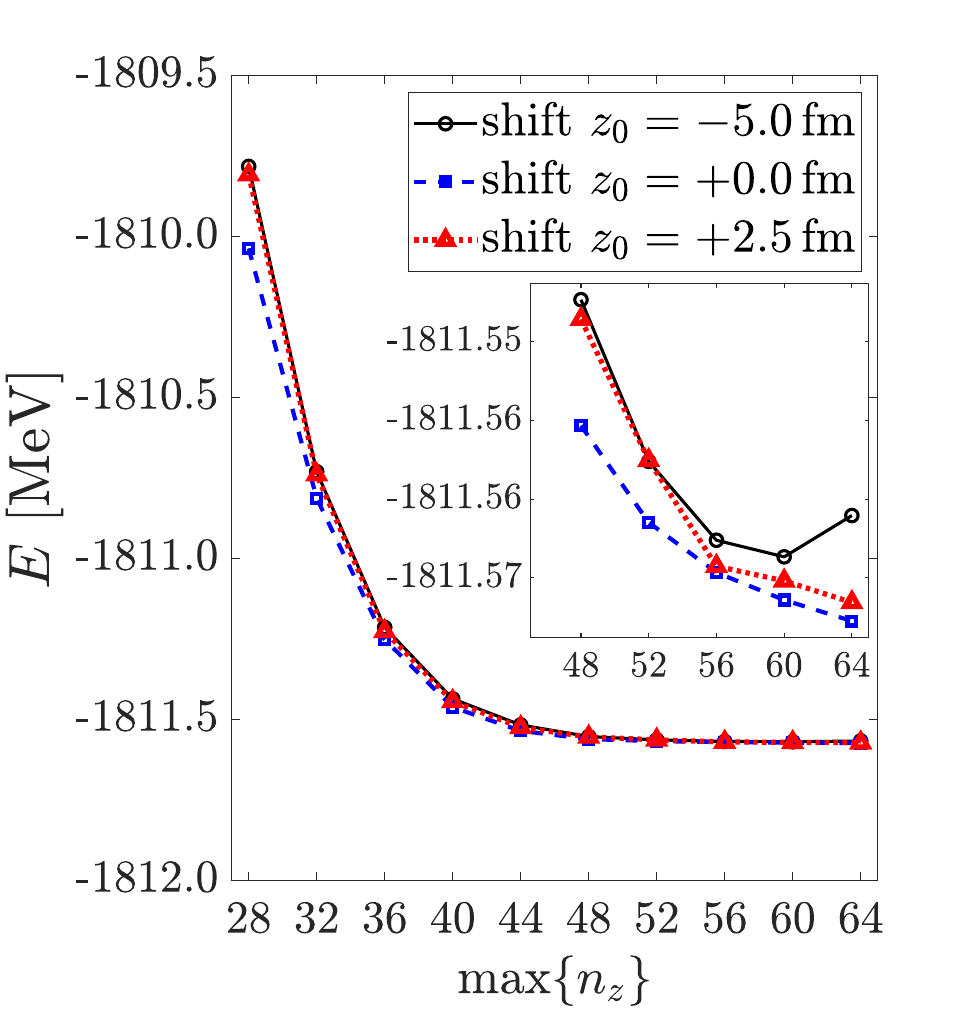}
    \caption{Convergence of total HFB energy as $\nzmax$ increases for three shift distances $z_0$ shown in Fig.~\ref{fig:DensitiesShifted}.}
    \label{fig:Evsmaxnz3shifts}
\end{figure}

Figure \ref{fig:DensitiesShifted} shows the density profiles for the three shifted cases. As expected, the density profiles look identical, up to an axial shift by distance $z_0$. In Figure~\ref{fig:Evsmaxnz3shifts} we show the convergence of total HFB energy when $\nzmax$ is increased. For all three shift distances, the total energies differ by only $\sim7\,\mathrm{keV}$ when a very large basis with $\nzmax=64$ is used. However, it is worth noting that if one uses a large shift distance, e.g. $z_0=\pm15\,\mathrm{fm}$, such that the density profile cannot fit within a cylinder of height $2Z_\mathrm{box}=50\,\mathrm{fm}$ and diameter $2R_\mathrm{box}=25\,\mathrm{fm}$, the converged energy will differ from the ones shown in Fig.~\ref{fig:Evsmaxnz3shifts}.
We also confirm that other observables, e.g. unconstrained multipole moments, converge to the same value with similar relative difference.

\section{Convergence with respect to time step}
\label{app:time}

In this Appendix, we study how to choose an appropriate time step for the time integration of the TDHFB equation \eqref{eq:TDHFB}. As discussed in Sec.~\ref{subsec:TimeEvolution}, introducing a quasiparticle energy cutoff inevitably breaks energy conservation—not only numerically, but also at the formal level of the theory. Therefore, for the purpose of this convergence study, we artificially include all quasiparticle states in the calculation. While this results in unrealistically large pairing correlations, it allows us to isolate numerical effects without the complications introduced by quasiparticle truncation.

As was noticed in Ref.~\cite{avez2008pairing}, pairing functionals $\mathcal{E}_{\rm pair}$ of the form
\begin{equation}
    \mathcal{E}_\mathrm{pair}(\rvec) = 
    \sum_{q\in\{n,p\}} \frac{V_0^{(q)}}{4}\left[1- \eta^{(q)}\left(\frac{\rho(\rvec)}{\rho_c}\right)^\gamma\right]
    \tilde{\rho}^{(q)}(\rvec),
\end{equation}
ensure conservation over time of the average number of particle, i.e. $\partial_t \braket{N^{(q)}} = \partial_t \operatorname{Tr}\left[\rho^{(q)}\right] = 0$. It is worth noting that the average particle number is also conserved even
when a quasiparticle cutoff is introduced.

Our criteria for choosing the time step are threefold: (i) how well the total energy is conserved, (ii) how well the neutron number is conserved, and (iii) how well the proton number is conserved throughout the TDHFB evolution. Since both energy and particle number are conserved in the full theory, deviations from these quantities provide a clear diagnostic of numerical accuracy.

For this test, we use a basis with $Z_\mathrm{box}=25\,\mathrm{fm}$, $R_\mathrm{box}=12.5\,\mathrm{fm}$ and $\nzmax=24$, and start the time evolution with an initial condition having deformation $(q_{20},q_{30})=(175\,\mathrm{b},15\,\mathrm{b}^{3/2})$. The same ${}^{240}$Pu nucleus with SkM$^{*}$ EDF and mixed pairing like in Sec.~\ref{subsec:conv_HFB} is considered, but without a quasiparticle cutoff. We use two time steps: one reasonably large $\Delta t = 0.5\,\mathrm{fm}/c$, and one very small $\Delta t = 0.025\,\mathrm{fm}/c$, for the Runge-Kutta RK4 time integration scheme.

\begin{figure}[!htb]
    \centering
    \includegraphics[width=1.0\columnwidth]{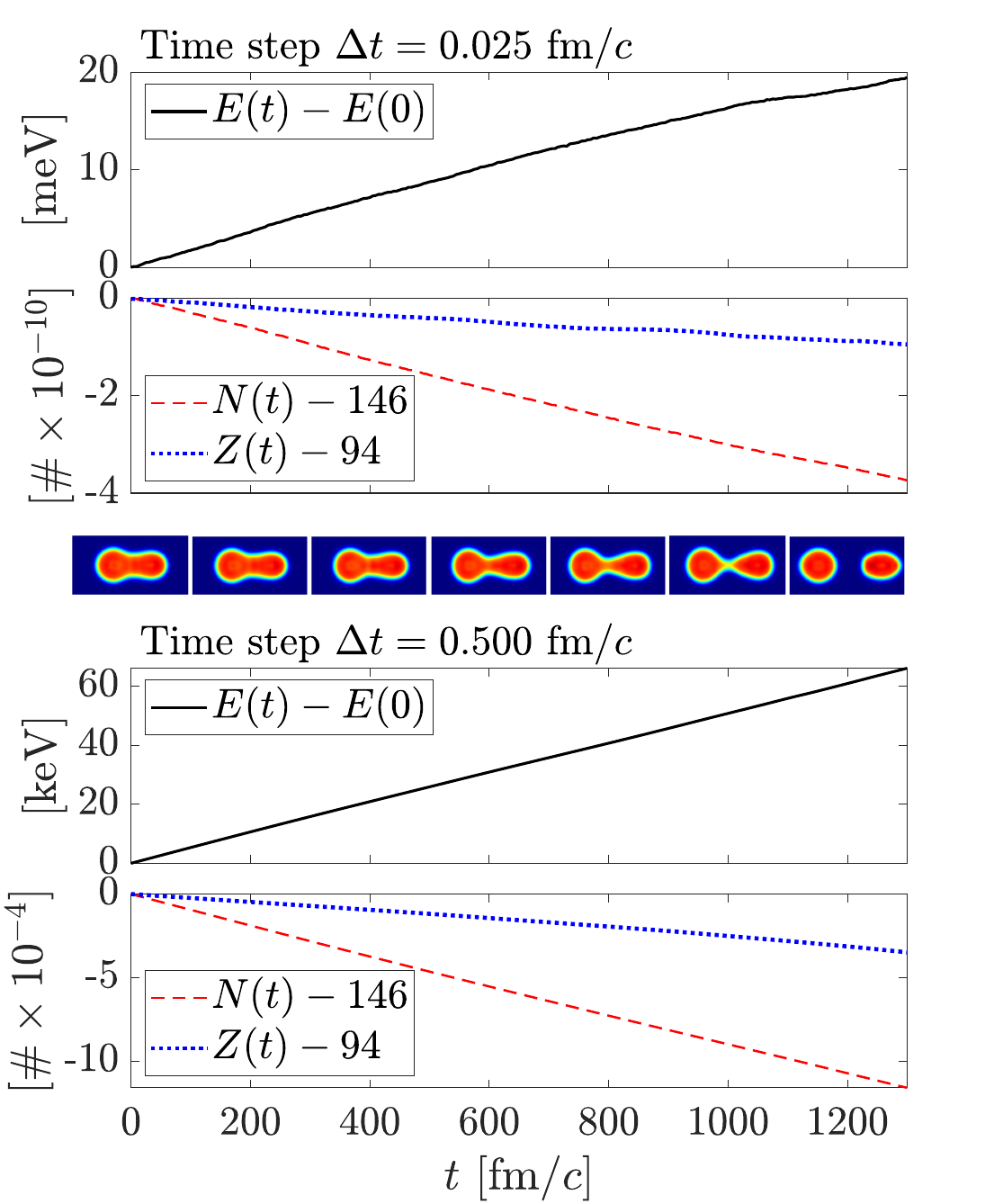}
    \caption{Numerical test of energy and particle number conservation for two time steps, $\Delta t = 0.025\,\mathrm{fm}/c$ and $\Delta t = 0.5\,\mathrm{fm}/c$, starting from a deformed configuration with $(q_{20},q_{30}) = (175\,\mathrm{b},15\,\mathrm{b}^{3/2})$ and no quasiparticle cutoff.
   Time evolution of the density is shown in snapshots separated by time intervals of $200\,\mathrm{fm}/c$. The color scale is the same as in Fig.~\ref{fig:4configurations}.
   }
    \label{fig:TimeSteps}
\end{figure}

Figure~\ref{fig:TimeSteps} shows the time evolution of the total energy $E(t)$ and particle numbers $N(t)$ and $Z(t)$ for two different time steps. For the smaller time step $\Delta t = 0.025\,\mathrm{fm}/c$, the total energy is conserved to within $20\,\mathrm{meV} = 20 \times 10^{-9}\,\mathrm{MeV}$ from the initial configuration to scission, while the particle numbers are conserved to within $4 \times 10^{-10}$ nucleons. These correspond to relative errors on the order of $10^{-12}$ for both energy and particle number, demonstrating the excellent numerical stability of the \texttt{AxialHOHFB} code.

For the larger time step $\Delta t = 0.5\,\mathrm{fm}/c$, the total energy is conserved to within $60\,\mathrm{keV}$, and the particle numbers to within $10^{-3}$ nucleons. While less precise, this level of conservation remains acceptable for practical calculations, as it provides a reasonable compromise between computational efficiency and numerical accuracy.

\section{Convergence of the continuity equation}
\label{app:continuity}

In this Appendix, we study how well the continuity equation relating the density and current is numerically satisfied, namely, the validity of the equation
\begin{equation}
\label{eq:ContinuityEq}
    {\partial_t\rho(\rvec,t)} + \frac{\hbar}{m}\nabla \cdot \jvec(\rvec,t) = 0.
\end{equation}
To simplify the analysis, we first consider the time-dependent Hartree-Fock (TDHF) case, i.e., without pairing correlations, and we ignore the spin dependence of the basis functions and the spin-dependent terms of the energy density functional. The full, non-local TDHF equation in coordinate space reads
\begin{equation}
\label{eq:TDHFcoordinate}
    i\hbar \,{\partial_t\rho}(\rvec,\rvec') = [\, h(\rvec,\rvec') \,,\, \rho(\rvec,\rvec')\,],
\end{equation}
where the non-local density matrix is
\begin{equation}
    \rho(\rvec,\rvec') = \sum_{ij}\rho_{ij}\, \phi_{i}(\rvec) \phi_{j}^*(\rvec').
\end{equation}

It was shown in Ref.~\cite{engel1975timedependent} that the diagonal part $\rvec = \rvec'$ of the TDHF equation~\eqref{eq:TDHFcoordinate} is equivalent to the continuity equation~\eqref{eq:ContinuityEq}. In other words, if the potential in the nuclear Hamiltonian is local, $V(\rvec_1,\rvec_2,\rvec'_1, \rvec'_2) = V(\rvec_1,\rvec_2) \delta(\rvec'_1 - \rvec_1)\delta(\rvec'_2 - \rvec_2)$, then the continuity equation is satisfied \cite{blaizot1985quantum}. When working with EDFs that are not strictly derived from a Hamiltonian, ensuring the validity of the continuity equation requires additional constraints on some of the terms of the functional. For instance, in the case of Skyrme functionals, terms such as $\rho\tau$ and $\jvec^2$ must appear in the EDF as the specific gauge invariant combination: $\rho\tau - \jvec^2$ ~\cite{dobaczewski1995timeodd}. Analogous constraints apply to the time-even, spin-dependent terms and on the time-odd parameters of the Skyrme EDF. The continuity equation is also satisfied in the TDHFB case for local pairing interaction, since in this case the extra term $\Delta\kappa^* - \Delta^*\kappa$ appearing on the right-hand side of \eqref{eq:TDHFcoordinate} also vanishes \cite{blaizot1985quantum}. Recall that the continuity equation does not hold in the special case of TDHF+BCS as shown in \cite{scamps2012pairing}.

However, when solving the TDHF equation in configuration space,
\begin{equation}
\label{eq:TDHFbasis}
    i\hbar \,{\partial_t\rho_{ij}} = \left[ h, \rho \right]_{ij},
\end{equation}
the continuity equation is not satisfied whenever a truncated basis is used. To demonstrate this, we focus for simplicity only on the derivative-free part of the Skyrme potential. In that case, the matrix elements of the single-particle Hamiltonian are given by
\begin{equation}
\label{eq:hk1k2Skyrme}
    h_{ij} = \int d\mathbf{x} \int d\mathbf{x}' \, \phi_{i}^*(\mathbf{x}) \, U(\mathbf{x}) \delta(\mathbf{x}-\mathbf{x}') \, \phi_{j}(\mathbf{x}'),
\end{equation}
for some real Skyrme potential $U(\rvec)$. Multiplying Eq.~\eqref{eq:TDHFbasis} by $\phi_{i}(\rvec) \phi_{j}^*(\rvec')$ and summing over the $i,j$ indices yields
\begin{equation}
    i\hbar\,{\partial_t\rho}(\rvec,\rvec') = \sum_{ijk}
    \left( h_{ik}\rho_{kj} - \rho_{ik}h_{kj} \right) \phi_{i}(\rvec) \phi_{j}^*(\rvec').
\end{equation}
Substituting Eq.~\eqref{eq:hk1k2Skyrme} into the above expression, it is straightforward to show
\begin{align}
i\hbar\,{\partial_t\rho}(\rvec,\rvec') 
=&  +\sum_{i} \phi_{i}(\rvec) \int d\mathbf{x}\, U(\mathbf{x})\rho(\mathbf{x},\rvec')\,\phi_{i}^*(\mathbf{x})
\nonumber \\
& - \sum_{j} \phi_{j}^*(\rvec') \int d\mathbf{x}\, U(\mathbf{x})\rho(\rvec,\mathbf{x})\,\phi_{j}(\mathbf{x}).
\end{align}
Using the property $\rho^{*}(\rvec,\rvec') = \rho(\rvec',\rvec)$ and the completeness relation for the basis functions, one obtains
\begin{equation}
    i\hbar \,{\partial_t\rho}(\rvec,\rvec') = \left[
    \, U(\rvec) \delta(\rvec-\rvec')\, , \, \rho(\rvec,\rvec')\,
    \right].
\end{equation}
Similar steps can be followed for the other terms in the Skyrme potential. Altogether, one can show that Eq.~\eqref{eq:TDHFbasis} implies Eq.~\eqref{eq:TDHFcoordinate}, and consequently the continuity equation~\eqref{eq:ContinuityEq}. A similar derivation applies to the TDHFB equation with a zero-range pairing force.

In summary, if the TDHFB equation is solved in configuration space using a complete basis, the continuity equation is satisfied exactly. However, the derivation relies on the completeness relation of the single-particle basis, which only applies for full, infinite bases of the Hilbert space of squere-integrable functions. Whenever the basis is truncated, the continuity equation is only approximately satisfied.

It is worth noting that the constant $\frac{\hbar}{m}$ in the continuity equation~\eqref{eq:ContinuityEq} originates from the kinetic energy term in the energy density functional, which takes the form $\frac{\hbar^2}{2m}\tau$. Since it is common practice to perform a one-body center-of-mass correction by replacing the kinetic term with $\left(1 - \frac{1}{A}\right)\frac{\hbar^2}{2m}\tau$, the constant in the continuity equation should accordingly be modified. In that case, the prefactor $\frac{\hbar}{m}$ needs to be replaced with $\left(1 - \frac{1}{A}\right)\frac{\hbar}{m}$.

\begin{figure}[!htb]
    \centering
    \includegraphics[width=1.0\columnwidth]{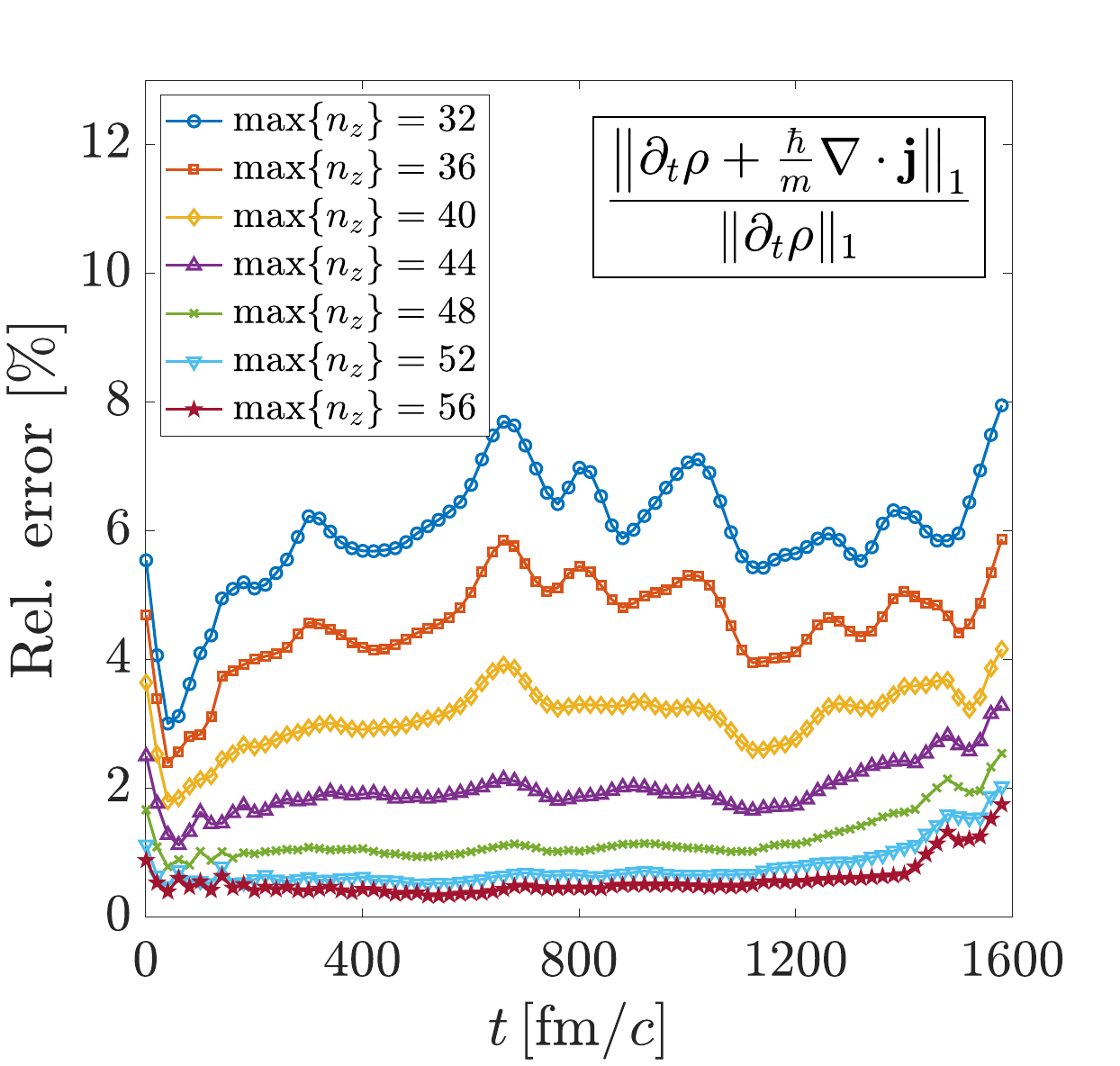}
    \caption{Numerical test of the continuity equation corresponding to trajectory \#1 (described in Sec.~\ref{subsec:conv_TDHFB}), shown as a function of the increasing basis size parameter $\nzmax$. Scission occurs at time $t=1400\,\mathrm{fm}/c$.  }
    \label{fig:ContinuityEq}
\end{figure}

To assess how well the continuity equation is numerically verified, we use the same setup as in the trajectory \#1 from Sec.~\ref{subsec:conv_TDHFB} and compute the relative error in the $L_1$ norm
\begin{equation}
    \frac{\left\| \partial_t \rho(\rvec,t) + \frac{\hbar}{m} \nabla \cdot \jvec(\rvec,t) \right\|_{1}}{\left\| \partial_t \rho(\rvec,t) \right\|_{1}},
\end{equation}
as a function of basis size, by increasing the parameter $\nzmax$ from 32 to 56. The time derivative $\partial_t\rho$ is approximated by a five-point central finite difference formula. Figure \ref{fig:ContinuityEq} shows the results. As expected, the continuity equation is only approximately satisfied when a finite basis is used, with the accuracy improving as the basis size increases. For instance, using a basis with $\nzmax=56$ (corresponding to approximately $N_{\mathrm{total}}=25{,}000$ basis vectors) results in a relative $L_1$ error below $1\%$ from saddle to scisssion.

\bibliography{zotero_output,books,others}

\end{document}